\documentclass[mnsc,nonblindrev]{informs3}

\OneAndAHalfSpacedXI 

\usepackage{comment}
\usepackage{mathtools}
\usepackage{custom_informs}
\usepackage[printwatermark]{xwatermark}
\usepackage{array}
\usepackage{wrapfig}
\usepackage{multirow}
\usepackage{tabu}
\usepackage{float}
\usepackage[normalem]{ulem}
\useunder{\uline}{\ul}{}
\usepackage{dcolumn}
\usepackage{xcolor}
\newcolumntype{V}[1]{>{\centering\arraybackslash}m{#1}}
\newcolumntype{L}[1]{>{\arraybackslash}m{#1}}
\usepackage{pdfpages}

\usepackage{subfigure}
\usepackage{graphicx}
\usepackage[justification=centering]{caption}

\newcolumntype{N}{@{}m{0pt}@{}}
\usepackage{booktabs}
\usepackage{threeparttablex} 
\usepackage{epstopdf}

\usepackage{natbib}
 \bibpunct[, ]{(}{)}{,}{a}{}{,}%
 \def\bibfont{\small}%
\TheoremsNumberedThrough     
\ECRepeatTheorems

\EquationsNumberedThrough    

\MANUSCRIPTNO{}

\usepackage{xr}
\begin{document}


\RUNAUTHOR{Chan et al.} 

\RUNTITLE{Inverse Optimization for Clinical Pathway Concordance}



\TITLE{An Inverse Optimization Approach to Measuring Clinical Pathway Concordance}

\ARTICLEAUTHORS{%

\AUTHOR{Timothy C. Y. Chan}
\AFF{Department of Mechanical and Industrial Engineering, University of Toronto, Toronto, Ontario M5S3G8, Canada, 
\EMAIL{tcychan@mie.utoronto.ca}}

\AUTHOR{Maria Eberg}
\AFF{IQVIA, Kirkland, Quebec H9H 5M3, Canada, 
\EMAIL{Maria.eberg@mail.McGill.ca}}

\AUTHOR{Katharina Forster, Claire Holloway, Luciano Ieraci }
\AFF{Ontario Health (Cancer Care Ontario), Toronto, Ontario M5G2C1, Canada, 
\EMAIL{\{Katharina.Forster, Claire.Holloway, Luciano.Ieraci\}@cancercare.on.ca}}

\AUTHOR{Yusuf Shalaby, Nasrin Yousefi}
\AFF{Department of Mechanical and Industrial Engineering, University of Toronto, Toronto, Ontario M5S3G8, Canada, 
\EMAIL{\{yusuf.shalaby, nasrin.yousefi\}@mail.utoronto.ca}}

} 

\ABSTRACT{%
Clinical pathways outline standardized processes in the delivery of care for a specific disease. Patient journeys through the healthcare system, though, can deviate substantially from these pathways. Given the positive benefits of clinical pathways, it is important to measure the concordance of patient pathways so that variations in health system performance or bottlenecks in the delivery of care can be detected, monitored, and acted upon. This paper proposes the first data-driven inverse optimization approach to measuring pathway concordance in any problem context. Our specific application considers clinical pathway concordance for stage III colon cancer. We develop a novel concordance metric and demonstrate using real patient data from Ontario, Canada that it has a statistically significant association with survival. Our methodological approach considers a patient's journey as a walk in a directed graph, where the costs on the arcs are derived by solving an inverse shortest path problem. The inverse optimization model uses two sources of information to find the arc costs: reference pathways developed by a provincial cancer agency (primary) and data from real-world patient-related activity from patients with both positive and negative clinical outcomes (secondary). Thus, our inverse optimization framework extends existing models by including data points of both varying ``primacy'' and ``alignment''. Data primacy is addressed through a two-stage approach to imputing the cost vector, while data alignment is addressed by a hybrid objective function that aims to minimize and maximize suboptimality error for different subsets of input data.
}%


\maketitle

%


\section{Introduction}
\label{sec:intro}

Clinical pathways, also known as disease pathways or integrated care pathways, are care plans outlining standardized processes in the delivery of care for a specific cohort of patients over a specific period of time \citep{campbell1998integrated,de2006defining,kinsman2010clinical}. They prescribe a sequence of steps for managing a clinical process with the aim of optimizing patient or population-level outcomes such as survival, cost, and wait times. Generally, they are defined by medical experts and multidisciplinary teams based on ``evidence from local, national and international clinical practice guidelines'' \citep{aboutmaps}. They support multidisciplinary care and evidence-based clinical practice and facilitate variance management \citep{vanhaecht2006clinical}. Clinical pathways are widely used across a range of medical domains including cancer \citep{evans2013improving,schmidt2018national,ling2018declining}, mental health \citep{samokhvalov2018outcomes,courtney2020way}, and surgical recovery \citep{van2018development}. With applicability to screening, diagnostic, and treatment processes, they have been shown to be effective in improving patient survival and satisfaction, wait times, in-hospital complications, hospital length of stay, and cost of care \citep{rotter2010clinical,rotter2012effects, panella2003reducing,schmidt2018national,vanounou2007deviation,van2013dynamic}.
Given the importance of clinical pathways in promoting best practices and standardizing or streamlining care, there is significant interest in developing quantitative metrics to measure the concordance of patient-traversed pathways against the recommended pathways, which we will refer to as \emph{reference pathways}. Such metrics assist in monitoring variations in the health system and across the care continuum, identifying bottlenecks or changes in the delivery of care, and ultimately providing data-driven evidence to inform health policy decisions.

Existing methodologies for measuring similarity between pathways have primarily leveraged process mining, a family of techniques that connect business process management with data mining algorithms \citep{bose2010trace,adriansyah2011conformance,yang2017data,yang2016duration}. General similarity metrics are typically based on edit distance algorithms \citep{van2010measuring,forestier2012classification,yan2018aligning,williams2014using}, which compare two processes represented as a sequence of distinct activities by counting the minimum number of operations needed to transform one into the other. Existing approaches in the literature are either unweighted, which implicitly assume all deviations are equally disadvantageous, or subjectively weighted, which is sensitive to misspecification. Given these limitations, our focus is on developing a weighted concordance metric with objectively derived or data-driven weights.

There are two sources of information that can be used to derive weights for a concordance metric: reference pathways and data from real-world patient activity. Reference pathways should be the primary information source since they are derived from best available medical evidence and the goal is to measure system performance against recommendations. Furthermore, concordance measurement in a health system is ideally optimized for a hierarchy of outcomes, with clinical outcomes usually given priority. Thus, reference pathways should be used directly to derive weights for the concordance metric and for the metric to be strongly associated with outcomes. Real-world patient activity, in the form of individual patient pathways, can serve as a secondary data source to fine tune the weights. However, patient pathways should not be the primary data source since there can be significant deviations from recommendations, and outcomes typically depend on many other factors besides pathway concordance. Overall, the weights should maximize concordance when a reference pathway is measured against itself while scoring discordant pathways lower. The problem of finding such weights can thus be posed naturally as an \emph{inverse optimization} problem.

In this paper, we use inverse optimization to develop a data-driven concordance metric for measuring clinical pathway concordance that has a statistically significant association with clinical outcomes. Our modeling approach progresses in two phases: the first phase identifies a cost vector that maximizes fitness against the reference pathways; the second phase fine-tunes the cost vector so it further maximizes fitness against the patient pathways with positive outcomes and minimizes fitness against those with negative outcomes.

We emphasize that our goal is to measure performance of the health system, rather than appropriateness of individual patient decisions. Thus, our metric measures the extent to which patients are concordant with reference pathways. Once we have this measure for different subpopulations (geographic, demographic, intervention-based, etc), policy makers can investigate the roots of discordance. 
For example, issues like congestion should be investigated after system performance is quantified. It should not be accounted for a priori since that would inaccurately characterize system performance by giving a more optimistic picture of health system performance than otherwise exists. Similarly, patient-specific discordance should be investigated post-hoc to see whether the discordance was inappropriate or not. Overall, our metric can help pinpoint discordant activity in a complex health system, which can support policy makers in crafting informed, data-driven decisions about health system interventions, prioritizing them for implementation, and modeling the impact of these interventions on concordance. 

We model a patient's journey through the healthcare system as a walk in a directed graph in which each node represents an activity that the patient can undertake. Patients incur ``costs'' by visiting nodes and traversing arcs through the network, which allows us to model both the costs associated with undertaking (or missing) certain activities, as well as sequencing costs for progressing from a specific activity to the next. A reference pathway can be considered an optimal solution of an appropriately formulated minimum cost network flow problem, a shortest path problem in particular. Given that the ``forward'' problem is a network flow problem, we can formulate the ``inverse'' problem tractably using linear programming duality \citep{ahuja2001inverse}. The goal of the inverse optimization model is to find arc costs such that all reference pathways are optimal for the resulting network. If there does not exist a cost vector that makes all reference pathways simultaneously optimal, then the costs should be such that a measure of aggregate error (e.g., optimality gap) is minimized. Once the arc costs are determined, any input pathway can be scored based on the cost of the associated directed walk through the network. In other words, these arc costs form the weights used in the concordance metric. 

For a practical application of our inverse optimization framework, we study clinical pathway concordance for stage III colon cancer pathways developed by Cancer Care Ontario. Colon cancer is one of the most commonly diagnosed cancers and a leading cause of death from cancer worldwide for both men and women \citep{colonthird}. Cancer Care Ontario, a business unit of Ontario Health, is the Ontario government's principal advisor on cancer care. It directs and oversees healthcare funds for hospitals and other cancer care providers, enabling them to deliver high-quality, timely services and improved access to care. As such, Cancer Care Ontario's responsibilities include developing guidelines and standards related to the delivery of cancer care, as well as monitoring the performance of the cancer system in Ontario, Canada. Indeed, one of the stated initiatives of Cancer Care Ontario's Disease Pathway Management program is to ``measure whether care follows the pathway maps'' \citep{dpm} -- this is the problem that our paper addresses.

\subsection{Literature Review}

\subsubsection{Inverse Optimization.}

Modern inverse optimization methodologies aim to impute unknown parameters of an optimization problem to make observed decisions minimally suboptimal in either offline settings \citep{troutt2006behavioral, keshavarz2011imputing, chan2014generalized, bertsimas2015data, aswani2018inverse, esfahani2018data, chan2018inverse, chan2018multiple} or online settings where data arrives sequentially  \citep{dong2018generalized,barmann2018online,dong2020inverse}. In our clinical pathway concordance problem, all data are available a priori in the form of reference pathways and patient pathways. Thus, we use an ``offline'' inverse optimization model.

Inverse network flow problems are well-studied in the literature \citep{zhang1996network, yang1997inverse, zhang1998inverse, ahuja2001inverse,ahuja2002combinatorial}, in particular inverse shortest path problems \citep{burton1992instance, xu1995inverse, zhang1995column, burton1997inverse}. However, they primarily focus on a single, noiseless observed decision for which it is possible to determine parameters that make the decision exactly optimal. Inverse network flow problems with multiple noisy observations have received only limited attention to date \citep{farago2003inverse,zhao2015learning}. Due to the equality constraints corresponding to flow balance, the feasible region in network flow problems is not full dimensional, which means there may exist non-zero cost vectors that are orthogonal to the entire feasible region, making all feasible solutions optimal. The lower dimensionality of the feasible region has not been discussed in prior literature. Part of the reason might be that previous papers have assumed that the distance to some prior cost vector should be minimized \citep{burton1992instance, xu1995inverse, zhang1995column, burton1997inverse, ahuja2001inverse}, and this objective may discourage the inverse optimization model from identifying a cost vector that is orthogonal to the entire feasible region. In inverse optimization problems with multiple noisy observations, a more relevant objective is to minimize some measure of error in the model-data fit, and as a result, we need to directly account for the lower dimensionality of the feasible region in the constraints. 

Many previous papers focus on the case of non-negative cost vectors, which either facilitates the solution algorithm  or is required for a particular application \citep{farago2003inverse, burton1992instance, xu1995inverse, zhang1995column, burton1997inverse}. In our case, we do not restrict the sign of the cost vector, since activities can be either rewarding or costly in terms of pathway concordance. Additionally, the idea of a two-stage model, with different input datasets and a subset of data to minimize fitness against, has not been proposed in the inverse optimization literature.

\subsubsection{Measuring Pathway Concordance.}
Clinical pathway concordance measurement is often limited to a binary approach where patient pathways are audited against a set of predefined criteria or sequences of activities and are classified as either concordant or discordant \citep{cheng2012integrated,bergin2020concordance,forster2020can}. However, the selection of such criteria or sequences can be subjective and certain types of discordance may not be captured. In addition, the selected criteria may vary in the strength of their relationship to the measured outcome, and a binary application of concordance/discordance would not incorporate this important clinical information. Other studies that explore the effectiveness of the clinical pathways compared to traditional care have further simplified concordance measurement by dividing patients into two groups, before and after clinical pathway implementation, and measure the differences of the two groups in terms of the desired outcomes \citep{vanounou2007deviation,delaney2003prospective,panella2003reducing}. 

More granular methodologies for measuring similarity between pathways have primarily leveraged edit distance as a metric. Edit distance quantifies the similarity between two strings by counting the minimum number of operations required to transform one string to the other. There are many different variations of edit distance, distinguished by the operations they include \citep{navarro2001guided}. For example, the Longest Common Subsequence distance considers only insertions and deletions \citep{apostolico1987longest}. Levenshtein distance is the most common edit distance metric and allows insertion, deletion, and substitution \citep{levenshtein1966binary}. Damerau–Levenshtein distance allows insertion, deletion, substitution, and the transposition of two adjacent activities \citep{damerau1964technique}. All these edit distance metrics have been used to measure clinical pathway concordance \citep{van2010measuring,yan2018aligning,williams2014using}.


Finally, we note that these studies have focused on a single intervention at a single institution \citep{rotter2012quality}, although they could potentially be used at the population level. 

\subsection{Contributions}

Our main contributions in this paper are as follows:

\begin{enumerate}
    \item We propose the first inverse optimization-based approach for measuring pathway concordance in any problem context. Methodologically, our approach considers the novel setup where there are primary (reference pathways) and secondary (patient pathways) datasets, which are given differential consideration when imputing the cost vector. Furthermore, our approach is unique in that our data contains subsets that differ in their intended fitness with the cost vector: one that the cost vector aims to fit well and another that it does not.
    
    \item We develop a metric to assess patient pathway concordance. This metric uses the inversely optimized cost vector to calculate a normalized concordance score for a given flow vector (patient pathway). The concordance metric is simple to compute and easily interpretable, taking values between 0 and 1, with 1 representing perfect concordance.
    
    \item We provide an in-depth case study applying our inverse optimization-based models and metrics to real patient data from stage III colon cancer patients. In particular, we:
    \begin{enumerate}
        \item Undertake a rigorous survival analysis to show that there is a statistically significant association between pathway concordance and survival, validating the meaningfulness of our concordance metric;
        \item Validate that our survival analysis results are robust with respect to an unnormalized version of the metric;
        \item Demonstrate that our results are not highly sensitive to potential unobserved confounders;
        \item Demonstrate the value of our approach by showing that a metric developed using only reference pathways can yield an insignificant association between concordance and survival or even a significant association in the incorrect direction;
        \item Demonstrate that our inverse optimization-based concordance metric outperforms several baseline concordance measures based on edit distance.
   \end{enumerate}
    
    \item We develop a rigorous framework to fully apportion the discordance of any patient pathway to its ``detours'' from the reference pathways. This approach allows us to characterize the contribution of different types of detours, as well as the individual activities within those detours, to total discordance at the individual and population level. Applied to our data, we identify extra imaging and ED visits as primary sources of discordance, consistent with the literature. 
    We also go deeper to provide new insight into how discordance evolves over the course of a patient's journey: detours tend to be shorter earlier in the pathway and more complex with many additional discordant activities later in the pathway. Finally, we show how discordance in two different subpopulations, repeatedly screened versus unscreened patients, differs from population-level discordance.
\end{enumerate}

Although we focus on colon cancer specifically in this paper, our inverse optimization-based approach is generalizable to other patient cohorts and even beyond the healthcare context to general process management applications. Proofs are included in the Electronic Companion.

\section{Motivating a Graph-Based Inverse Optimization Approach}
\label{sec:comparison}

In this paper, we leverage network optimization and the corresponding graph-based representation of the problem to measure pathway concordance. To motivate our approach, we briefly highlight the differences between edit distance-based methods \citep{navarro2001guided} and our proposed approach. In particular, we show that our approach provides more granularity and modeling flexibility in capturing different types of discordance. 

First, consider unweighted edit distance algorithms, which measure distance as the minimum number of operations required to transform a source pattern into a target pattern \citep{van2010measuring,forestier2012classification,yan2018aligning,williams2014using}. The operations are defined on the activities in the pathway and usually include insertions, deletions, substitutions, and transpositions, although some algorithms may count a transposition as two substitutions. In contrast, our graph-based approach measures distance using the difference in the total cost along the arcs traversed by the two pathways. Note that edit distance has been extended to graph structures where the operations are applied to nodes and arcs \citep{gao2010survey}.
A key difference is that graph edit distance compares two pathways represented as two distinct graphs, whereas in our approach the two pathways are distinct walks in a single, pre-defined graph.

Our approach improves on edit distance algorithms by better characterizing pathway deviations and optimally estimating weights for those deviations. We refer to arcs that are in a reference pathway as ``concordant'' if they connect the desired activities in the order indicated according to clinical guidelines. In contrast, ``discordant'' arcs are those that do not connect activities contained in the reference pathways, or connect such activities out of sequence. The concordance of an observed pathway is then a function of the discordant arcs present and the concordant arcs absent. 

To illustrate the ability of our algorithm to better characterize pathway variation, we present concrete examples shown in Table \ref{tab:operations}. Consider a network with an artificial start node $S$ and end node $E$ that bookend all pathways and let $A-B-C$ be a reference pathway represented by the sequence of arcs $(S,A)$, $(A,B)$, $(B,C)$ and $(C,E)$. Now consider five example pathways listed in Table \ref{tab:operations}, all of which have an edit distance of one from $A-B-C$. Despite having the same edit distance, these pathways have different numbers of concordant and discordant arcs. Concordance measurement based on absence of concordant arcs and presence of discordant arcs implies that each pathway can have a different graph-based distance from the reference pathway. Furthermore, when considering pathways with an edit distance of two or more from the reference pathway, differences with graph-based distance measurement are magnified. For example, two adjacent deletions are different from two nonadjacent deletions in terms of the number of concordant and discordant arcs.

\begin{table}[!ht]
\caption{Comparing concordant and discordant arcs in pathways that have edit distance of one from the reference pathway}
\label{tab:operations}
\centering
\begin{tabular}{V{1.6cm}V{2.2cm}V{1.9cm}V{1.9cm}V{7cm}}
\toprule
  Pathway &  Operation     & Missing concordant arcs 
  & Extra discordant arcs & Pathway representation \\ \bottomrule
\multirow{4}{*}{A-B-C }  &  & & \\
   & -            & 0                                                                & 0                                                              & \raisebox{-0.3\totalheight}{\includegraphics[width=0.4\textwidth, height=9mm]{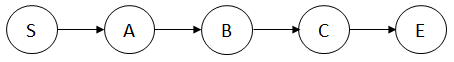}  }  \\ 
\hline
 \multirow{5}{*}{A-A-B-C} &  & & \\
 
 & Insertion (duplicated activity)      & 0                                                                & 1                                                              & \raisebox{-0.1\totalheight}{\includegraphics[width=0.4\textwidth, height=14mm]{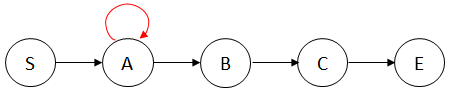}  }   \\ 
 &  & & \\
 \hline
 \multirow{5}{*}{X-A-B-C} &  & & \\
 & Insertion (discordant activity)    & 1                                                                & 2                                                              & \raisebox{-0.3\totalheight}{\includegraphics[width=0.4\textwidth, height=19.5mm]{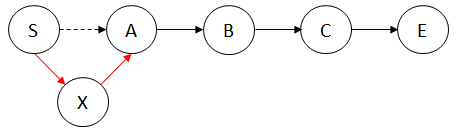}   }  \\ 

 \hline
 \multirow{4}{*}{B-C}   &  & & \\  
 & Deletion      & 2                                                                & 1                                                              & \raisebox{-0.4\totalheight}{\includegraphics[width=0.4\textwidth, height=11.5mm]{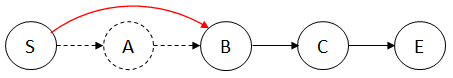} }    \\ 
 \hline
\multirow{5}{*}{X-B-C}    &  & & \\
& Substitution  & 2                                                                & 2                                                              & \raisebox{-0.4\totalheight}{\includegraphics[width=0.4\textwidth, height=19mm]{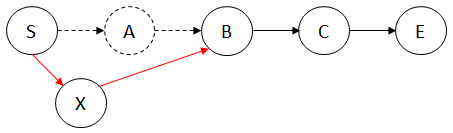}   }   \\  
\hline
 \multirow{4}{*}{B-A-C}    &  & & \\
 & Transposition & 3                                                                & 3                                                              & \raisebox{-0.5\totalheight}{\includegraphics[width=0.4\textwidth, height=15mm]{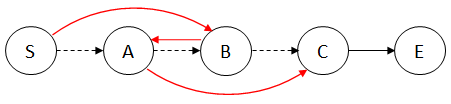} }    \\ 

 \bottomrule
\end{tabular}
\mbox{}\\{\footnotesize\emph{Note.} Missing concordant arcs are dashed. Extra discordant arcs are red.}
\end{table}

The preceding discussion assumes unweighted distance measurement. If we now consider a weighted distance metric, an issue arises where the cost of operations would need to be determined a priori and exogenously from the reference pathways. In weighted edit distance, weights or costs are associated with each type of operation. The distance between pathways then becomes the minimum cost of operations required to transform one into the other. Determining the costs of each operation based on a set of reference and non-reference pathways is complicated by the dependency of the operations on the costs. Expectation maximization can be used, but this approach does not guarantee convergence to a global optimum \citep{neuhaus2004probabilistic,ristad1998learning}. 
In contrast, our inverse optimization approach avoids this dependency and can determine optimal costs for each arc by assuming pathways are optimal solutions to an appropriately formulated optimization problem. The ability to optimally infer arc costs or weights, together with finer characterization of pathway variation using concordant and discordant arcs, are the main strengths of our inverse optimization approach.

\section{Inverse Optimization for Pathway Concordance Measurement}
\label{sec:inversemodel}

We first present a model of the clinical activity network representing the activities that a cancer patient can undertake during his or her cancer care journey. A patient journey is modeled as a walk through the network, accumulating costs along each arc traversed. Then, we present an inverse optimization approach to determining arc costs such that reference pathways are optimal solutions to an appropriate optimization problem defined on this network. 

\subsection{The Clinical Activity Network} \label{sec:network}

Let $\mathcal{N}$ be a set of $m$ nodes and let $\mathcal{A} \subseteq \mathcal{N} \times \mathcal{N}$ be a set of $n$ directed arcs in the network. Nodes represent clinical activities that can be undertaken by a patient. To model the cost of undertaking a specific activity $i$, we split each activity node into two nodes, $i_s$ and $i_e$, representing the start and end, respectively, of activity $i$, and then add an arc $(i_s,i_e)$ with cost $c_{i_s i_e}$. The remaining arcs connect the end node of some activity $i$ with the start node of another activity $j$. Each arc $(i_e,j_s)$ has an associated cost $c_{i_e j_s}$, representing the cost of undertaking activity $j$ immediately after activity $i$. The arc costs are unrestricted in sign. The set $\mathcal{A}$ also includes arcs of the form $(i_e,i_s)$, which allows the network to model a patient journey that involves consecutive repetitions of activity $i$. Finally, we add two artificial nodes representing the common start and end of each patient journey. Thus, $m$ is equal to two times the number of clinical activities plus two.


A reference pathway is represented as the solution of a shortest path problem on this network. We write this shortest path problem as a standard form linear optimization problem representing the corresponding network flow model: $\underset{\bx}{\text{min}} \{ \bc'\bx :  \bA \bx = \bb, \bx \ge \bzero\}$. We denote this model the ``forward'' model and will refer to it as $\mathbf{FO}(\bc)$. The decision vector $\bx \in \R^{n}$ describes the flow on each arc. The matrix $\bA \in \R^{(m-1) \times n}$ is the node-arc incidence matrix, but without the $m$th row, which corresponds to the flow balance constraint of the artificial end node. Thus, it is full rank. The parameter $\bb \in \R^{m-1}$ is a vector of zeros except with a $+1$ at the artificial start node. The objective function quantifies the minimum cost path from the artificial start to the artificial end node.

A patient journey will be represented as a walk on this network. Note that it may not necessarily be a path since patients may traverse an arc several times. Each journey through the network can be represented by a flow vector $\bx \in \R^n$, where the value of $x_{ij}$ is the number of times arc $(i,j)$ is traversed in the walk. In keeping with the medical terminology, we will refer to a patient journey as a ``pathway'', which in general should be thought of as a walk through the network. We will reserve the use of the phrase ``path'' for the mathematical concept of a walk without repeated nodes.

Note that with the current setup, costs are ``path-independent''. That is, the cost of traversing arc $(i,j)$ is equal to $c_{ij}$ regardless of how the walk arrives at node $i$. To model path-dependent costs, where the cost of completing activity $j$ immediately following activity $i$ depends on what activities were completed prior to $i$, we could modify the nodes in the network from representing a single activity to representing the sequence of activities from the artificial start node to the current activity. That is, imagine a network organized in ``layers'', where the first layer of nodes comprises all activities that can be reached from the start node in one step, the second layer of nodes comprises all sequences of activities that can be reached in two steps, etc. While the path-dependent approach provides richer modeling possibilities, the downside is an exponential explosion in the number of nodes in the network. Thus, we proceed with the network as currently defined (i.e., with path-independent costs) to maintain a parsimonious model, and instead integrate path-dependent behavior into the constraints of the inverse optimization model. 

\subsection{Inverse Optimization Model}
\label{sec:model}

Our goal is to identify a cost vector $\bc \in \R^n$ that minimizes the aggregate suboptimality of given reference pathways, assuming they represent shortest paths. We first present a basic inverse optimization model that provides an exact formulation of this problem. Then, we introduce a second formulation that uses patient pathways to refine the optimal cost vector returned by the first model.

\subsubsection{A Model Using Only Reference Pathways.}
\label{sec:model_ref}

Let $\hat{\mathcal{X}}^r=\{\hat\bx^r_1, \ldots, \hat\bx^r_R\}$ denote the set of $R$ reference pathways, each assumed to be feasible flow vectors for the forward problem. We follow a standard approach to formulating the inverse optimization problem using duality of linear optimization.  
Let $\bp \in \R^{m-1}$ be the dual vector associated with flow balance constraints. The following formulation is an inverse optimization model that minimizes the sum of squared absolute duality gaps induced by the cost vector $\bc$ and the $R$ reference pathways:
\begin{equation} \label{model:reference}
\begin{alignedat}{1}
\mathbf{IO}^{\textrm{ref}}(\hat{\mathcal{X}}^r): \quad \underset{\bc, \bp, \bepsilon^r}{\text{minimize}} & \quad \sum_{q=1}^{R}(\epsilon^r_q)^2\\
\textrm{subject to} & \quad \bA' \bp \le \bc,\\
& \quad \bc'\hat\bx^r_q=\bb'\bp+\epsilon^r_q, \quad  q=1, \ldots, R, \\
& \quad \|\bc\|_{\infty}= 1,\\
& \quad \bA \bc= \bzero.
\end{alignedat}
\end{equation}

The first constraint represents dual feasibility. The second constraint defines the duality gap $\epsilon^r_q$ for each reference pathway $\hat\bx^r_q$ as the difference between the primal and dual objective values. Each duality gap variable is nonnegative since the reference pathways are feasible for the forward problem. The third constraint is a normalization constraint to ensure the cost vector is not trivial (i.e., not zero). The fourth constraint is needed to ensure that $\bc$ lies in the lower dimensional space defined by the constraints $\bA\bx = \bb$. Since $\bc$ lies in the space defined by the intersection of the equality constraints ($\ba_i'\bx = b_i, \; i = 1, \ldots, m-1$), it must be orthogonal to the vectors defining those constraints ($\ba_i$). Without the constraint $\bA\bc = \bzero$, an optimal solution (i.e., cost vector) to formulation~\eqref{model:reference} could be orthogonal to the entire forward feasible region, which would render $\bc$ uninformative since every feasible flow vector would be optimal. 
Note that the constraint $\bA\bc=\bzero$ means that the optimal cost vector will be a circulation. 

To model additional application-specific considerations, such as user-specified rankings between certain activities or preferences for how a patient should begin or end her pathway, we will add a set of constraints on the cost vector, $\bc \in \mathcal{C}$, to formulation~\eqref{model:reference}. Specific examples of such constraints are provided in Section~\ref{sec:DPC}.

First, we prove that formulation~\eqref{model:reference} has an optimal solution for any non-trivial network.

\begin{lemma}
\label{lem:feasible}
$\mathbf{IO}^{\emph{ref}}(\hat{\mathcal{X}}^r)$ has an optimal solution if and only if there are at least two distinct paths from the start node to the end node.
\end{lemma}

Going forward, we assume that the network is non-trivial.
Next, we present an important property of an optimal solution to $\mathbf{IO}^{\textrm{ref}}(\hat{\mathcal{X}}^r)$.

\begin{proposition}
\label{prop:boundedness}
Let $(\bc^*,\bp^*, \bepsilon^{r*})$ be an optimal solution to formulation~\eqref{model:reference}. Then, the network with arc costs $\bc^*$ does not contain any directed negative cost cycles.
\end{proposition}

This result has an intuitive real-world interpretation. It says that a patient traversing the cancer care network cannot accrue a positive benefit through repeating any sequence of activities. This will certainly be the case if the outcome is measured in terms of monetary cost of the activities. But even viewed through the lens of survival, it makes sense that such negative cost cycles do not exist. Otherwise, there would be a sequence of activities that guarantees improved survival, which is not realistic. We also note that a finite optimal value for~\eqref{model:reference} is important for the concordance metric, which we define later. 

Formulation~\eqref{model:reference} is non-convex because of the normalization constraint. However, it can be solved using polyhedral decomposition: solving $2n$ quadratic optimization problems where the normalization constraint is replaced with $c_{ij} \in [-1,1]$ for all $(i,j) \in \mathcal{A}$ and with one additional constraint on one $c_{ij}$ being set to either $+1$ or $-1$:
\begin{equation} \label{model:decomposition}
\begin{alignedat}{1}
\mathbf{IO}^{\textrm{ref}}_{ij}(\hat{\mathcal{X}}^r): \quad \underset{\bc, \bp, \bepsilon^r}{\text{minimize}} & \quad \sum_{q=1}^{R}(\epsilon^r_q)^2\\
\textrm{subject to} & \quad \bA' \bp \le \bc,\\
& \quad \bc'\hat\bx^r_q=\bb'\bp+\epsilon^r_q, \quad q=1, \ldots, R, \\
& \quad c_{ij} = -1 \lor c_{ij} = 1,\\
& \quad -1 \le c_{ij} \le 1, \quad (i,j) \in \mathcal{A},\\
& \quad \bA \bc= \bzero.\\
\end{alignedat}
\end{equation}
Furthermore, if in a particular problem context it is clear which arc should be either the most heavily rewarded (set $c_{ij} = -1$) or heavily penalized (set $c_{ij} = +1$), then only one of these $2n$ problems needs to be solved. As we demonstrate later, this is the case for our clinical problem.

\subsubsection{Refining the Solution using Patient Data.}
\label{sec:model_patient}

Let $\hat{\mathcal{X}}^s=\{ \hat\bx^s_1, \ldots,\hat\bx^s_S\}$ denote a dataset of $S$ patient pathways with positive clinical outcomes (i.e., survived) and $\hat{\mathcal{X}}^d=\{ \hat\bx^d_1, \ldots,\hat\bx^d_D\}$ denote a dataset of $D$ patient pathways with negative clinical outcomes (i.e., died), all assumed to be feasible flow vectors for the forward problem. If formulation~\eqref{model:reference} has multiple optimal solutions, we can use this patient data to determine a cost vector that not only maximizes fit with the reference pathways, but also provides ``separation'' between $\hat{\mathcal{X}}^s$ and $\hat{\mathcal{X}}^d$. Once formulation~\eqref{model:reference} is solved and an optimal duality gap vector, $\bepsilon^{r*}$, is generated, we use $\bepsilon^{r*}$ as input into the following inverse optimization model:

\begin{equation} \label{model:patientref}
\begin{alignedat}{1}
\mathbf{IO}^{\textrm{pat}}(\hat{\mathcal{X}}^s,\hat{\mathcal{X}}^d,\bepsilon^{r*}): \quad \underset{\bc, \bp, \bepsilon^s, \bepsilon^d}{\text{minimize}} & \quad  \frac{D}{S}\sum_{q=1}^{S}\epsilon^s_q-\sum_{q=1}^{D}\epsilon^d_q \\
\textrm{subject to} & \quad \bA' \bp \le \bc,\\
& \quad \bc'\hat\bx^r_q=\bb'\bp+\epsilon_q^{r*}, \quad  q=1, \ldots, R, \\
& \quad \bc'\hat\bx^s_q=\bb'\bp+\epsilon^s_q, \quad  q=1, \ldots, S, \\
& \quad \bc'\hat\bx^d_q=\bb'\bp+\epsilon^d_q, \quad  q=1, \ldots, D, \\
& \quad \|\bc\|_{\infty}= 1,\\
& \quad \bA \bc= \bzero.\\
\end{alignedat}
\end{equation}

This model treats $\hat{\mathcal{X}}^s$ like input to traditional inverse optimization models (and like the reference pathways), where the goal is to minimize the aggregate suboptimality of these data points. In contrast, the optimal cost vector from~\eqref{model:patientref} purposefully aims to not fit the data from $\hat{\mathcal{X}}^d$. Instead, the cost vector should generate data points as far away as possible from those in $\hat{\mathcal{X}}^d$ when solving the forward problem, which is why in formulation~\eqref{model:patientref} the objective maximizes suboptimality with respect to $\hat{\mathcal{X}}^d$. The weight $D/S$ simply scales the two sub-objectives to account for the difference in the number of data points in the two groups. The second constraint forces the cost vector to achieve the optimal duality gap found in model~\eqref{model:reference}. In other words, all feasible solutions of~\eqref{model:patientref} are optimal solutions to~\eqref{model:reference}. The third and fourth constraints define the duality gaps with respect to the pathways in $\hat{\mathcal{X}}^s$ and $\hat{\mathcal{X}}^d$, respectively. Again, the duality gaps are nonnegative since all patient pathways are feasible for the forward problem. The remaining constraints are as defined in~\eqref{model:reference}. Formulation~\eqref{model:patientref} can be solved using the same polyhedral decomposition technique described previously. The difference is that each subproblem is a linear optimization problem, since the objective function is now linear. A linear objective was chosen for formulation~\eqref{model:patientref} to facilitate the combination of duality gaps that needed to be minimized and maximized in the same objective.

Next, we establish that formulation \eqref{model:patientref} will always return an optimal cost vector. 

\begin{proposition}
\label{prop:optsol2}
$\mathbf{IO}^{\emph{pat}}(\hat{\mathcal{X}}^s,\hat{\mathcal{X}}^d,\bepsilon^{r*})$ has an optimal solution.
\end{proposition}

Looking ahead, our numerical results will use two distinct patient datasets. One dataset comprises both $\hat{\mathcal{X}}^s$ and $\hat{\mathcal{X}}^d$ and is input to~\eqref{model:patientref} to find an optimal cost vector (see Sections~\ref{sec:iospecial} and~\ref{sec:ioresults}). Using this cost vector, we calculate concordance scores for patients in the second dataset. These out-of-sample concordant scores form the basis of the subsequent numerical experiments, including the survival analysis where they constitute an independent variable.

\section{Concordance Metric}
\label{sec:omega}
In this section, we develop a metric that measures the concordance of a given patient pathway against the reference pathways. It should be applied to patient pathways that have yet to be observed and that are not in the ``training data'' that was used to generate the optimal cost vector. 

Using the cost vector $\bc^*$ generated by inverse optimization, our concordance metric $\omega(\hat\bx)$ measures the cost of a patient pathway $\hat\bx$ against the cost of a shortest path with respect to $\bc^*$ as
\begin{equation}\label{eq:omega}
    \omega(\hat\bx)= 1-\frac{\bc^*{'}\hat\bx-\bc^*{'}\bx^*}{M(\hat\bx)-\bc^*{'}\bx^*}
\end{equation}
where $M(\hat\bx) = \underset{\bx}{\text{max}} \{\bc^*{'}\bx : \bA \bx= \bb, \bx \ge \bzero, \|\bx\|_1\le \|\hat\bx\|_1, \bx \textrm{ is a walk} \}$ is the cost of a longest walk with up to $\|\hat\bx\|_1$ steps, which can be determined via dynamic programming (see~\ref{app:longestwalk}).  
Normalizing in this fashion implies that a walk with more steps could be, paradoxically, more concordant than a shorter walk, something that would not occur if the normalization term was a constant. However, we choose to normalize using $M(\hat\bx)$ for several reasons. First, this normalization allows the computation of $\omega(\hat\bx)$ to depend only on the given pathway $\hat\bx$ (and $\bc^*$) and not any other pathway. Normalizing by a fixed amount, $M$, while maintaining the property that $\omega(\hat\bx) \ge 0$ for all $\hat\bx$ would require that $M$ be longer than all patient pathways that would be considered by $\omega(\hat\bx)$, an amount that could not be known a priori. Second, this approach aligns with common normalization approaches for edit distance metrics that are based on string length \citep{yujian2007normalized}. Later, in our numerical results, we show that using $M(\hat\bx)$ helps to ``un-skew'' the distribution of duality gaps over the patient pathways, resulting in an approximately normal distribution of $\omega(\hat\bx)$ for our dataset.

Note that we use the cost of the shortest path, $\bc^*{'}\bx^*$, instead of the cost of a reference pathway when measuring concordance. Since a reference pathway may not actually be a shortest path under $\bc^*$, and since there may be multiple reference pathways with different costs under $\bc^*$, we define the concordance metric using $\bc^*{'}\bx^*$ to ensure a consistent baseline. Note that if the inverse optimization model finds a cost vector that perfectly fits all of the reference pathways, then the cost of all reference pathways will be equal to $\bc^*{'}\bx^*$, and $\omega(\hat\bx)$ will be measuring concordance directly against the reference pathways.

It follows from the definition of $\omega(\hat\bx)$ that it is unitless metric between 0 and 1, with 1 representing a perfectly concordant pathway (proof omitted).

\begin{proposition}
\label{pro:omega_properties}
Given any feasible solution $\hat\bx$ to the forward problem: 
\begin{enumerate} 
    \item $\omega(\hat\bx) \in [0,1]$,
    \item $\omega(\hat\bx)=1$ if and only if $\bc^*{'}\hat\bx=\bc^*{'}\bx^*$,
    \item $\omega(\hat\bx)=0$ if and only if $\hat\bx \in \underset{\bx}{\arg\max} \{ \bc^*{'}\bx : \bA \bx= \bb, \bx \ge 0, \|\bx\|_1 \le \|\hat\bx\|_1, \bx \textrm{ is a walk} \}$.
\end{enumerate}
      
\end{proposition}

Finally, we note that the unnormalized cost difference between the patient pathway ($\hat\bx$) and a shortest path ($\bx^*$) using the optimal cost vector ($\bc^*$), which is exactly the duality gap $\epsilon$ associated with the patient pathway, is another possible measure of pathway concordance. However, we believe the unnormalized metric is less interpretable, since it requires the decision maker to put it in context of the underlying network and arc costs. By definition, the normalized metric does this automatically, which facilitates the comparison of concordance across different patient cohorts or different diseases that may be modeled using different networks. We believe the interpretability of the normalized metric makes it an attractive tool for non-experts. Nevertheless, we will demonstrate robustness of our survival analysis results with respect to the unnormalized metric.

\section{Application to Stage III Colon Cancer}
\label{sec:DPC}
In this section, we demonstrate the application of our inverse optimization framework to measure clinical pathway concordance, focusing on stage III colon cancer.

\subsection{Problem Background}
\label{sec:dpcbackground}
\subsubsection{Disease Pathway Management at Cancer Care Ontario.}
The Disease Pathway Management (DPM) program at Cancer Care Ontario takes a systems view of setting priorities for cancer control, planning cancer services and improving the quality of cancer care in Ontario. DPM views patient journeys across the entire cancer continuum as part of an integrated process, as opposed to evaluating individual points of care in isolation. Clinicians are engaged to establish evidence-based best practices for the sequence of activities patients should take through the care network. These best practices are captured in \emph{pathway maps} \citep{mapall}, which reflect clinical guidelines for the care that patients should receive to optimize clinical outcomes. A pathway map is essentially a flowchart that provides a high level view of the care pathways for a specific disease and shows all the different ways a patient can navigate the system. A single pathway map may encompass many clinical pathways. Pathway maps cover points of care spanning disease prevention, screening, diagnosis, treatment, recovery and end-of-life palliative care. 

\subsubsection{Stage III Colon Cancer and Pathway Map.}
Colon cancer is a malignant tumor of the colon. 
In 2018, roughly 180,000 people were diagnosed with colorectal cancer in North America and there were 64,000 deaths due to this cancer \citep{stat}; colon cancer constitutes approximately 70\% of the colorectal cancer cases \citep{statcolon}. Stage III colon cancer is characterized by metastasis of the cancer to the lymph nodes near the colon or by direct invasion into nearby organs. Chemotherapy is recommended as adjuvant therapy for surgically resected stage III disease. 

From a systems perspective, stage III colon cancer patients are considered a uniform cohort and the optimal management of stage III colon cancer follows a single pathway map \citep{map}. This pathway map spans diagnosis and treatment (see~\ref{app:pathway_maps}), and contains the following major categories of activities: 1) clinical consultations, 2) endoscopy, 3) diagnostic imaging, 4) surgery, 5) adjuvant chemotherapy. Clinical consultations include consultations with a gastroenterologist (GI), surgeon and medical oncologist (MO). Diagnostic imaging includes imaging of the abdomen, pelvis and chest with any of the following modalities: computed tomography (CT), magnetic resonance imaging  (MRI), ultrasound  (US), or x-ray. The set of concordant pathways for stage III colon cancer are derived from its pathway map.

All concordant pathways must start with an endoscopy followed by abdominal imaging, chest imaging, surgical resection of the tumor, a MO consultation and finally chemotherapy. The patient will have had a GI or surgical consultation prior to the endoscopy, and a surgical consultation at some point before the surgical resection. Similarly, a MO consultation must occur before chemotherapy, preferably after the surgical resection. Abdominal imaging may take place as either an abdomen CT scan followed by a pelvis CT scan, or as an abdominal ultrasound. Similarly, chest CT scans or chest x-rays constitute chest imaging. Chemotherapy is considered concordant if the patient pathway includes at least six treatment cycles. All of these different combinations are captured in the pathway map. Enumerating all of the possible concordant pathways from the pathway map results in 108 distinct reference pathways for stage III colon cancer.

\subsection{Data}
\label{sec:data}
\subsubsection{Data Sources.}
\label{sec:datasources}
In Ontario, health care encounters are recorded in administrative databases using standardized methods. Using encrypted person-specific identifiers, we linked multiple sources to select the study population, define patient characteristics, and record activity related to screening, imaging, treatments, emergency department (ED) visits, and death. The data sources used in this study are listed in~\ref{app:data}.

\subsubsection{Study Population.}
\label{sec:studypop}

We gathered two datasets for our numerical experiments, which were used as ``training'' and out-of-sample ``testing'' sets. The first one is a population-based cohort of individuals newly diagnosed with stage III colon cancer in 2010 in Ontario. Their clinical activities were recorded from six months prior to cancer diagnosis until either death or censoring at four years following diagnosis. Included patients had surgical resection within one year of diagnosis. Patients were excluded based on the following exclusion criteria: no Ontario health insurance number, residence outside of Ontario, non-incidental cancer diagnosis, any prior history of cancer or identification of a new primary cancer during the follow-up period, cancers diagnosed based on death records, unknown cancer stage, and unknown tumor histology. The final cohort contained 763 patients, 257 of whom died (from any cause) within the four-year post-diagnosis observation period. The second dataset is similar to the first one, except that it is for newly diagnosed patients from 2012-2016. After applying the same inclusion criteria, the final 2012-2016 cohort included 4826 patients, 1464 of whom died (from any cause) within their follow-up period.

Each dataset consists of event logs for each patient in the cohort that describe the activities undertaken and their corresponding times relative to the date of diagnosis. However, not all events in the event log are relevant to the diagnostic and treatment phases of care, which is our focus. Therefore we refine each patient's event log by filtering out events earlier than 30 days prior to diagnosis and following one year post-diagnosis, since patients are expected to complete diagnostic and treatment activities within one year. If the patient completed chemotherapy within the year, we remove all events after their last chemotherapy cycle. We also remove activities that have no bearing on concordance, such as minor procedures like barium enemas. All data refinements were implemented in consultation with our clinician team. 

\subsection{Network Design}
\label{sec:networkdesign}
Next, we form the clinical activity network based on activities from the pathway map, observed patient pathways, and discussions with clinical experts from the DPM team. In clinical practice, GI, surgery and MO consultations always precede endoscopies, surgical resection and chemotherapy treatment, respectively. We therefore do not include nodes for these consultations and assume that they occur with the corresponding procedure. However, it is possible for patients to have additional consultations, which are costly for the system and deemed discordant. These discordant consultations, like visits to the emergency department, were quite prevalent in our dataset, so we included EXTRA CONSULT and ED VISIT as nodes in the graph. Another reason to include the latter is because reducing unnecessary visits to the ED is a Cancer Care Ontario priority.

Imaging events are distinct from consultations in that they are not implicit with other events, so we kept them as separate standalone activities. Imaging activities that were deemed functionally equivalent from a concordance perspective were merged into a single node. For example, abdomen MRI and ultrasound were combined into the node ABDOMEN MRI/US. We formed a single node for PELVIS MRI/US and CHEST IMAGING (which encompasses chest x-rays and CT scans) for the same reason. CT is preferred over the other imaging modalities for abdomen and pelvis, so we defined separate nodes for abdomen CT and pelvis CT.

\begin{figure}[t]
\centering
\includegraphics[width=0.8\textwidth]{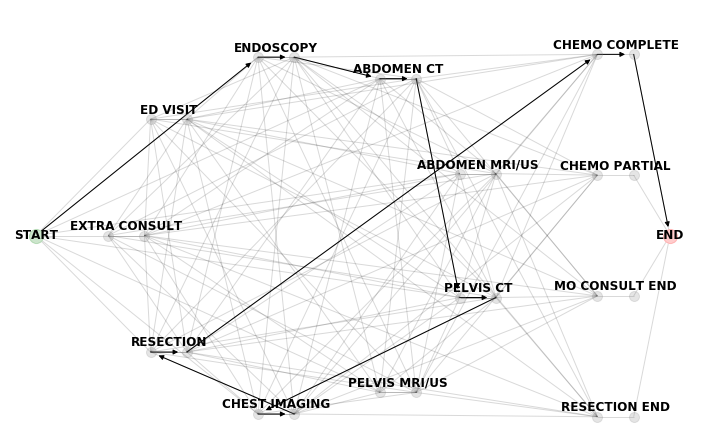}
\caption{Stage III colon cancer clinical activity network with a reference pathway highlighted }\label{fig:graph_modified}
\end{figure}

Finally, our discussions with experts revealed that the type of treatment received by the patient should have extra bearing on their concordance score. To model the different possible ways each pathway can end, we include an ``outcome layer'' with four distinct activities; each pathway must exit the graph via one of these four. Patients who receive six chemotherapy treatments exit through the node CHEMO COMPLETE. Receiving between one and five chemotherapy treatment cycles leads to an exit through the node CHEMO PARTIAL. The remaining pathways (no chemotherapy) exit either through the nodes MO CONSULT END (if the patient receives an MO consultation) or RESECTION END (if the patient does not). Recall that our patient cohort only includes patients who had surgical resection. The final network design used in our model is shown in Figure~\ref{fig:graph_modified}.

\subsection{Inverse Optimization Model Specification}
\label{sec:iospecial}

Since our final network design merged together several activities specified in the Cancer Care Ontario pathway map, the set of 108 reference pathways were reduced to two distinct reference pathways. One reference pathway is shown in Figure~\ref{fig:graph_modified}. The other reference pathway replaces ABDOMEN CT and PELVIS CT with ABDOMEN MRI/US. The network, reference pathways, survived and died patient pathways from the 2010 dataset constitute the $(\bA, \bb),\hat{\mathcal{X}}^r:=\{\hat\bx_1^r, \hat\bx_2^r\}$, $\hat{\mathcal{X}}^s$ and $\hat{\mathcal{X}}^d$, respectively, in formulations \eqref{model:reference} and~\eqref{model:patientref}. The value of the $(i,j)$-th component of $\hat\bx_q^r$ is 1 if reference pathway $q$ includes arc $(i,j)$ and 0 otherwise. Patient pathways are mapped to the graph similarly, except that each component of the vector represents the number of times that arc is traversed in the pathway. Finally, with respect to $\mathcal{C}$, we included three sets of application-specific constraints: 1) activity ranking constraints, 2) subpath constraints, and 3) penalty constraints.

Activity ranking constraints enforce a relative importance between pairs of activities. For example, clinical experts deemed ABDOMEN CT to be a more important activity than PELVIS CT. To account for this relationship in the model, we introduced a constraint that enforced the cost of completing ABDOMEN CT to be at most the cost of completing PELVIS CT. If $i$ represents ABDOMEN CT and $j$ represents PELVIS CT, then the constraint would be $c_{i_s i_e} \le c_{j_s j_e}$. There were eight such constraints in total between the nine non-outcome layer nodes. 

Subpath rankings generalize activity ranking constraints to a sequence of nodes. For example, the subpath of ENDOSCOPY to ABDOMEN CT to PELVIS CT to CHEST IMAGING was preferred to the subpath of ENDOSCOPY to ABDOMEN MRI/US to CHEST IMAGING. Therefore, the sum of the arc costs along the first subpath (starting from the end node of ENDOSCOPY and ending with the start node of CHEST IMAGING) was constrained to be at most the sum of the arc costs along the second subpath. There were four such constraints in total. The first is the subpath ranking above. The remaining three rank the subpaths exiting the graph through the four outcome layer nodes, starting each subpath at RESECTION. The nodes in the outcome layer as depicted in Figure~\ref{fig:graph_modified} are arranged from top to bottom in order of most preferred to least preferred outcome. For example, if $i$ represents RESECTION, $j$ represents RESECTION END, $k$ represents MO CONSULT END, and $E$ represents the final end node, then the subpath ranking between RESECTION END and MO CONSULT END would be $c_{i_e j_s} + c_{j_s j_e} + c_{j_e E} \ge  c_{i_e k_s} + c_{k_s k_e} + c_{k_e E}$. Similar subpath rankings were generated for MO CONSULT END versus CHEMO PARTIAL and CHEMO PARTIAL versus CHEMO COMPLETE.

Finally, since CHEMO COMPLETE is the best outcome activity through which to exit the graph, we set the cost of the CHEMO COMPLETE to END arc to $-1$, anchoring all the arc costs relative to this one. This constraint is also beneficial from a model tractability perspective because instead of using polyhedral decomposition to solve~\eqref{model:reference}, only a single optimization problem is solved (see Section~\ref{sec:model_ref}).

\subsection{Inverse Optimization Results}
\label{sec:ioresults}

\begin{figure}[t]
\centering
\includegraphics[width=0.7\textwidth]{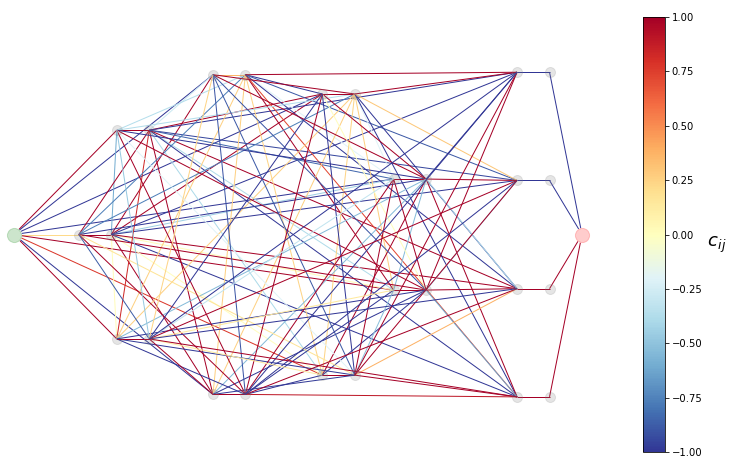}
\caption{Stage III colon cancer clinical activity network with optimal cost vector overlaid }\label{fig:optimalgraph}
\end{figure}

We obtained an optimal cost vector, $\bc^*$, by first solving formulation~\eqref{model:reference} followed by formulation~\eqref{model:patientref} with the inputs described in Section~\ref{sec:iospecial}. This cost vector comprises all arc costs and by definition minimizes the total suboptimality error of the two reference pathways, $\hat{\bx}^r_1$ and $\hat{\bx}^r_2$, and secondarily minimizes (maximizes) the suboptimality error of the patient pathways from survived (died) patients in $\hat{\mathcal{X}}^s$ ($\hat{\mathcal{X}}^d$). Figure \ref{fig:optimalgraph} illustrates the $\bc^*$ values on the graph. Recall that since $\bc^*$ satisfies $\bA\bc^*=\bzero$, the sum of the costs on the incoming and outgoing arcs for each node are equal. As a consequence, the associated cost of passing through each node in the outcome layer is doubled, since each node in that layer has only one outgoing arc (i.e., the intra-activity arc $(i_s,i_e)$ followed by the arc from the outcome node $i_e$ to END).

\begin{figure}[t]
\centering
\includegraphics[width=0.6\textwidth]{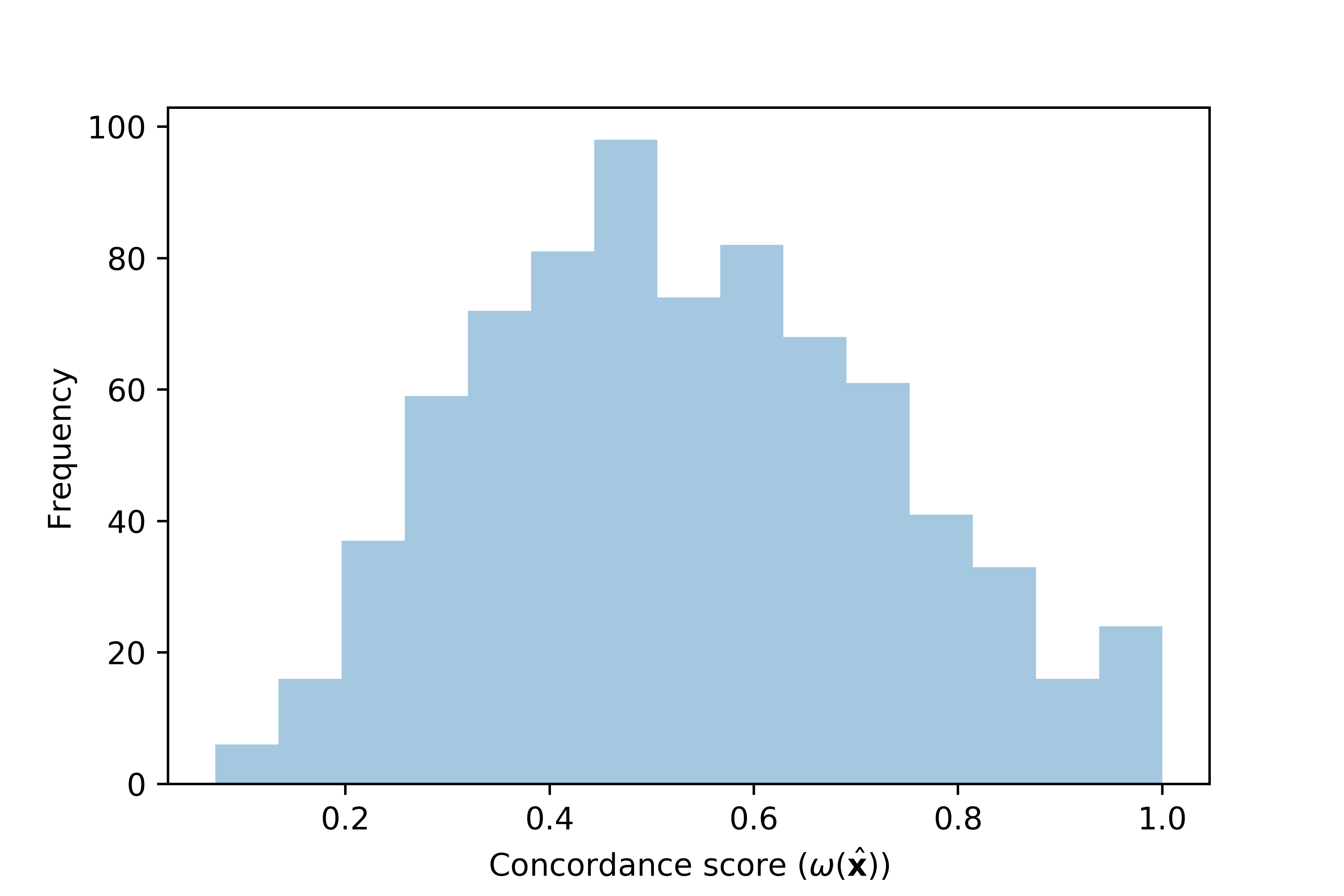}
\caption{Distribution of concordance score over patient cohort}
\label{fig:omegapdf}
\end{figure}

We calculated the (in-sample) concordance score of each patient pathway from the 2010 dataset using \eqref{eq:omega}. The distribution of concordance scores is shown in Figure \ref{fig:omegapdf}. 
The fact that this distribution is approximately normal stems from the choice of normalization in $\omega(\hat\bx)$. As Figure \ref{fig:epsilondist} shows, the distributions of the duality gap $\epsilon$ and pathway length $\|\hat\bx\|_1$ are both skewed similarly, resulting in an approximately linear relationship between the two quantities. In other words, the lack of fitness of a given patient pathway is associated with its corresponding length, and dividing one by the other creates the shape of the distribution seen in Figure \ref{fig:omegapdf}.

%
%

\begin{figure}[t]
\centering
\includegraphics[width=0.6\textwidth]{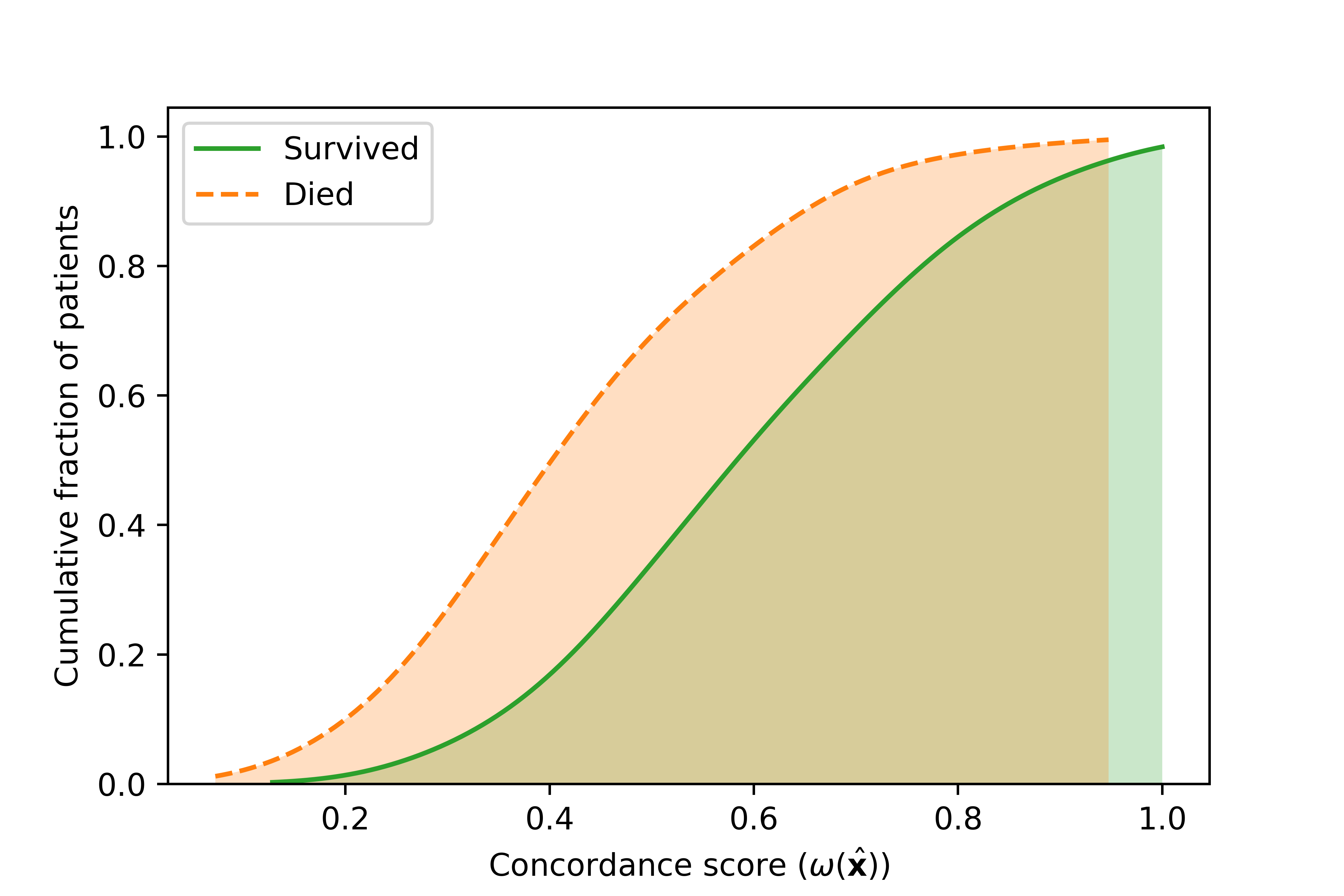}
\caption{Cumulative distribution of concordance score over 2010 dataset stratified by patient outcome}
\label{fig:graph_cumdists}
\end{figure}

Figure \ref{fig:graph_cumdists} shows smoothed cumulative distributions of the concordance scores, stratified by survival outcome, depicting first-order stochastic dominance. That is, for any given concordance score $\omega_0 \in [0,1]$, the probability of scoring below $\omega_0$ is higher for patients who died. Formal validation of this observation requires a rigorous analysis of the relationship between concordance and survival, which we carry out in the next section.


\section{Validating the Concordance Metric against Survival} 
\label{sec:validation2012}
In this section, we investigate the association between concordance and survival using the 2012-2016 dataset. However, we use the concordance metric developed using the 2010 dataset. As such, this analysis provides out-of-sample validation.

\subsection{Methods}
\label{sec:survivalmethods2012}
Patient characteristics were summarized as counts with proportions for categorical data and means with standard deviations for continuous data. We binned the patient concordance scores into terciles and difference in covariate distributions among concordance score terciles was assessed with a chi-squared test for categorical variables and a one-way ANOVA for means and standard deviations. A Kaplan-Meier estimator was used for graphical presentation of (unadjusted) survival over time by the terciles of concordance scores. Difference in survival probabilities was assessed using the log-rank test statistic.

A semi-parametric Cox proportional hazards model was implemented to evaluate the association between concordance scores and mortality \citep{cox}. Cox regression provides estimates of association between each variable and the hazard of event, i.e., the instantaneous rate of event. A unit increase in a covariate is multiplicative with respect to the hazard. An estimate of the hazard ratio (HR) above $1$ indicates a variable that is positively associated with the risk of mortality, i.e., negatively associated with survival.

A range of known potential confounders and predictors of survival was assessed for inclusion in the model. These variables consist of demographic characteristics (age, sex, rural residency, quintiles of neighbourhood median income, terciles of immigrant population), Charlson comorbidity score \citep{charlson1987new,quan2005coding}, cancer diagnostic and treatment characteristics (screening category, stage, tumor grade, emergent surgery status, length of hospital stay following the surgical treatment), and healthcare utilization a year prior to diagnosis (number of outpatient visits, number of in-patient admissions, number of ED visits), which is a proxy for overall patient health. Screening category refers to frequency and timing of blood tests \citep{james2018repeated}. Stage specifies the cancer sub-stage, which denotes severity. The emergent surgery indicator refers to patients who had an urgent in-hospital admission when the surgical resection was performed. 

We employed preliminary variable selection to control for important confounding effects while excluding covariates that would compromise model efficiency. Variable importance in predicting mortality was evaluated using an ensemble tree method for analysis of right-censored survival data \citep{ishwaran2008random}.

For the survival model, assumption of a linear relationship between continuous independent variables and the outcome was evaluated by plotting the residuals against each covariate. If nonlinearity was detected either from the plots of Martingale residuals or Schoenfeld residuals, restricted cubic splines were used to model the corresponding continuous predictor \citep{stone1985additive}. 

Variance of the estimated association between the concordance score and survival can increase in case of overfitting. Bootstrap model validation with 5000 resamples was performed to evaluate the degree of overfitting and potential overadjustment for confounders \citep{harrell2015regression}. 

Patients were excluded from the final survival model if they did not have all required covariate values, other than tumor grade, recorded. Patients who died within one year of diagnosis were also excluded, as they did not have the opportunity to complete their treatment journey in accordance with the clinical guidelines. 

\subsection{Results}
\label{sec:survivalresults2012}

A total of 4242 individuals met the inclusion criteria for survival analysis. The mean age at diagnosis was 67.3 years (standard deviation = 13.1). Of the included patients, 880 (20.7\%) died. A higher proportion of deaths occurred among those with low concordance scores (35.9\%, p-value $<0.0001$). 

Table \ref{tab:descriptive2012} lists the characteristics of the patients in the 2012-2016 dataset and by concordance terciles. Significant differences between the terciles were observed for age, sex, immigrant population, comorbidity score, healthcare utilization, screening group, cancer stage, tumor grade, and length of hospital stay following the surgery. Rural residency and neighbourhood income were not significantly different across terciles at the 0.05 level.

\begin{figure}[t]
\centering
\includegraphics[width=0.6\textwidth]{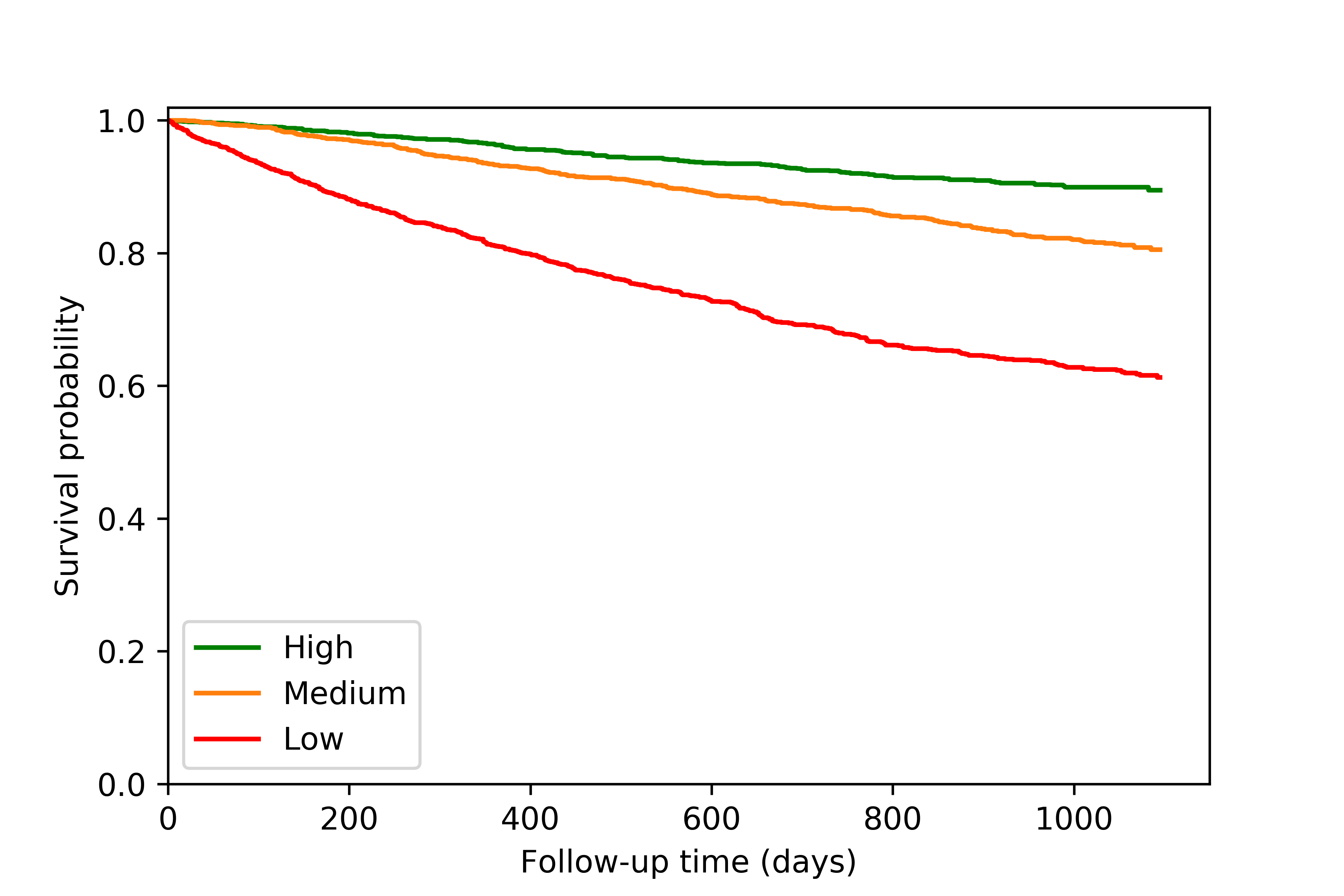}
\caption{Kaplan-Meier curves for each concordance score tercile}\label{fig:km2}
\end{figure}

Statistically significant differences in (unadjusted) survival over time were observed for concordance score terciles (p-value $<0.0001$, Figure \ref{fig:km2}). 


Table \ref{tab:results2012} summarizes the association between concordance and survival for both unadjusted and adjusted Cox models with both categorical and continuous measures for the concordance score. The unadjusted Cox model found a significant effect of concordance on survival in both the continuous (HR = 0.70, 95\% CI = (0.68, 0.73)) and categorical models (HR = 0.41, 95\% CI = (0.35, 0.48) for Medium vs. Low tercile; HR = 0.22, 95\% CI = (0.18, 0.26) for High vs. Low tercile). Note that a unit change in the continuous model is scaled to represent a change of 0.1 in $\omega(\hat{\bx})$.

Along with concordance score, the adjusted Cox regression included age, sex, rural residency, number of outpatient visits a year prior to diagnosis, comorbidity score, screening group, cancer stage and grade, emergent surgery status, and length of hospital stay following the surgery. A nonlinear relationship between two covariates (age and length of hospital stay following the surgical treatment) was observed, so restricted cubic splines with three knots were used to model these covariates. No violation of model assumptions was detected. Results of the adjusted analyses showed statistical significance for the association between the concordance score and mortality. Similar to the unadjusted model, there was a significant effect of concordance on survival, with decreased risk of mortality associated with higher values of the concordance score in the continuous model (HR = 0.80, 95\% CI = (0.77, 0.84)) and categorical models (HR = 0.61, 95\% CI = (0.52, 0.72) for Medium vs. Low tercile; HR = 0.38, 95\% CI = (0.31, 0.47) for High vs. Low tercile).
\begin{table}[t]
\centering
\caption{Risk of mortality associated  with concordance score for categorical and continuous models}
\label{tab:results2012}
\begin{tabular}{llllllll}
\toprule
                             &               & \multicolumn{3}{c}{Unadjusted}                                                             & \multicolumn{3}{c}{Adjusted}                                                      \\ \cmidrule(r){3-5} \cmidrule(l){6-8}

                     &       & \multicolumn{1}{c}{HR}   & \multicolumn{1}{c}{95\% CI} & \multicolumn{1}{c}{p-value} & \multicolumn{1}{c}{HR}   & \multicolumn{1}{c}{95\% CI} & \multicolumn{1}{c}{p-value} \\ \midrule
\multirow{2}{*}{Categorical} & $\omega$ (Med. vs Low) & \multicolumn{1}{c}{0.41} & \multicolumn{1}{c}{(0.35, 0.48)}  & \multicolumn{1}{c}{\textless{}0.0001}   & \multicolumn{1}{c}{0.61} & \multicolumn{1}{c}{(0.52, 0.72)}  & \multicolumn{1}{c}{\textless{}0.0001}   \\
                             & $\omega$ (High vs Low)   & \multicolumn{1}{c}{0.22} & \multicolumn{1}{c}{(0.18, 0.26)}  & \multicolumn{1}{c}{\textless{}0.0001}   & \multicolumn{1}{c}{0.38} & \multicolumn{1}{c}{(0.31, 0.47)}  & \multicolumn{1}{c}{\textless{}0.0001}   \\ \midrule
Continuous                   & $\omega$ (unit change)  & \multicolumn{1}{c}{0.70} & \multicolumn{1}{c}{(0.68, 0.73)}  & \multicolumn{1}{c}{\textless{}0.0001}   & \multicolumn{1}{c}{0.80} & \multicolumn{1}{c}{(0.77, 0.84)}  & \multicolumn{1}{c}{\textless{}0.0001}   \\ \bottomrule
\end{tabular}
\end{table}

\begin{figure}[t]
\centering
\includegraphics[width=0.8\textwidth]{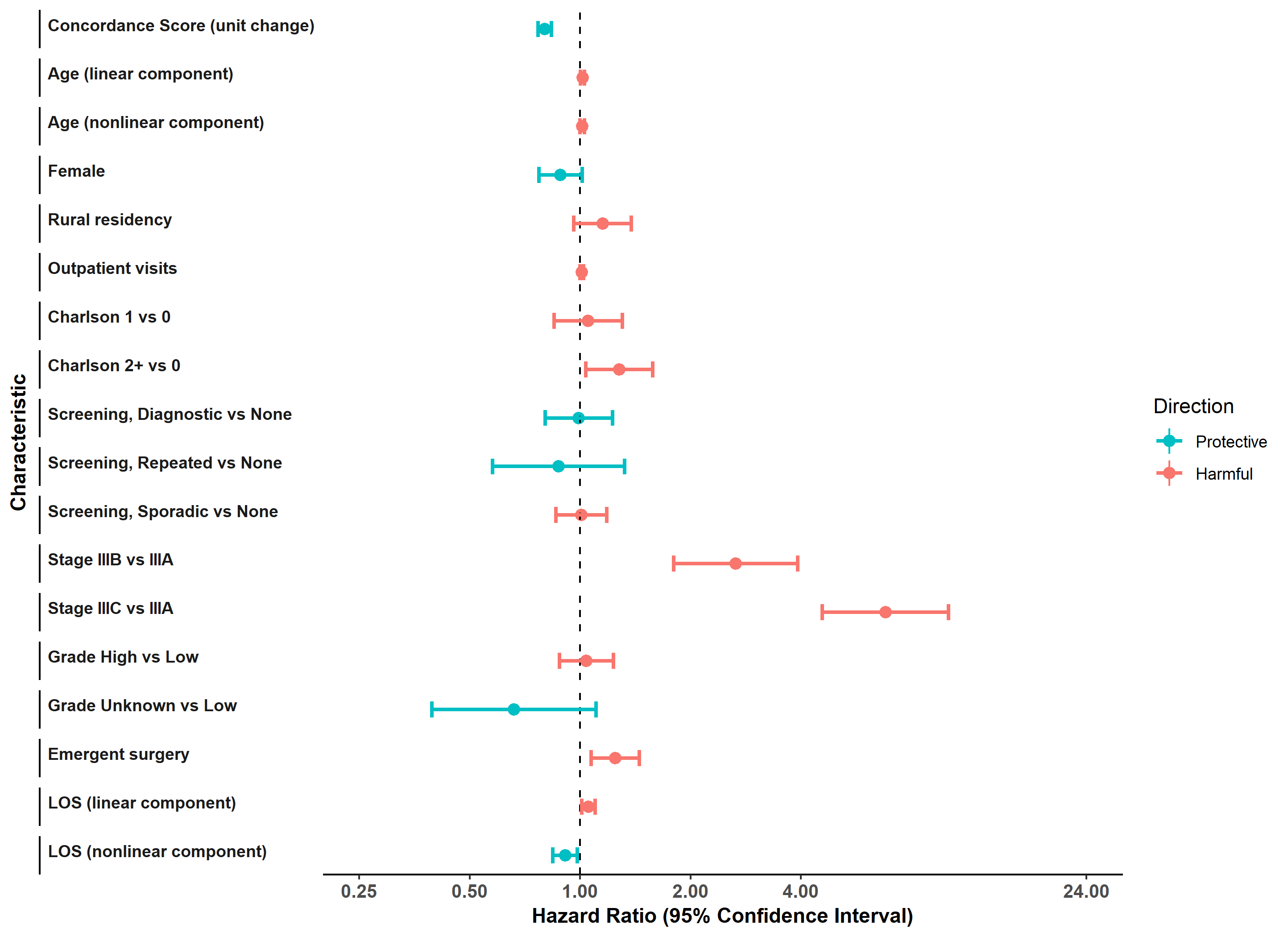}
\caption{Forest plot of variable effects in continuous Cox regression model with 95\% confidence intervals}\label{fig:fp2012}
\end{figure}

Figure \ref{fig:fp2012} summarizes the influence of each covariate in the adjusted continuous Cox model on survival. A hazard ratio below (above) 1 indicates a beneficial (harmful) effect on survival. If the confidence interval contains 1, the effect is not statistically significant at the 0.05 level. Concordance score has a statistically significant beneficial effect on survival, while age, comorbidity score, cancer stage, emergent surgery, and length of hospital stay following the surgery have significant harmful effects. The corresponding figure for the adjusted categorical Cox model is shown in~\ref{app:fp}.

Model validation with 5000 bootstrap resamples showed acceptable coefficient shrinkage (0.95) and model optimism (C-statistic reduction by 0.005), indicating no evidence of overfitting. Based on the average of the bootstrap samples, we estimated a bias of $-$0.0011 for the concordance score coefficient (log(HR) = $-$0.2213 or HR = 0.8015 in our model versus log(HR) = $-$0.2224 or HR = 0.8006 for the average of the bootstrap samples). The empirical confidence intervals of the concordance score estimates had similar coverage (0.77, 0.84) as the confidence intervals from the original model. The results of resampling imply model consistency in estimating the effect of the inverse optimization-based concordance score.

\subsection{The Value of Patient Data}
\label{sec:valueofdata2012}
The purpose of model~\eqref{model:patientref} is to identify an optimal cost vector (with respect to model~\eqref{model:reference}) that results in a significant association between concordance and survival. Without using the patient data, the association may be insignificant, or significant but with an incorrect directionality of effect. To illustrate this last point, we repeat the survival analysis using a different optimal solution for model~\eqref{model:reference}, one that is generated by considering the extreme case where the two datasets in model~\eqref{model:patientref} are reversed (recall that all solutions to~\eqref{model:patientref} are optimal for~\eqref{model:reference}).

Table \ref{tab:results2012_ref} summarizes the association between concordance and survival. Similar to before, the unadjusted Cox models all find a significant association between pathway concordance and survival. However, the effect is going in the \emph{opposite} direction now, with a higher concordance score associated with an \emph{increased} mortality risk (e.g., continuous model HR = 1.35, 95\% CI = (1.26, 1.44)).
After adjusting for other relevant covariates, the association is either insignificant, or remains significant but again in the opposite direction. 
The effects of the other covariates remain unchanged. For illustrative purposes, we include the forest plot of the continuous model in~\ref{app:fp}.

\begin{table}[t]
\centering
\caption{Risk of mortality associated with concordance score for categorical and continuous models using a cost vector from the model with patient data $(\mathbf{IO}^\textrm{pat}(\hat{\mathcal{X}}^r))$ where the sets $\hat{\mathcal{X}}^s$ and $\hat{\mathcal{X}}^d$ are swapped}
\label{tab:results2012_ref}
\begin{tabular}{llllllll}
\toprule
                             &               & \multicolumn{3}{c}{Unadjusted}                                                             & \multicolumn{3}{c}{Adjusted}                                                      \\ \cmidrule(r){3-5} \cmidrule(l){6-8}

                     &       & \multicolumn{1}{c}{HR}   & \multicolumn{1}{c}{95\% CI} & \multicolumn{1}{c}{p-value} & \multicolumn{1}{c}{HR}   & \multicolumn{1}{c}{95\% CI} & \multicolumn{1}{c}{p-value} \\ \midrule
\multirow{2}{*}{Categorical} & $\omega$ (Med. vs Low) & \multicolumn{1}{c}{1.21} & \multicolumn{1}{c}{(1.01, 1.44)}  & \multicolumn{1}{c}{0.0409}   & \multicolumn{1}{c}{1.03} & \multicolumn{1}{c}{(0.86, 1.24)}  & \multicolumn{1}{c}{0.7139}   \\
                             & $\omega$ (High vs Low)   & \multicolumn{1}{c}{1.96} & \multicolumn{1}{c}{(1.67, 2.32)}  & \multicolumn{1}{c}{\textless{}0.0001}   & \multicolumn{1}{c}{1.34} & \multicolumn{1}{c}{(1.13, 1.59)}  & \multicolumn{1}{c}{0.0009}   \\ \midrule
Continuous                   & $\omega$ (unit change)  & \multicolumn{1}{c}{1.35} & \multicolumn{1}{c}{(1.26, 1.44)}  & \multicolumn{1}{c}{\textless{}0.0001}   & \multicolumn{1}{c}{1.15} & \multicolumn{1}{c}{(1.08, 1.24)}  & \multicolumn{1}{c}{\textless{}0.0001}   \\ \bottomrule
\end{tabular}
\end{table}

\subsection{Robustness Checks}

In \ref{sec:epsilon2012} and \ref{sec:sensitivity}, we include two robustness checks. The first repeats the survival analysis with unnormalized concordance scores. The second tests for sensitivity to potential unobserved confounders. Overall, we find that the results presented above are robust.

\section{A Comparison with Edit Distance}
\label{sec:benchmark}
In this section, we compare our inverse optimization-based concordance metric, $\omega$, against three baselines for measuring pathway concordance from the literature \citep{van2010measuring,forestier2012classification,yan2018aligning,williams2014using}, all based on edit distance. 

As baselines, we consider Longest Common Subsequence distance (only insertion and deletion operations), Levenshtein distance (insertions, deletions, and substitutions) and Damerau-Levenshtein distance (insertions, deletions, substitutions, and adjacent transpositions). We refer to these three distance metrics as LCSD, LD, and DLD, respectively. As is most common in the literature, we assume all edits have unit cost. 


Next, we repeated the survival analysis from Section \ref{sec:survivalmethods2012} with the same covariates for each metric. We use the normalized scores, based on the number of activities in both the patient and reference pathways, for all baselines so their values reside in $[0,1]$, similar to $\omega$. Table \ref{tab:compare} summarizes the results of survival analysis for both adjusted and unadjusted continuous models. While there is a significant association between concordance and survival for each approach, the hazard ratio is lower for $\omega$ and the difference to all other baselines is statistically significant (p-value $<0.0001$), based on a bootstrap calculation with 1000 samples.

\begin{table}[t]
\centering
\caption{Risk of mortality associated with $\omega$ and three baseline edit distance-based concordance scores for continuous model}
\label{tab:compare}
\begin{tabular}{lllllll}
\toprule
                                            & \multicolumn{3}{c}{Unadjusted}                                                             & \multicolumn{3}{c}{Adjusted}                                                      \\ \cmidrule(r){2-4} \cmidrule(l){5-7}

                            & \multicolumn{1}{c}{HR}   & \multicolumn{1}{c}{95\% CI} & \multicolumn{1}{c}{p-value} & \multicolumn{1}{c}{HR}   & \multicolumn{1}{c}{95\% CI} & \multicolumn{1}{c}{p-value} \\ \midrule
  $\omega$   & \multicolumn{1}{c}{0.70} & \multicolumn{1}{c}{(0.68, 0.73)}  & \multicolumn{1}{c}{\textless{}0.0001}   & \multicolumn{1}{c}{0.80} & \multicolumn{1}{c}{(0.77, 0.84)}  & \multicolumn{1}{c}{\textless{}0.0001}   \\ 
  
  LCSD   & \multicolumn{1}{c}{0.73} & \multicolumn{1}{c}{(0.70, 0.76)}  & \multicolumn{1}{c}{\textless{}0.0001}   & \multicolumn{1}{c}{0.83} & \multicolumn{1}{c}{(0.80, 0.87)}  & \multicolumn{1}{c}{\textless{}0.0001}   \\ 
                              LD  & \multicolumn{1}{c}{0.75} & \multicolumn{1}{c}{(0.72, 0.78)}  & \multicolumn{1}{c}{\textless{}0.0001}   & \multicolumn{1}{c}{0.85} & \multicolumn{1}{c}{(0.82, 0.89)}  & \multicolumn{1}{c}{\textless{}0.0001}   \\ 
                     DLD & \multicolumn{1}{c}{0.76} & \multicolumn{1}{c}{(0.73, 0.79)}  & \multicolumn{1}{c}{\textless{}0.0001}   & \multicolumn{1}{c}{0.86} & \multicolumn{1}{c}{(0.83, 0.90)}  & \multicolumn{1}{c}{\textless{}0.0001}   \\ \bottomrule
\end{tabular}
\end{table}

\begin{figure}[t]
\centering  
\subfigure[HR]{\includegraphics[width=0.47\linewidth]{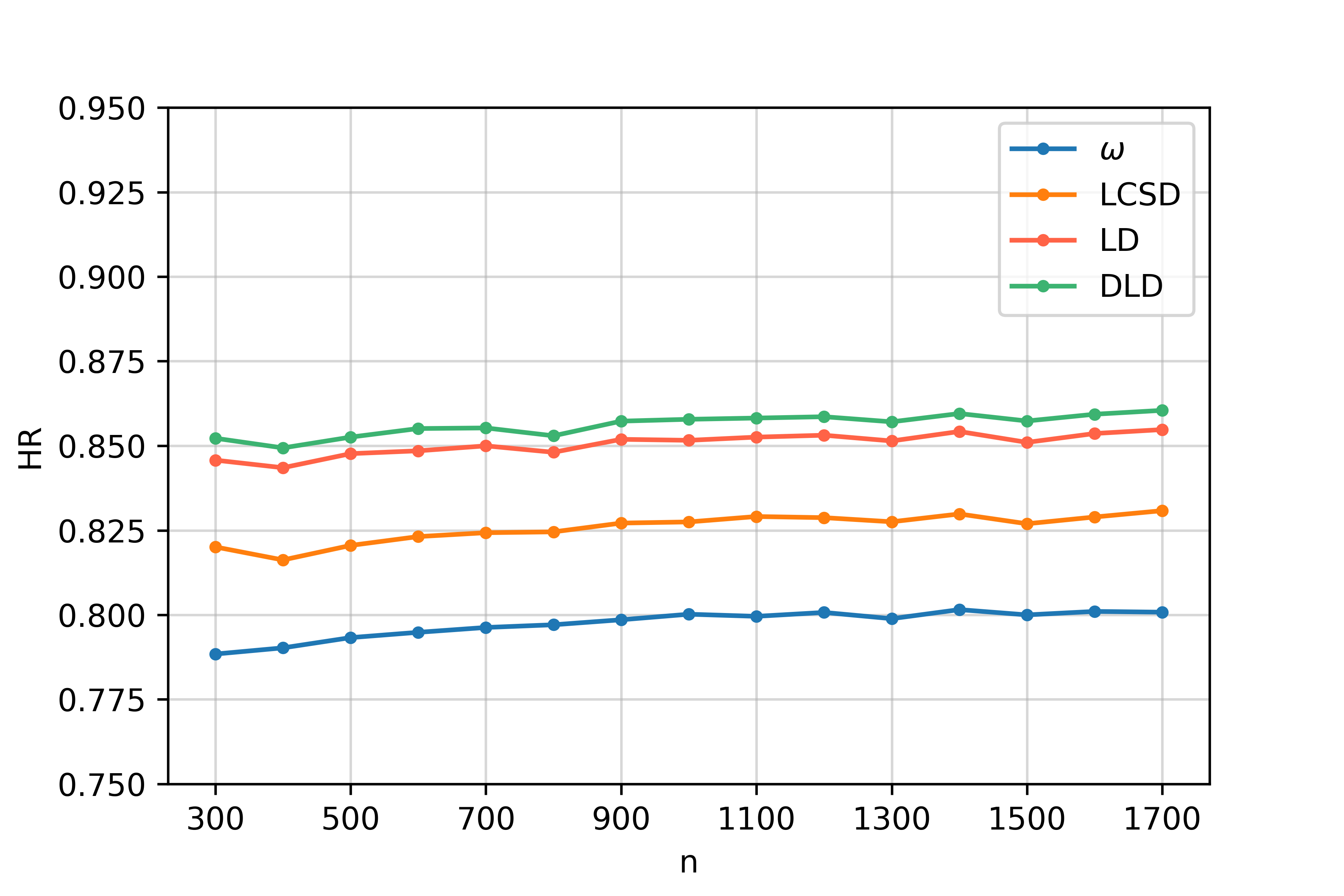}}
\subfigure[p-value]{\includegraphics[width=0.47\linewidth]{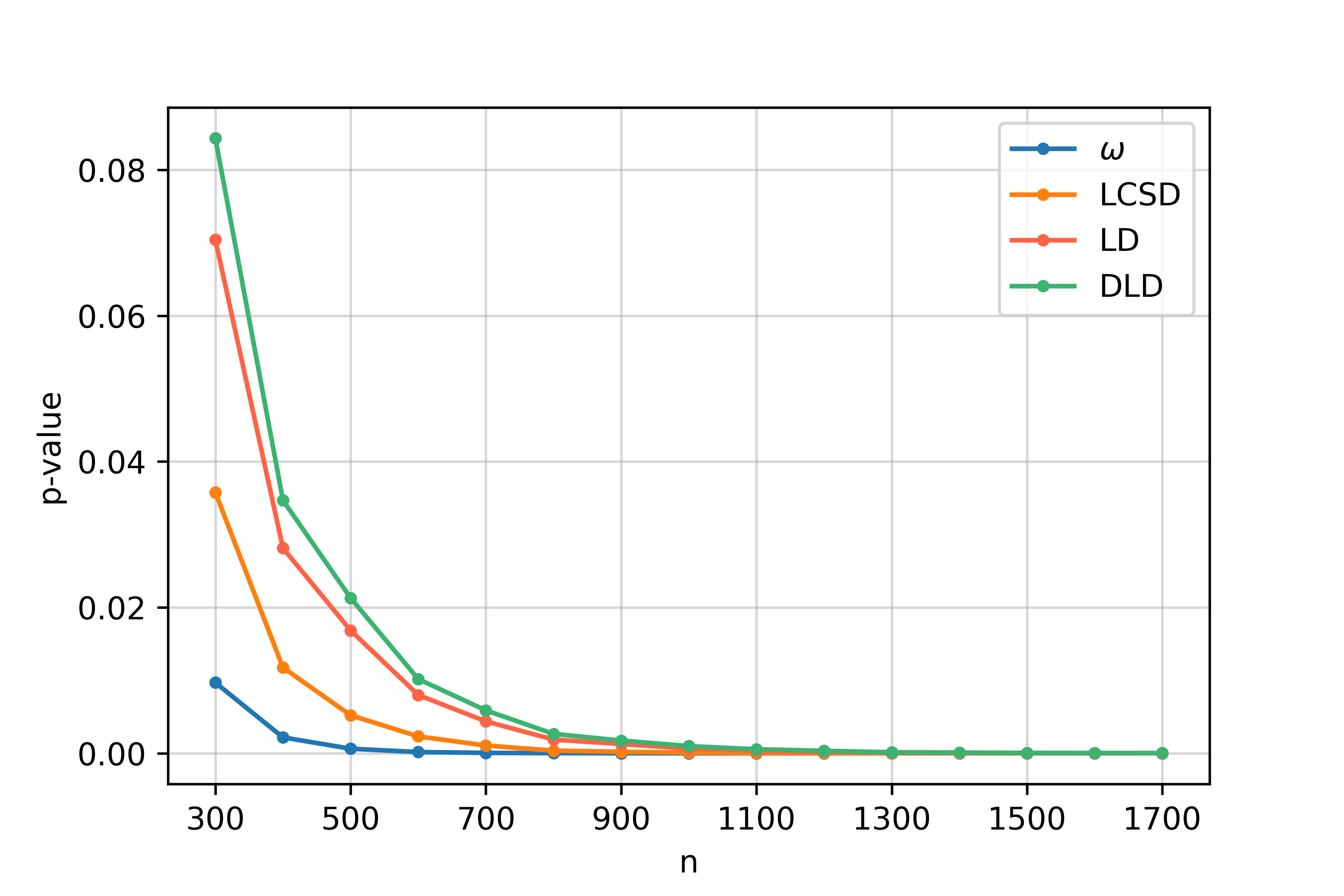}}
\caption{Comparison of survival analysis result for $\omega$ and three baselines}
\label{fig:compare}
\end{figure}

Given this result, we hypothesized that as the sample size is decreased, the baselines would lose their statistically significant association with survival before $\omega$. To study this issue, we report the median HR and p-values based on 1000 samples of the original dataset for sample sizes ranging from 300 to 1700 patients (Figure \ref{fig:compare}). For each sample size, $\omega$ has the lowest HR and lowest p-value, indicating a stronger association with survival. In addition, at 300 patients, the association for the LD and DLD metrics becomes insignificant at the 0.05 level. If we consider a stricter measure for statistical significance such as 0.01, then sample sizes above 600 are needed for LD and DLD, and 400 for LCSD, whereas 300 patients remain sufficient for $\omega$.

Overall, we believe our concordance metric has modest advantages over existing approaches based on the survival analysis. Next, we demonstrate an additional advantage of $\omega$. Namely, we show that our approach provides a mathematically rigorous link between a pathway's concordance score and its decomposition into discordant subpathways.

\section{Points of Discordance}
\label{sec:pod}

In this section, we demonstrate how our framework allows for the rigorous decomposition of a patient pathway into its constitute concordant and discordant parts, which in turn allows for a clean attribution of a pathway's concordance score to interpretable subpathways. When aggregated over a population, we can identify key activities or sequences of activities that are the root of major discordances in the system. We can also compare subpopulations and investigate differences in their points of discordance. Our analyses are conducted using the 2012-2016 dataset. 

\subsection{Pathway Decomposition}
\label{sec:pod_decomposition}

Let $\bx^*$ be a reference pathway, assumed to be a shortest path in the clinical activity network, defined by $K$ ordered concordant activities $\gamma_1, \ldots, \gamma_K$. We represent $\bx^*$ as $(\gamma_0,\gamma_1, \gamma_2, \ldots, \gamma_K, \gamma_{K+1})$, where $\gamma_0$ and $\gamma_{K+1}$ are the artificial START and END nodes of the network. Let $\hat\bx$ be an arbitrary patient pathway written as $(\gamma_0, \Delta_0, \Gamma_1, \Delta_1, \Gamma_2, \Delta_2, \ldots, \Gamma_K, \Delta_{K}, \gamma_{K+1})$, where $\Gamma_i$ represents potential concordant activities and $\Delta_i$ represents potential discordant activities. Each $\Gamma_i, i = 1, \ldots, K$, is either $\emptyset$ or $\gamma_i$. It is $\emptyset$ if concordant activity $\gamma_i$ is missing or if a downstream concordant activity $\gamma_j, j > i$, is completed before activity $\gamma_i$ is visited for the first time. Our modeling framework adheres to the principle of optimality, so anytime a concordant activity $\gamma_j$ is reached in a pathway, then the reference pathway from $\gamma_j$ onwards is the lowest cost subpathway to END, even if concordant activities $\gamma_i, i < j$, are missing. 
In other words, it costs more to go back and complete missed activities than it does to simply continue from $\gamma_j$ onwards. Each $\Delta_i$ is either $\emptyset$ or an ordered set of discordant activities, which may include discordant activities like an emergency department visit or repeated imaging. As a convention, we assume that $\Delta_i = \emptyset$ if $\Gamma_i = \emptyset$, to avoid having to make a subjective choice of which set, $\Delta_i$ or $\Delta_{i-1}$, is assigned potential discordant activities if $\gamma_i$ is missing.

A \emph{detour} is formed any time the patient pathway deviates from the reference pathway. The detour ends when it rejoins the reference pathway at a later concordant activity (or the END node). The simplest example of a detour would be when there is a set of discordant activities $\Delta_i \ne \emptyset$ between concordant activities $\gamma_i$ and $\gamma_{i+1}.$ The detour comprises the extra arcs between $\gamma_i$ and $\gamma_{i+1}$. Other examples of detours include missing a concordant activity $\gamma_i$ ($\Gamma_i = \Delta_i = \Delta_{i-1} = \emptyset$) or combinations of missing and extra steps (e.g., $\Gamma_i = \Delta_i = \emptyset$ but $\Delta_{i-1} \ne \emptyset$).

Next, we define the cost of a detour. Given a sequence of activities $(\nu_1, \ldots, \nu_k)$, we define
\begin{equation*}
    \phi((\nu_1, \ldots, \nu_k))= \sum_{i=1}^{k-1} c^*_{\nu^e_{i} \nu^s_{i+1}} + \sum_{i=1}^{k} c^*_{\nu^s_{i} \nu^e_{i}}
\end{equation*}
to be the cost along all arcs formed by this sequence using the optimal cost vector, where $\nu_i^s$ and $\nu_i^e$ are the start and end node for activity $\nu_i$ (recall construction of clinical activity network in Section~\ref{sec:network}). Now, consider a detour $\theta$ that starts at concordant activity $\gamma_i$, ends at concordant activity $\gamma_{i+l}$ (skipping activities $\gamma_{i+1}, \ldots, \gamma_{i+l-1}$), and includes discordant activities $\Delta_{i} = (\delta_1, \ldots, \delta_k)$ in between. We define the cost of this detour, $C(\theta)$, based on the cycle that is induced by the extra discordant activities and missing concordant activities between $\gamma_i$ and $\gamma_{i+l}$. That is,

\begin{equation} 
\label{eq:detourcost}
    C(\theta) = \frac{  \overbrace{c^*_{\gamma^e_i \delta^s_1}+ \phi(\Delta_{i}) + c^*_{\delta^e_k \gamma^s_{i+l}}}^{\text{extra discordant arcs}}-\overbrace{\left( c^*_{\gamma^e_i \gamma^s_{i+1}}+\phi((\gamma_{i+1}, \ldots, \gamma_{i+l-1}))+ c^*_{\gamma^e_{i+l-1} \gamma^s_{i+l}} \right)}^{\text{missing concordant arcs}}  } {M(\hat\bx)-\bc^*{'}\bx^*}.
\end{equation}

If detour $\theta$ has only missing concordant activities but no extra discordant activities, then the cost of extra discordant arcs is simplified to $c^*_{\gamma_i^e \gamma_{i+l}^s}$. On the other hand, if detour $\theta$ has no missing concordant activities and only extra discordant activities following activity $\gamma_i$ ($l=1$), then the cost of missing concordant arcs is simplified to $c^*_{\gamma_i^e \gamma_{i+1}^s}$. 

With this definition of detour cost, we can fully partition the discordance of any patient pathway $\hat\bx$, which is captured by its duality gap $\epsilon$, into the cost of its detours.

\begin{theorem}
\label{prop:detours}
Let $\Theta$ be the set of all detours in patient pathway $\hat\bx$. Then $\omega(\hat\bx) = 1-\sum_{\theta \in \Theta} C(\theta).$
\end{theorem}

Theorem \ref{prop:detours} states that the discordance associated with a patient pathway is entirely attributable to its detours. In a similar fashion, detour costs can be further broken down by the individual missing and/or extra activities in the detour. 

In the next two subsections, we illustrate how this concept of detour costs can be used to identify the major sources of discordance in the population and between different subpopulations. We note that some of the subsequent analysis, particularly descriptive statistics around prevalence of discordant activities and detours, can be accomplished without inverse optimization. However, these results are firmly grounded in our network-based approach to modeling the concordance problem. Modeling the concordance problem as an inverse shortest path problem allows us to leverage the graph structure to very precisely define detours. And as shown in Theorem 1, discordance is exactly equivalent to detour costs. All of this leverages the principle of optimality, which makes detours interpretable and gives the concordance metric an additive property over the detours. Interpretability is a critical application of this work. The purpose of identifying points of discordance is to address them subsequently. To the extent that minimizing discordance necessitates change management in the clinical environment, stakeholder buy-in requires that the data prompting change is accurate and interpretable.

\subsection{Population-level Discordance Analysis}
\label{sec:pod_population}
\begin{figure}[t]
\centering
\includegraphics[width=\textwidth]{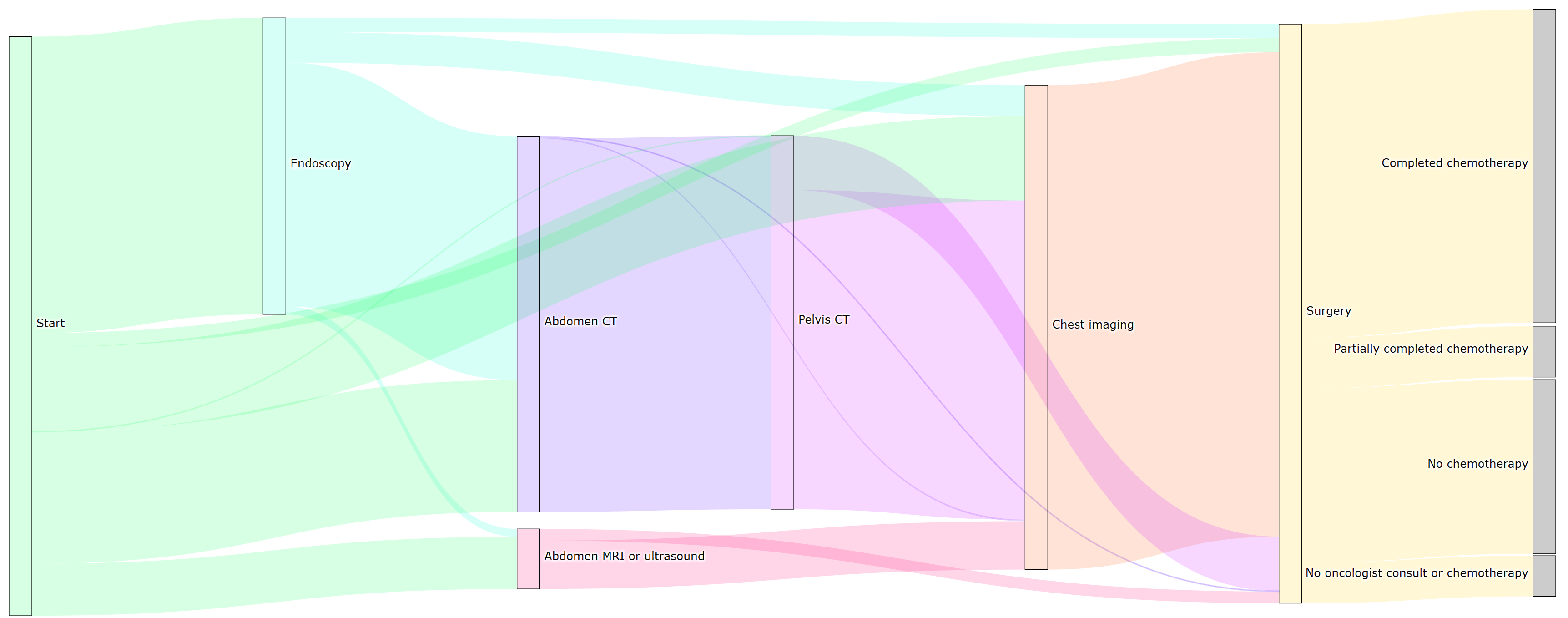}
\caption{Visualization of proportion of patients that flow between concordant activities in clinical activity network}
\label{fig:sankflow}
\end{figure}

Figure \ref{fig:sankflow} provides a high-level view of how patients flow through the clinical activity network. Roughly half 
of the patients start with the first concordant activity, endoscopy. The majority of the discordant individuals go directly to abdomen or chest imaging. 
About half of the patients complete chemotherapy and overall only 3\% of patients follow a shortest path. 

Figure \ref{fig:state_discordance} breaks down the proportion of the population discordance by the starting node of detours. We observe that the majority of discordance occurs following surgery. This means that detours after surgery are more prevalent and/or more costly (in terms of discordance score) than after other activities. The next two nodes where the greatest discordance originates is after chest imaging and at the START node (i.e., before completing a single concordant activity). We focus on detours originating from these three nodes in the remainder of the analysis, since they comprise about 94\% of the total population discordance.

Table \ref{tab:population1} shows that detours that originate closer to the start of the pathway tend to be shorter, while those that originate towards the end of the pathway, around the surgery event, tend to be longer and more complex. From the START node, 56\% of the patient pathways begin with a detour. These detours are relatively short, with at most five extra discordant activities, and many that only have missing activities with no extra activities. The chest imaging node, which is visited by 84\% of the patient population, generates a mix of concordant transitions, short detours (i.e., fewer than five extra discordant activities), and a small fraction of long detours (i.e., six or more extra discordant activities). The most complex detours arise from the surgery node. After surgery, 41\% of patients take a detour with over five extra discordant activities. This behavior is nonexistent at the start of the pathway (0\%) and infrequent after chest imaging (11\%).

\begin{table}[t]
\centering

\centering
\caption{Percentage (\%) of detour types originating from different nodes in clinical activity network for the entire patient population}
\label{tab:population1}
\begin{tabular}{lccc}
\toprule
                & Start          & Chest imaging & Surgery        \\
\midrule
Concordant                  & 44           & 37          & 5           \\ 
Missing only                & 21          & 0          & 5           \\ 
Extra only (1-5)            & 7           & 52         & 32           \\ 
Missing and extra (1-5)     & 28           & 0          & 17           \\ 
Extra only (6-10)           & 0            & 8           & 12           \\ 
Missing and extra (6-10)    & 0            & 0           & 12           \\ 
Extra only (\textgreater{}10)        & 0            & 3           & 5            \\ 
Missing and extra (\textgreater{}10) & 0            & 0          & 12          \\ 
\midrule
\textbf{Total}              & \textbf{100} & \textbf{100} & \textbf{100} \\ 
\bottomrule
\end{tabular}
\end{table}

Finally, we decompose the cost of a detour into costs associated with individual missing and extra activities based on the proportion of the detour cost in the arcs associated with individual activities. Table \ref{tab:population2} clearly shows that the detour costs originating from the surgery node are dominant. Following surgery, discordance is primarily due to extra imaging activities, then missing chemotherapy, ED visits, and extra clinical consultations. Following a concordant chest imaging step, the largest detour costs are due to extra imaging, followed by extra consultations, ED visits, and extra endoscopies. Finally, from the START node, the two major contributors to discordance are ED visits and missing concordant activities (primarily endoscopies).

\begin{table}[t]
\centering
\caption{Attribution of discordance ($1-\omega$)  to missing and extra activities for the entire patient population}
\label{tab:population2}
\begin{tabular}{lccc}
\toprule
                & Start          & Chest imaging & Surgery        \\
\midrule

Missing activities           & 0.024 & 0.000         & 0.047   \\
ED visit                     & 0.028 & 0.007         & 0.039   \\
Extra consultation           & 0.002 & 0.012         & 0.037   \\
Extra endoscopy              & 0.000 & 0.006         & 0.003   \\
Extra abdomen/pelvis imaging & 0.000 & 0.020         & 0.071   \\
Extra chest imaging          & 0.000 & 0.017         & 0.091   \\
\midrule
\textbf{Total}               & \textbf{0.054} & \textbf{0.062} & \textbf{0.288} \\
\bottomrule
\end{tabular}
\end{table}

A key insight from these results is that the distribution of detour types and discordant activities that comprise these detours (both missing and extra activities) varies as patients progress through the pathway map. While extra imaging as a source of discordance has been demonstrated previously \citep{simos2015physicians}, our results highlight that they primarily contribute to higher discordance scores after surgery has been completed. ED visits, which are generally undesirable from a health system perspective \citep{doan2019impact,yun2017cost,sun2013effect} account for a much higher proportion of total discordance at the diagnostic phase of patient pathways. As patients progress on their cancer journey, they tend to accumulate extra discordant activities, especially after surgery. This result suggests that extra follow-ups may be needed with patients following surgery, to make sure they are following the appropriate treatment course to complete their cancer care journey. Since the intended use of concordance measurement is quality improvement, monitoring variation in patient pathways is critical to identifying and eventually reducing detours.

Absence of endoscopy should be investigated as a potential quality gap. Policy makers can then tailor quality improvement initiatives based on the possible reasons for missing endoscopy. For example, if the absence of endoscopy is due to emergency presentation, then initiatives to increase screening participation should be considered. If the reason is difficulty in accessing endoscopy services, then process or capacity improvement initiatives should be undertaken. If the reason is an education gap (e.g., a CT scan was requested to evaluate symptoms rather than an endoscopy), the initiative should include improved knowledge translation and exchange for primary care providers.

ED visits are another major point of discordance. To address this quality gap, the first step is to determine which ED visits are avoidable. ED visits that did not lead to hospital admission could potentially be managed in another setting like outpatient clinics. Identifying the presenting symptoms leading to an ED visit may yield targeted quality improvement opportunities. For example, a telephone triage service or nurse-led clinics could be tapped to handle non-emergent issues arising between scheduled follow-up visits. Moreover, additional points of discordance (e.g., extra imaging) may occur as a direct result of ED attendance. Identifying these subsequent discordant activities is critical in assessing the total impact of interventions to reduce ED visits. For example, suppose it was determined that a large number of ED visits could be avoided with the establishment of a telephone triage service. Our concordance metric can be used to model the impact of that service on the measure of concordance as well as the net change in cost from reduced ED visits, including avoidance of subsequent additional imaging that is a direct result of those ED visits.

\subsection{Comparing Screened and Unscreened Patients}
\label{sec:pod_screening}
This final subsection studies how two subpopulations of patients -- those who get repeated screening versus those who do not get any screening -- differ in their discordance. Comparing these two particular subpopulations is meaningful because organized screening programs are associated with a decreased risk of developing advanced stages of cancer \citep{james2018repeated}. Patients with repeated screening, defined as those with ``two or more FOBTs [fecal occult blood tests] where the latest FOBT was performed within nine months of diagnosis and a second was performed 12–24 months prior to the latest FOBT'' were significantly less likely to present with stage IV colorectal cancer later. 
In both groups, we only include patients with stage IIIB colon cancer (controlling for substage with the largest of the three substages), no comorbidities, and that are between 60 and 79 years of age. After applying these inclusion criteria, there were 76 patients in the screened group and 593 patients in the unscreened group.

Table \ref{tab:screening1} summarizes pathways of the screened and unscreened groups following the start, chest imaging, and surgery nodes. From the start node, 68\% of patients in the screened group are concordant compared to only 41\% in the unscreened group; unscreened patients have significantly more detours with both missing concordant activities and extra discordant activities. Unscreened patients also tend to have longer detours following chest imaging and surgery. Overall, the pathways of patients in the unscreened group are less likely to complete endoscopy and chemotherapy, and are more likely to have extra activities.

\begin{table}[t]
\centering
\caption{Percentage (\%) of detour types originating from different nodes in clinical activity network for screened versus unscreened patients
}
\label{tab:screening1}
\begin{tabular}{l c c c c c c c c c}
\toprule
                       & \multicolumn{3}{c}{Start}                   & \multicolumn{3}{c}{Chest imaging}                 & \multicolumn{3}{c}{Surgery}  
 \\ \cmidrule(r){2-4} \cmidrule(l){5-7} \cmidrule(l){8-10}
        & Unscr          & Scr  &  Diff        & Unscr          & Scr &  Diff        & Unscr          & Scr &  Diff      \\ 
\midrule
Concordant                  & 41           & 68          & -27            & 39           & 29        & 10            & 6        & 7           & -1         \\ 
Missing only                & 19           & 18          & 1            & 0           & 0           & 0            & 4          & 3           & 1          \\ 
Extra only (1-5)            & 8            & 5           & 3           & 51            & 67          & -16           & 35          & 42         & -7         \\ 
Missing and extra (1-5)     & 32            & 9          & 23           & 0           & 0           & 0           & 12         & 11           & 1          \\ 
Extra only (6-10)           & 0            & 0           & 0           & 8            & 4           & 4           & 17          & 17           & 0          \\ 
Missing and extra (6-10)    & 0            & 0           & 0            & 0            & 0           & 0            & 9          & 7           & 2          \\ 
Extra only (\textgreater{}10) & 0          & 0           & 0            & 2            & 0           & 2            & 6          & 5           & 1          \\ 
Missing and extra (\textgreater{}10) & 0    & 0           & 0            & 0            & 0           & 0           & 11          & 8           & 3          \\ 
\midrule
\textbf{Total}              & \textbf{100} & \textbf{100} & \textbf{0} & \textbf{100} & \textbf{100} & \textbf{0} & \textbf{100} & \textbf{100} & \textbf{0} \\ 
\bottomrule

\end{tabular}
\end{table}

Table \ref{tab:screening2} breaks down discordance scores $(1-\omega)$ for the two subpopulations. For each node, the average value of $1-\omega$ is greater in the unscreened group than the screened group. At the start node, the difference (0.060 vs. 0.029) is largely due to two factors: unscreened patients are more likely to miss concordant activities (19\% of the net difference in $1-\omega$) and unscreened patients are incurring more ED visits (74\% of the net difference in $1-\omega$). The latter appears to be a key difference between these two groups. ED visits followed by missing endoscopy are the two major contributors to discordance at the start node in the general population (Table~\ref{tab:population2}) as well as the unscreened group. However, in the screened group, the discordance due to both are significantly diminished, especially ED visits, and their order is swapped. From chest imaging, the higher discordance of the unscreened group (0.063 vs. 0.058) is largely due to extra imaging (60\% of the net difference in $1-\omega$). Finally, at the surgery node, the higher discordance of the unscreened group (0.257 vs. 0.232) is largely due to extra imaging activities (32\% of the net difference in $1-\omega$) and missing chemotherapy (40\%) with additional discordance primarily attributed to ED visits and extra consultations.

\begin{table}[t]
\centering
\caption{Attribution of discordance ($1-\omega$) to missing and extra activities in screened versus unscreened patients}
\label{tab:screening2}
\begin{tabular}{l c c c c c c c c c}
\toprule
                       & \multicolumn{3}{c}{Start}                   & \multicolumn{3}{c}{Chest imaging}                 & \multicolumn{3}{c}{Surgery}  
 \\ \cmidrule(r){2-4} \cmidrule(l){5-7} \cmidrule(l){8-10}
        & Unscr          & Scr  &  Diff        & Unscr          & Scr &  Diff        & Unscr          & Scr &  Diff      \\ 
\midrule

Missing activities           & 0.025 & 0.019 & 0.006 & 0.000 & 0.000 & 0.000  & 0.035 & 0.025 & 0.010   \\
ED visit                     & 0.032 & 0.009 & 0.023 & 0.007 & 0.004 & 0.003  & 0.033 & 0.030 & 0.003  \\
Extra consultation           & 0.003 & 0.000 & 0.003 & 0.012 & 0.014 & -0.002 & 0.036 & 0.034 & 0.002  \\
Extra endoscopy              & 0.000 & 0.000 & 0.000 & 0.007 & 0.006 & 0.001  & 0.003 & 0.001 & 0.002  \\
Extra abdomen/pelvis imaging & 0.000 & 0.001 & -0.001& 0.020 & 0.018 & 0.002  & 0.068 & 0.071 & -0.003 \\
Extra chest imaging          & 0.000 & 0.000 & 0.000 & 0.017 & 0.016 & 0.001  & 0.082 & 0.071 & 0.011  \\
\midrule
\textbf{Total}                        & \textbf{0.060} & \textbf{0.029} & \textbf{0.031} & \textbf{0.063} & \textbf{0.058} & \textbf{0.005 } & \textbf{0.257} &\textbf{ 0.232 }& \textbf{0.025}  \\

\bottomrule

\end{tabular}
\end{table}

\citet{james2018repeated} hypothesized that the reason for improved patient outcomes among patients who are part of an organized screening program is that they are more likely to have access to other program features and their associated benefits. These patients have better access to care and are more likely to seek it. Our results support this hypothesis and provide more granular insight into where the potential benefit arises at different parts of the pathway.

\section{Conclusions} 
This paper proposes the first data-driven inverse optimization approach to measuring pathway concordance in any problem context. Our specific development focuses on clinical pathway concordance for cancer patients. In contrast to previous work in clinical pathway concordance, which focus on single events or courses of treatment by patients in a single institution, our work focuses on population-level health system monitoring and reporting. Given its statistically significant association with clinical outcomes, our metric provides meaningful quantitative measures of system efficiency and variation. Using our metric to identify the points of discordance in a health system allows officials to evaluate complex practice patterns at the population level and describe clinical outcomes in relation to variation in care. Concordance metrics optimized for clinical outcomes are also useful in promoting and implementing best practice, helping providers focus on appropriate interventions and reduce inappropriate resource use. 

As a concrete application, we perform an in-depth case study using stage III colon cancer patient data, demonstrating how our concordance metric identifies population-level discordance, as well as points of discordance between patients who had repeated screening versus those who did not. Our framework rigorously attributes discordance to different sources throughout the pathway, ultimately providing targets for quality improvement initiatives and interventions to improve system performance. The development of our population-level metric is enabled by access to population-level administrative health data and pathway maps that are designed for the population, rather than individuals. Future work will focus on the application of the metric for identifying and prioritizing quality gaps that can be addressed with specific quality improvement initiatives, and on modeling the impact of different interventions on population concordance.

\ECSwitch
\ECHead{Electronic Companion}

\section{Proofs}
\label{app:proofs}
\proof{Proof of Lemma \ref{lem:feasible}.} Since the objective function has a lower bound of zero, it suffices to show that $\mathbf{IO}^{\textrm{ref}}(\hat{\mathcal{X}}^r)$ is feasible if and only if there are at least two distinct paths from the start node (denoted $s$) to the end node (denoted $t$).\\
($\Leftarrow$) Let $P_1$ and $P_2$ be two distinct paths from $s$ to $t$. Let $P_1\setminus P_2 =\{(i,j)\in \mathcal{A} \,|\, (i,j)\in P_1 \text{ and }(i,j)\notin P_2\}$ and $P_2\setminus P_1= \{(i,j)\in \mathcal{A} \,|\, (i,j)\in P_2 \text{ and }(i,j)\notin P_1\}$. If we set $c_{ij} = -1$ for $(i,j) \in P_1\setminus P_2$, $c_{ij} = 1$ for $(i,j) \in P_2\setminus P_1$, and $c_{ij} = 0$ for all the other arcs in the network, then $\bc$ satisfies $\|\bc\|_\infty = 1$ and $\bA\bc=\bzero$. Note that $P_1$ is the shortest path for the designed cost vector $\bc$ which has finite total cost. From strong duality, its dual is also optimal meaning that there exists a $\bp$ that satisfies constraint $\bA'\bp \le \bc$ for $\bc$. Finally, we build a feasible solution for $\mathbf{IO}^{\textrm{ref}}(\hat{\mathcal{X}}^r)$ by setting $\epsilon^r_q = \bc'\hat\bx^r_q - \bb'\bp$ for $q=1, \ldots, R$.  \\
($\Rightarrow$) Suppose to the contrary that there is only one path from $s$ to $t$, which means the entire network itself is the path from $s$ to $t$. Since there is a feasible $\bc$, there is some arc $(i,j)$ with cost $|\bc_{ij}| = 1$. The constraint $\bA\bc = \bzero$ implies that the incoming arc to $i$ and outgoing arc from $j$ also has cost equal to 1 in absolute value. Continuing this logic, the same will be true of the outgoing arc from $s$, but this violates $\bA\bc = \bzero$, meaning $\bc$ cannot be feasible.
\Halmos
\endproof

\proof{Proof of Proposition \ref{prop:boundedness}.}
The dual of $\mathbf{FO}(\bc^*)$ is feasible since $\bc^*$ and $\bp^*$ satisfy the dual feasibility conditions. Hence, $\mathbf{FO}(\bc^*)$ is not unbounded, implying no negative cost cycles.
 \Halmos 
\endproof

\proof{Proof of Proposition \ref{prop:optsol2}.}
Let $(\bc^*,\bp^*, \bepsilon^{r*})$ be an optimal solution to $\mathbf{IO}^{\textrm{ref}}(\hat{\mathcal{X}}^r)$. We can construct a feasible solution to $\mathbf{IO}^{\textrm{pat}}(\hat{\mathcal{X}}^s,\hat{\mathcal{X}}^d,\bepsilon^{r*})$ by setting $\epsilon^s_q = \bc^*{'}\hat\bx^s_q - \bb'\bp^*$ for $q=1, \ldots, S$ and $\epsilon^d_q = \bc^*{'}\hat\bx^d_q - \bb'\bp^*$ for $q=1, \ldots, D$. Hence, what remains is to show that the objective is bounded. Substituting the third and fourth constraints into the objective, we have

\begin{align*}
\frac{D}{S}\sum_{q=1}^{S}\epsilon^s_q-\sum_{q=1}^{D}\epsilon^d_q & = \frac{D}{S}\left(\sum_{q=1}^{S} \bc'\hat\bx^s_q - S \bb'\bp\right)-\left(\sum_{q=1}^{D} \bc'\hat\bx^d_q - D \bb'\bp\right) \\
& = \frac{D}{S} \sum_{q=1}^{S} \bc'\hat\bx^s_q - \sum_{q=1}^{D} \bc'\hat\bx^d_q,
\end{align*}
which is clearly bounded below since $\hat\bx^s_q$ and $\hat\bx^d_q$ are data and $\|\bc\|_\infty = 1$. In particular, if $M = \max_{q,q'}\{\|\hat\bx^s_q\|, \|\hat\bx^d_{q'}\|\}$, then by Holder's inequality the absolute value of the objective is bounded by $2DM$.
\Halmos 
\endproof

\proof{Proof of Theorem \ref{prop:detours}.}

Taking the patient pathway $\hat\bx$ and subtracting the reference pathway $\bx^*$ results in a set of disjoint cycles, where the forward arcs in each cycle correspond to extra discordant activities (in the patient pathway but not in the reference pathway) and the backward arcs correspond to missing concordant activities (in the reference pathway but not in the patient pathway). If there are no repeated discordant activities within a detour, then there is a 1-1 correspondence between detours and cycles. Otherwise, a detour will generate one cycle with forward and backward arcs, as well as one or more cycles with only forward arcs corresponding to the repeated discordant activities. By construction (see equation~\eqref{eq:detourcost}), the numerator in the cost of a detour $\theta$ is equal to the cost around the cycle(s) that it generates. Thus, the cost difference $\bc^*{'}\hat\bx - \bc^*{'}\bx^*$ equals the sum of the costs around all cycles. Since each cycle belongs to only one detour $\theta \in \Theta$, $\omega(\hat\bx) := 1- (\bc^*{'}\hat\bx - \bc^*{'}\bx^*) / (M(\hat\bx)-\bc^*{'}\bx^*) =1- \sum_{\theta\in\Theta}C(\theta)$. 
\Halmos 
\endproof

\section{Calculation of the Cost of Longest Walk ($M(\hat\bx)$)}
\label{app:longestwalk}
We use the Bellman-Ford equations to calculate the longest walk for walks with up to a given number of steps. Let the node set be $\mathcal{N}$, the start node be $s$, and the end node be $t$. Let $C(v,k)$ be the cost of the longest walk from $s$ to $v \in \mathcal{N}$ that takes $k$ steps, computed via backward induction over the equations
\begin{align*}
    & C(v,k) = \underset{u\in\mathcal{N}}{\text{max}} \{C(u,k-1)+c(u,v) \}, \quad k=2,\ldots, N, \quad v \in \mathcal{N},\\
    & C(s,1) = 0,\\
    & C(v,1) = \infty, \quad v\neq s,
\end{align*}
where $c(u,v)=c_{uv}$ if $(u,v) \in \mathcal{A}$, and $c(u,v)=\infty$ otherwise. We solve this system of equations once up to a value of $N$ that equals to the cost of the longest patient pathway ($\max_{q,q'}\{\|\hat\bx^s_q\|, \|\hat\bx^d_{q'}\|\}$). Then, the costs of the longest walks with the number of steps for each integer up to $N$ are stored and can simply be looked up when computing $\omega(\hat\bx)$ for each patient pathway.

\section{Data Sources}
\label{app:data}

\begin{itemize}
    \item Cancer Care Ontario
    \begin{itemize}
        \item Ontario Cancer Registry (OCR)
        \item Collaborative Stage Dataset (CSI)
        \item Colonoscopy Interim Reporting Tool (CIRT)
        \item Lab Reporting Tool (LRT)
        \item Screening HUB
        \item Activity Level Reporting (ALR)
    \end{itemize}
    \item Ministry of Health and Long-Term Care (MoHLTC)
    \begin{itemize}
        \item Registered Persons Database (RPDB)
        \item Ontario Health Insurance Plan (OHIP) Claims
    \end{itemize}
    \item Canadian Institute for Health Information (CIHI)
    \begin{itemize}
        \item National Ambulatory Care Reporting System (NACRS)
        \item Discharge Abstract Database (DAD)
    \end{itemize}
    \item Statistics Canada
    \begin{itemize}
        \item Postal Code Conversion file (PCCF+6B)
    \end{itemize}

\end{itemize}

\section{Supplementary Tables and Figures}
\label{app:fp}

Table \ref{tab:descriptive2012} lists the characteristics of the patients in the 2012-2016 dataset and by concordance terciles.

\begin{table}[t]
\centering
\caption{Characteristics of stage III colon cancer cohort included in the survival analysis}
\label{tab:descriptive2012}
\begin{tabular}{L{6cm} c c c c c}
\toprule
Characteristic                                        & \begin{tabular}[c]{@{}c@{}}Low \\ (n=1414)\end{tabular}            & \begin{tabular}[c]{@{}c@{}}Medium \\ (n=1414)\end{tabular}     & \begin{tabular}[c]{@{}c@{}}High \\ (n=1414)\end{tabular}             & \begin{tabular}[c]{@{}c@{}}Total \\ (n=4242)\end{tabular}            & p-value          \\ \bottomrule

Death, n (\%)                                         & 508 (35.9)                                                       & 241 (17.0)                                                         & 131 (9.3)                                                         & 880 (20.7)                                                        & \textless{}0.0001 \\ 
Concordance score, mean (SD)                          & 0.4 (0.1)                                                         & 0.6 (0.1)                                                           & 0.8 (0.1)                                                           & 0.6 (0.2)                                                           & \textless{}0.0001 \\ \midrule

\multicolumn{6}{c}{Patient characteristics}                                                                                                                                                                                                                                                                                                                  \\ \midrule
Age at diagnosis, mean (SD)                           & 72.4 (13.6)                                                       & 65.9 (12.8)                                                         & 63.7 (11.4)                                                         & 67.3 (13.1)                                                         & \textless{}0.0001 \\ 

Female, n (\%)                                        & 747 (52.8)                                                      & 680 (48.1)                                                        & 680 (48.1)                                                        & 2107 (49.7)                                                       & 0.0145           \\ 
Rural residency, n (\%)                               & 205 (14.5)                                                       & 201 (14.2)                                                          & 196 (13.9)                                                         & 602 (14.2)                                                         & 0.8886           \\ 
\multicolumn{6}{l}{Neighbourhood Income Quintile, n (\%)}                                                                                                                                                                                                                                                                                                    \\ 
\quad 1 (lowest)                                            & 239 (16.9)                                                       & 307 (21.7)                                                         & 289 (20.4)                                                         & 835 (19.7)                                                        & 0.0505           \\ 
\quad 2                                                     & 283 (20.0)                                                       & 283 (20.0)                                                         & 305(21.6)                                                         & 871 (20.5)                                                        &                  \\ 
\quad 3                                                     & 274 (19.4)                                                       & 295 (20.9)                                                         & 253 (17.9)                                                        & 822 (19.4)                                                        &                  \\ 
\quad 4                                                     & 299 (21.1)                                                       & 283 (20.0)                                                         & 303 (21.4)                                                         & 885 (20.9)                                                        &                  \\ 
\quad 5 (highest)                                           & 239 (16.9)                                                       & 307 (21.7)                                                         & 289 (20.4)                                                         & 835 (19.7)                                                        &                  \\ 
\multicolumn{6}{l}{Neighbourhood  Immigration Tercile, n (\%)}                                                                                                                                                                                                                                                                                               \\ 
\quad 1 (lowest)                                            & 862 (61.0)                                                      & 893 (63.2)                                                        & 824(58.3)                                                        & 2579 (60.8)                                                        & 0.0077           \\ 
\quad 2                                                     & 351 (24.8)	&301 (21.3)	&333 (23.6)	&985 (23.2)
                                                        &                  \\ 
\quad 3 (highest)                                           & 201 (14.2)	&220 (15.6)	&257 (18.2)	&678 (16.0)
                                                    &                  \\ 
                                                    
Charlson score = 0, n (\%)                            & 1046 (74.0)	&1187 (84.0)	&1246 (88.1)	&3479 (82.0)
                                                        & \textless{}0.0001           \\ 
Number of ED Visits in the year before cohort entry, mean (SD)                      & 0.7 (1.8)	& 0.4 (1.0)	& 0.3 (0.8)	& 0.5 (1.3)
                                                           & \textless{}0.0001           \\ 

Number of OHIP Outpatient Visits in the year before cohort entry, mean (SD)         & 5.1 (6.6)	& 3.8 (4.8)	& 3.4 (4.7)	& 4.1 (5.5)
                                                        & \textless{}0.0001 \\
Number of hospital admissions in the year before cohort entry, mean (SD)
& 0.2 (0.6)	& 0.1 (0.3)	& 0.1 (0.3)	& 0.1 (0.5) & \textless{}0.0001\\

                                                        \midrule

\multicolumn{6}{c}{Cancer-related characteristics}                                                                                                                                                                                                                                                                                                           \\ \midrule
\multicolumn{6}{l}{Screening group, n (\%)}                                                                                                                                                                                                                                                                                                                  \\ 
\quad None                                                  & 870 (61.5)	& 739 (52.3)	& 654 (46.3)	& 2263 (53.3)
                                                        & \textless{}0.0001           \\ 
\quad Repeated                                              & 34 (2.4)	& 58 (4.1)	& 69 (4.9)	& 161 (3.8)
                                                          &                  \\ 
\quad Diagnostic                                            & 169 (12.0)	& 205 (14.5)	& 216 (15.3)	& 590 (13.9)
                                                         &                  \\ 
\quad Sporadic                                              & 341 (24.1)	& 412 (29.1)	& 475 (33.6)	& 1228 (28.9)
                                                        &                  \\ 
\multicolumn{6}{l}{Stage at diagnosis, n (\%)}                                                                                                                                                                                                                                                                                                               \\
\quad IIIA (least advanced)                                                 & 113 (8.0)	& 163 (11.5)	& 213 (15.1)	& 489 (11.5)
                                                          & \textless{}0.0001           \\ 
\quad IIIB                                                  & 976 (69.0)	& 985 (69.7)	& 946 (66.9)	& 2907 (68.5)
                                                        &                  \\ 
\quad IIIC (most advanced)                                                  & 325 (23.0)	& 266 (18.8)	& 255 (18.0)	& 846 (19.9)
                                                        &                  \\ 
\multicolumn{6}{l}{Tumor Grade Category, n (\%)}                                                                                                                                                                                                                                                                                                            \\ 
\quad Low                                                   & 1130 (79.9)	&1183 (83.7)	& 1184 (83.7)	& 3497 (82.4)
                                                        & 0.0359           \\ 
\quad High                                                  & 248 (17.5)	& 196 (13.9)	& 195 (13.8)	& 639 (15.1)
                                                        &                  \\ 
\quad Unknown                                               & 36 (2.5)	& 35 (2.5)	& 35 (2.5)	& 106 (2.5)
                                                          &                  \\ 
Emergent surgery, n (\%)                              & 586 (41.4)	& 290 (20.5)	& 104 (7.4)	& 980 (23.1)
                                                        & \textless{}0.0001 \\ 
Length of in-hospital stay post surgery, mean (SD)    & 12.0 (23.3)	& 6.3 (5.1)	& 4.7 (2.3)	& 7.7 (14.2)
                                                          & \textless{}0.0001 \\ 

\bottomrule

\end{tabular}
\end{table}

\newpage

\begin{figure}[H]
\centering
\includegraphics[width=0.7\textwidth]{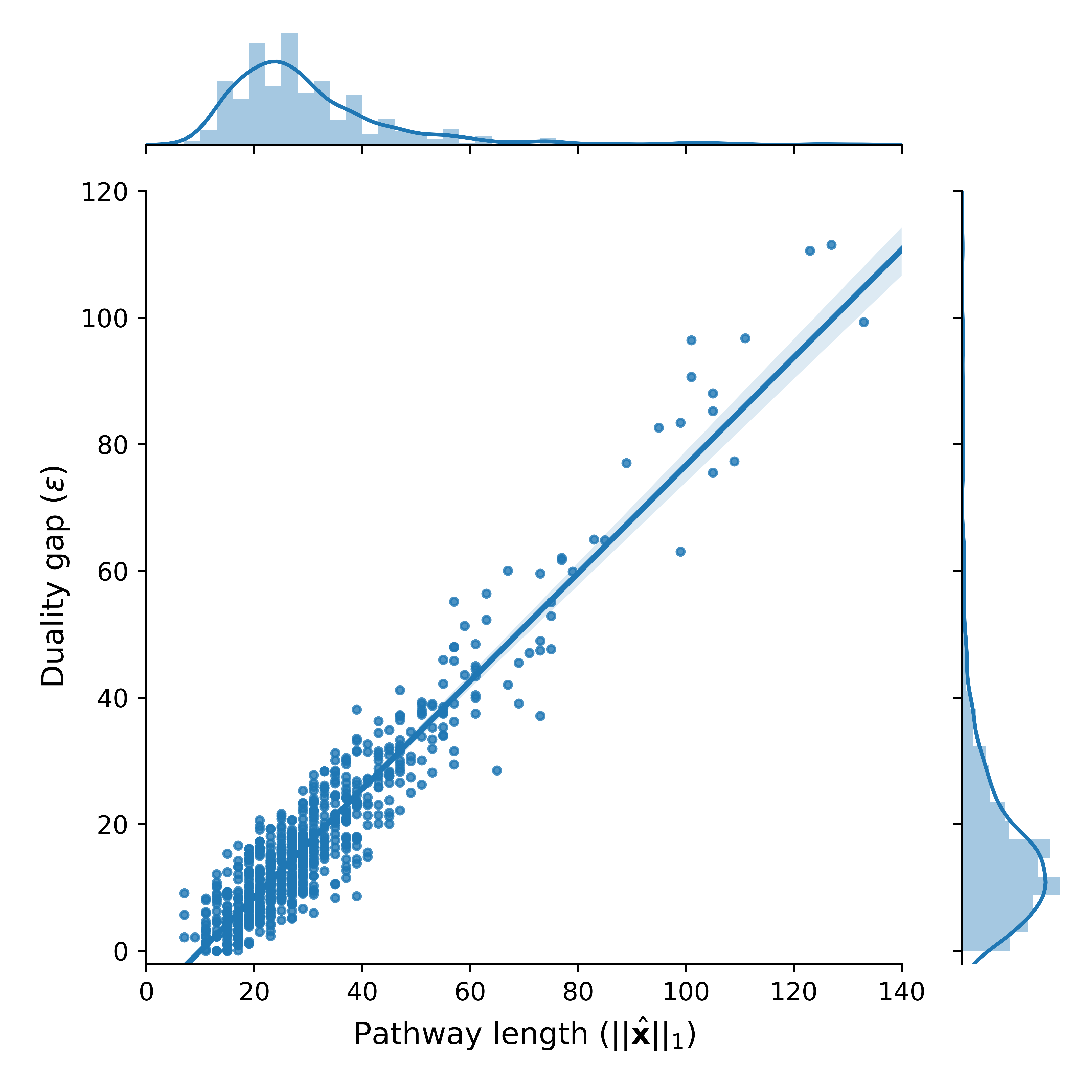}
\caption{Relationship between duality gap and pathway length}
\label{fig:epsilondist}
\end{figure}

\begin{figure}[H]
\centering
\includegraphics[scale=0.55]{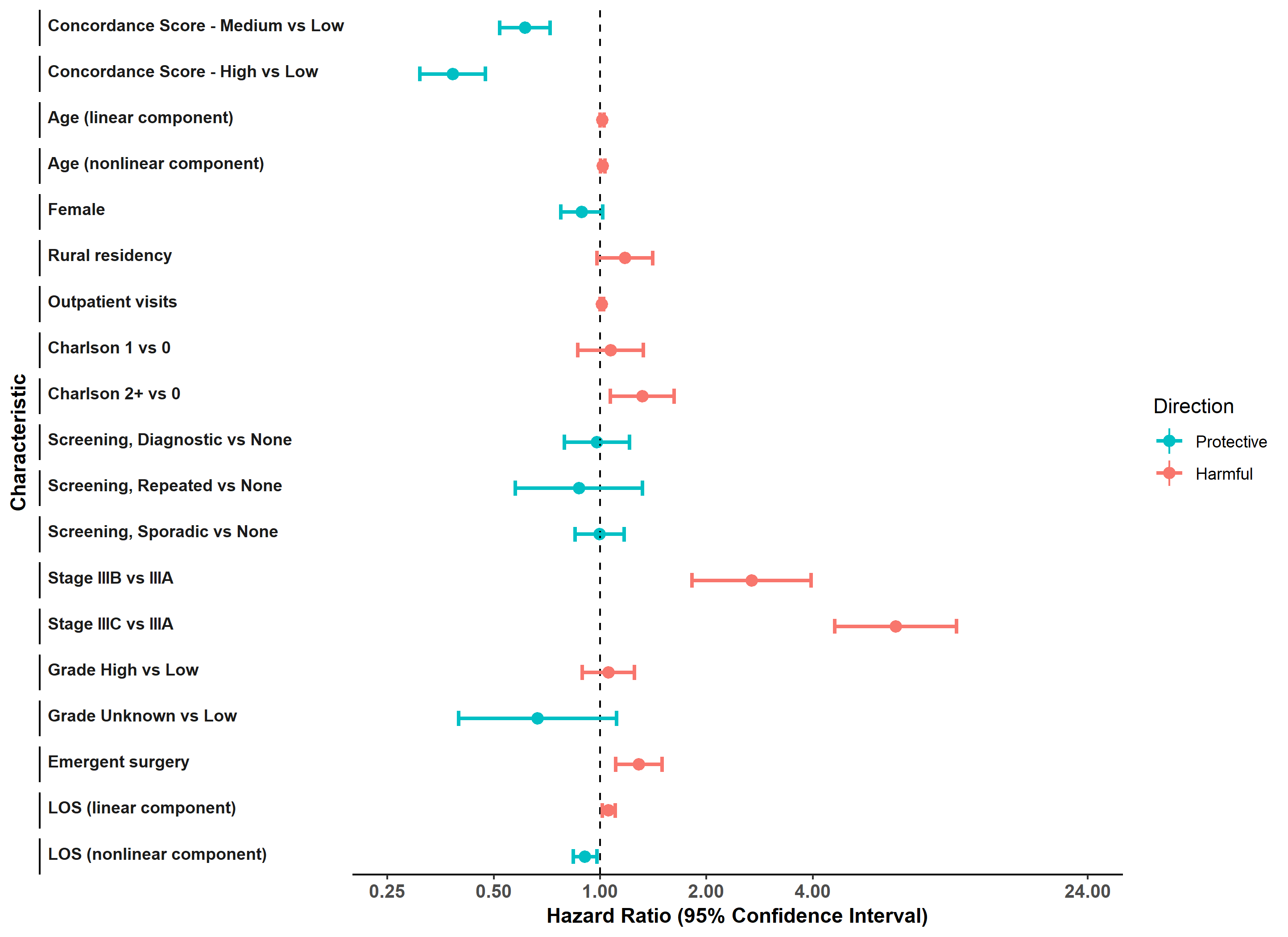}
\caption{Forest plot of variable effects in categorical Cox regression model with 95\% confidence intervals}
\label{fig:fp3}
\end{figure}

\begin{figure}[H]
\centering
\includegraphics[scale=0.55]{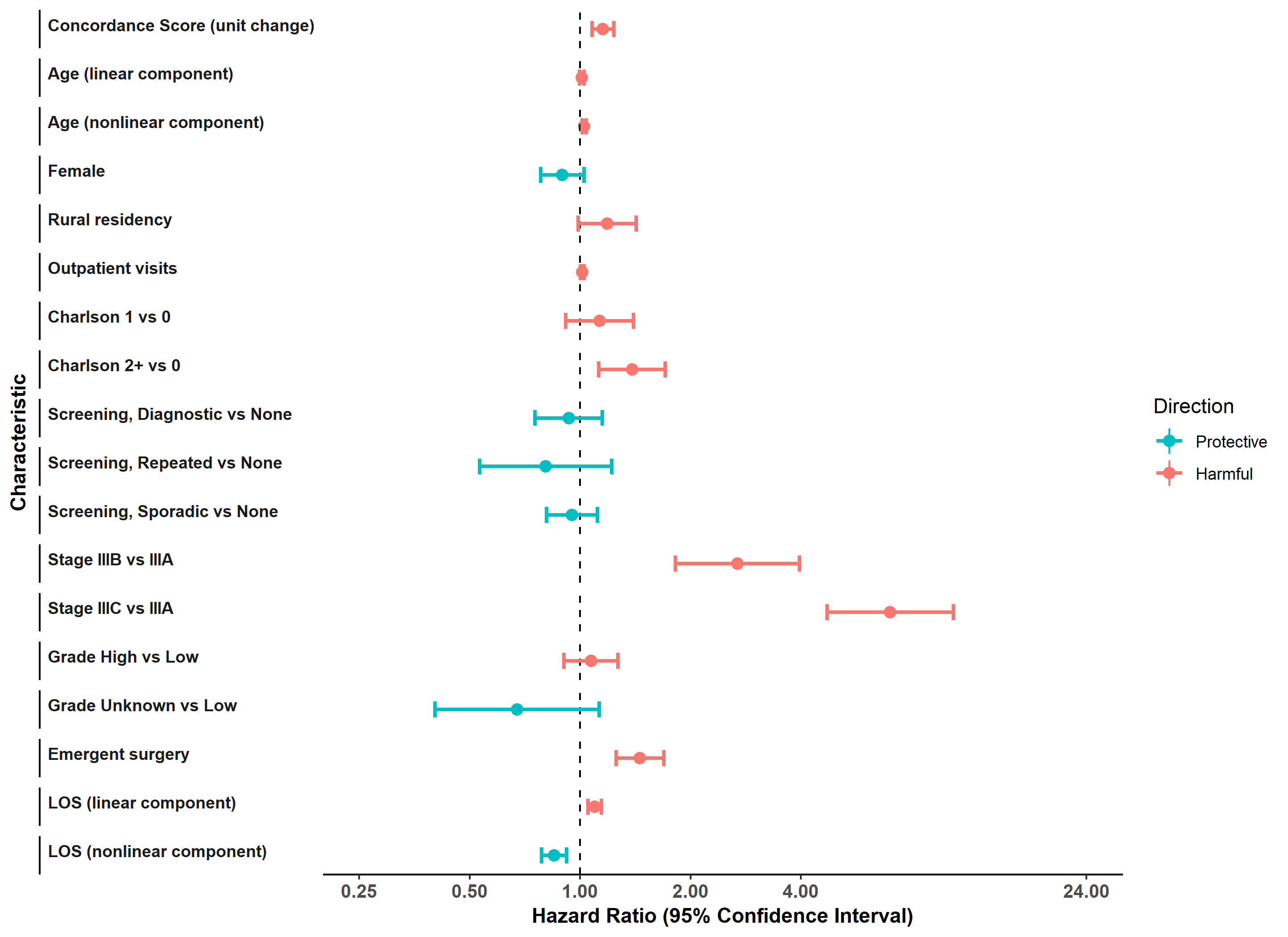}
\caption{Forest plot of variable effects in continuous Cox regression model with 95\% confidence intervals using concordance score based only on reference pathway data}
\label{fig:fp2_2012}
\end{figure}

\begin{figure}[h]
\centering
\includegraphics[scale=0.8]{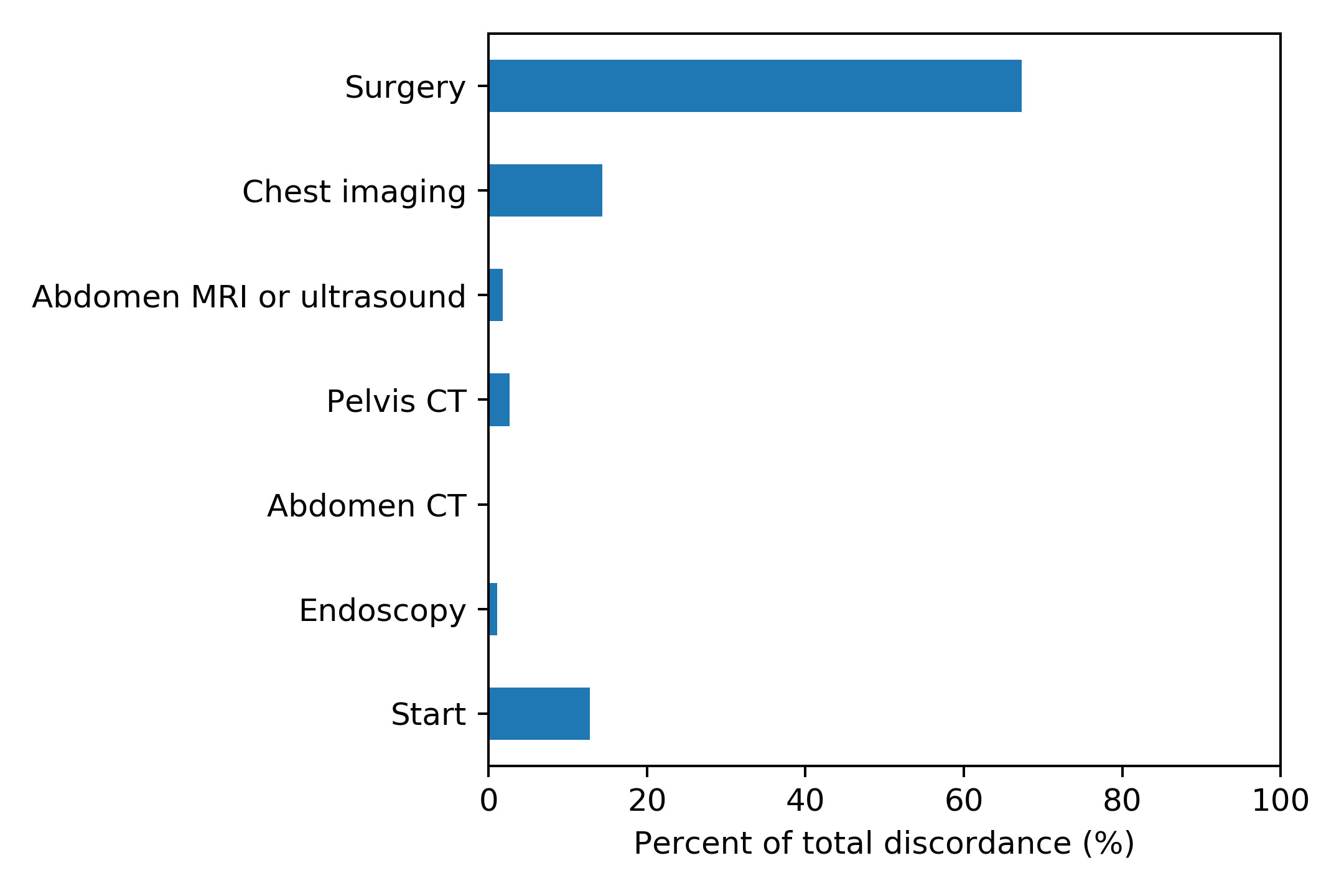}
\caption{Distribution of total patient discordance across concordant nodes}
\label{fig:state_discordance}
\end{figure}

Figure \ref{fig:epsilondist} shows the distributions of the duality gap and pathway length for 2010 dataset. Figure \ref{fig:fp3} summarizes the influence of each covariate in the adjusted categorical Cox model from Section~\ref{sec:survivalresults2012} on survival. Figure \ref{fig:fp2_2012} depicts a similar forest plot, but for the adjusted continuous Cox model with the concordance score based on the reference pathway-only model from Section~\ref{sec:valueofdata2012}. Figure \ref{fig:state_discordance} shows the distribution of discordance due to detours starting from different nodes in the pathway.

\section{Unnormalized Concordance Scores}
\label{sec:epsilon2012}
To test the robustness of our survival analysis results, we repeated the analysis using the duality gap $\epsilon(\hat\bx) := \bc^*{'}\hat\bx - \bc^*{'}\bx^*$ as an unnormalized concordance metric, instead of $\omega(\hat\bx)$. As shown in Table \ref{tab:epsilon}, we observe a significant association between concordance and survival. Note that smaller values of $\epsilon$ imply higher concordance, which is why an HR value greater than 1 indicates the appropriate association between concordance and survival. Furthermore, the same covariates that were significant in the original analysis remain significant now.

\begin{table}[t]
\centering
\caption{Risk of mortality associated with unnormalized concordance score for categorical and continuous models}
\label{tab:epsilon}
\begin{tabular}{llllllll}
\toprule
                             &               & \multicolumn{3}{c}{Unadjusted}                                                             & \multicolumn{3}{c}{Adjusted}                                                      \\ \cmidrule(r){3-5} \cmidrule(l){6-8}

                     &       & \multicolumn{1}{c}{HR}   & \multicolumn{1}{c}{95\% CI} & \multicolumn{1}{c}{p-value} & \multicolumn{1}{c}{HR}   & \multicolumn{1}{c}{95\% CI} & \multicolumn{1}{c}{p-value} \\ \midrule
\multirow{2}{*}{Categorical} & $\epsilon$ (Med. vs Low) & \multicolumn{1}{c}{2.07} & \multicolumn{1}{c}{(1.69, 2.53)}  & \multicolumn{1}{c}{\textless{}0.0001}   & \multicolumn{1}{c}{1.60} & \multicolumn{1}{c}{(1.30, 1.96)}  & \multicolumn{1}{c}{\textless{}0.0001}   \\
                             & $\epsilon$ (High vs Low)   & \multicolumn{1}{c}{3.73} & \multicolumn{1}{c}{(3.09, 4.50)}  & \multicolumn{1}{c}{\textless{}0.0001}   & \multicolumn{1}{c}{2.16} & \multicolumn{1}{c}{(1.75, 2.65)}  & \multicolumn{1}{c}{\textless{}0.0001}   \\ \midrule
Continuous                   & $\epsilon$ (unit change)  & \multicolumn{1}{c}{1.03} & \multicolumn{1}{c}{(1.02, 1.03)}  & \multicolumn{1}{c}{\textless{}0.0001}   & \multicolumn{1}{c}{1.02} & \multicolumn{1}{c}{(1.01, 1.02)}  & \multicolumn{1}{c}{\textless{}0.0001}   \\ \bottomrule
\end{tabular}
\end{table}

\section{Sensitivity Analysis}
\label{sec:sensitivity}

As a final sensitivity analysis, we investigate how the association between concordance and survival can be affected by a potential unobserved confounder. 
For example, patient behavioral characteristics are a well-known confounder \citep{mackian2004up,norman2015}. Patients with appropriate health-seeking behavior may be more likely to follow guidelines \citep{sewitch2007adherence} and thus have higher concordance. But they are also more likely to engage in healthy behaviours that may be associated with higher survival due to reductions in cancer and non-cancer-related risk factors such as physical activity, smoking and diet.

Our approach follows an established Monte Carlo simulation method \citep{lash2003semi,albert2012effectiveness,corrao2014statins}. We represent the unobserved confounder as a binary variable, $U$, where a value of 1 indicates a trait that affects both concordance and survival (e.g., health-seeking behavior). Each patient is placed in one of six ``concordance-mortality'' categories: a combination of the concordance ($\omega$) terciles from Section \ref{sec:survivalresults2012} (low, medium, and high) and the clinical outcome (survived or died). The prevalence of $U=1$ differs by category. 

We assume the prevalence of $U=1$ in patients who survived is $b$-fold higher than those who died. We also assume the prevalence of $U=1$ in high concordant (resp. medium concordant) patients is $RR$-fold higher than medium concordant (resp. low concordant) patients, where $RR$ denotes the Risk Ratio. A Risk Ratio of 1 is equivalent to no relationship between $U$ and concordance, representing a scenario when the estimated coefficient for the concordance score is unchanged.

We consider multiple scenarios, distinguished by the base prevalence assigned to the medium-survived category. The prevalence in the other categories is determined by $b$ and $RR$. In each scenario, we use the prevalence of $U=1$ in each category as the parameter of a Bernoulli random variable to generate a value of $U$ for each patient in that category. After generating $U$ for all the patients, we repeat the survival analysis from Section~\ref{sec:survivalresults2012} using the same models as before but with the additional covariate $U$. We report the HR for the concordance metric and the corresponding confidence intervals based on 100 simulations for each scenario. 

\begin{figure}[H]
\centering  
\subfigure[base prevalence = 0.1]{\includegraphics[width=0.47\linewidth]{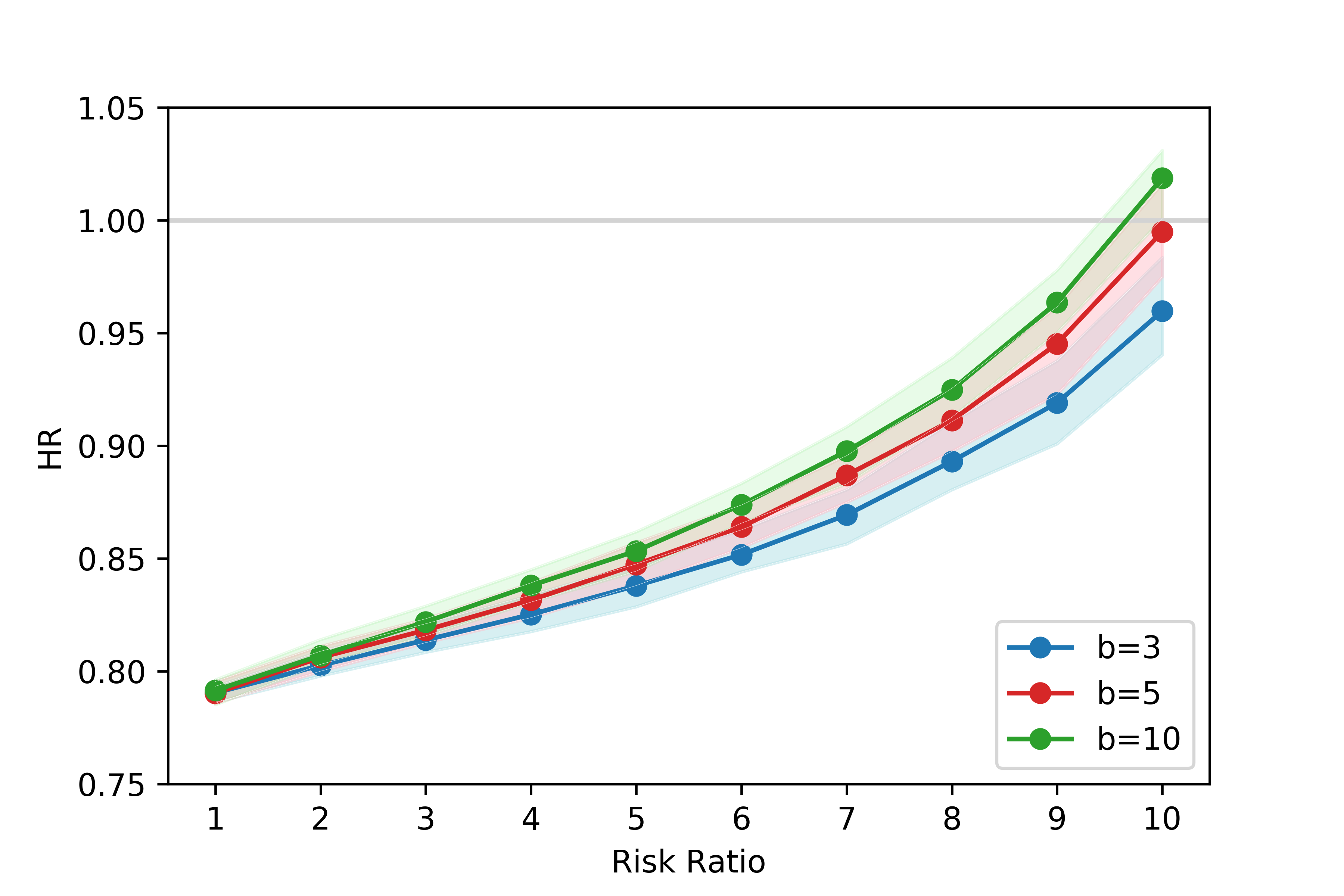}}
\subfigure[base prevalence = 0.2]{\includegraphics[width=0.47\linewidth]{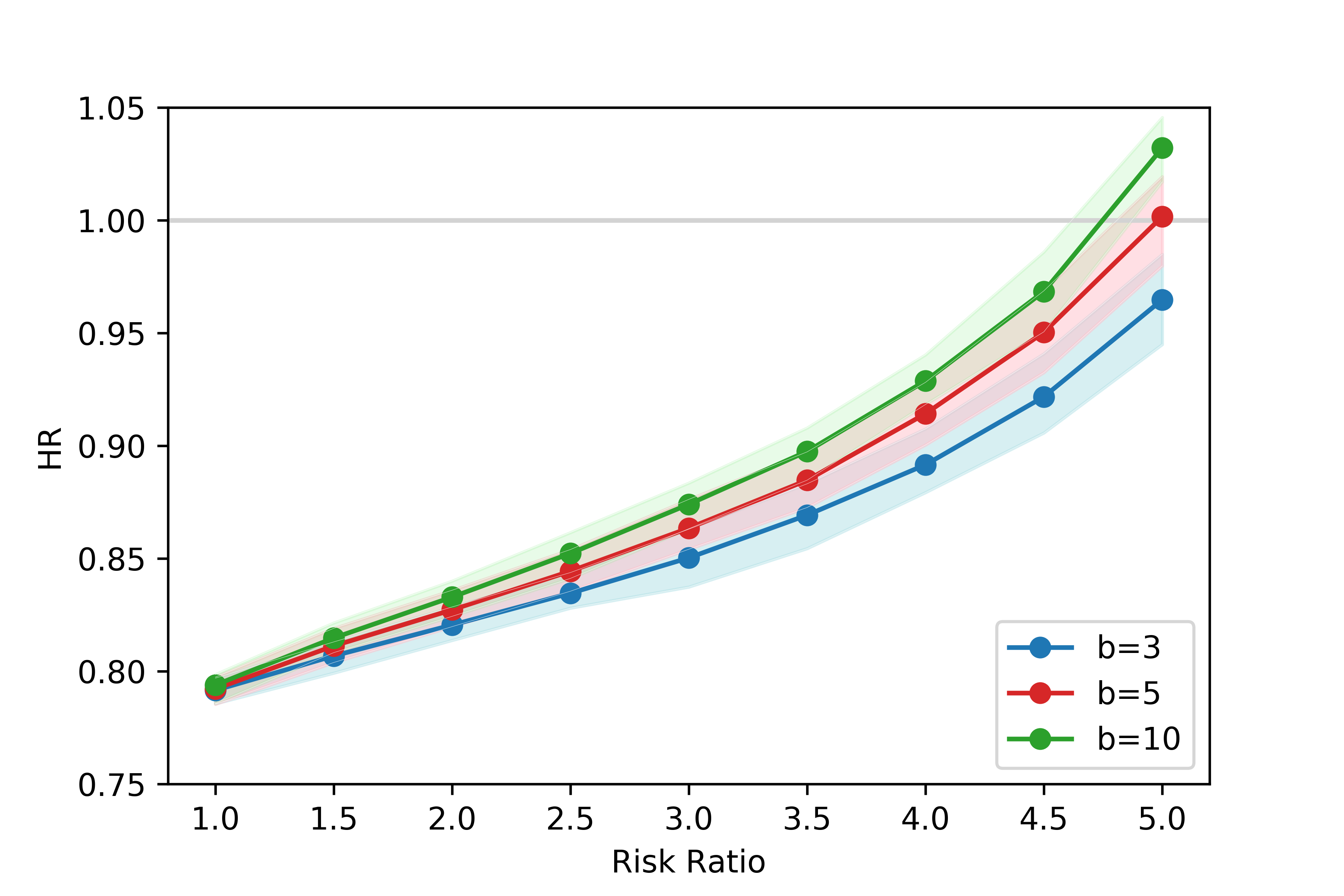}}
\subfigure[base prevalence = 0.3]{\includegraphics[width=0.47\linewidth]{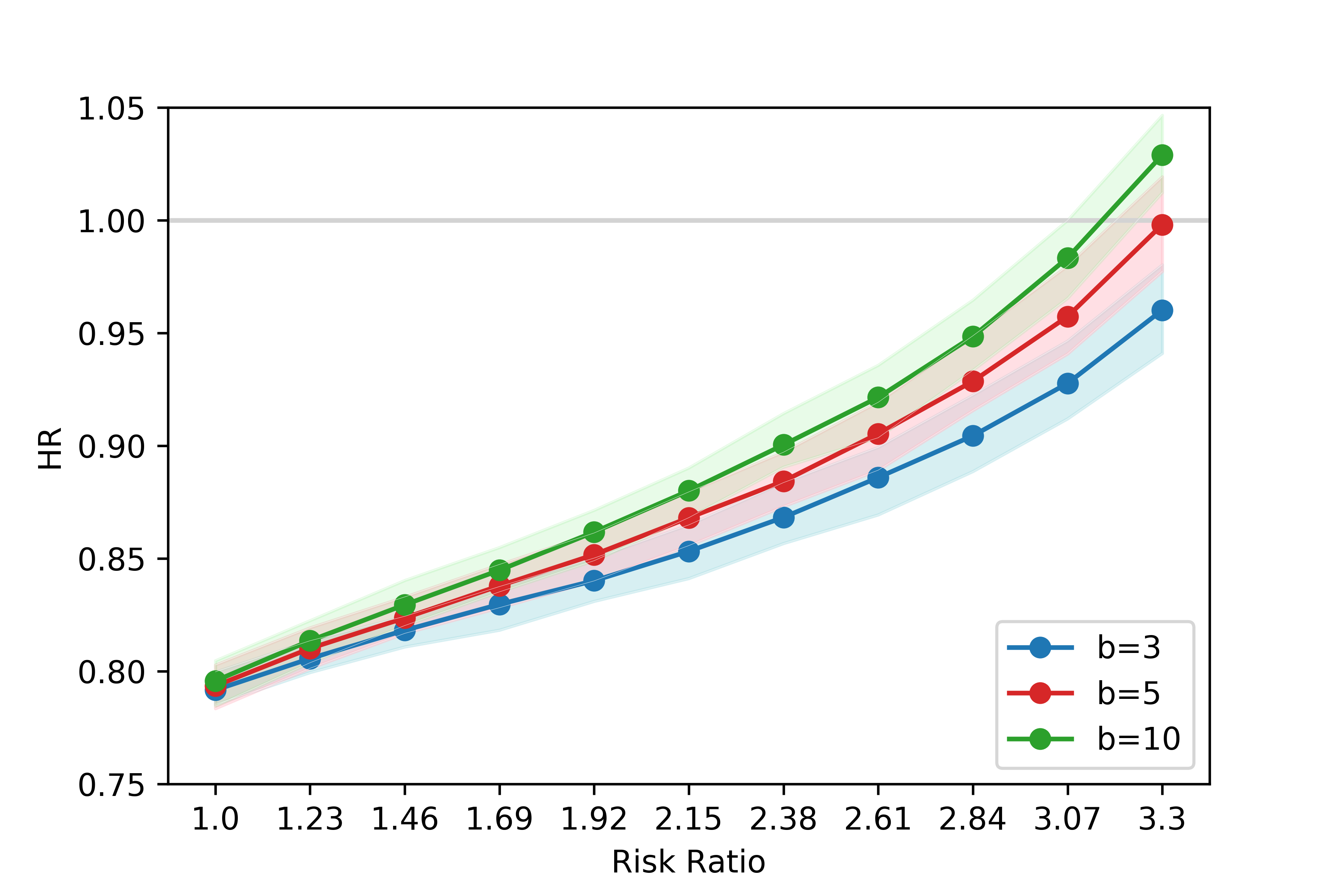}}
\subfigure[base prevalence = 0.4]{\includegraphics[width=0.47\linewidth]{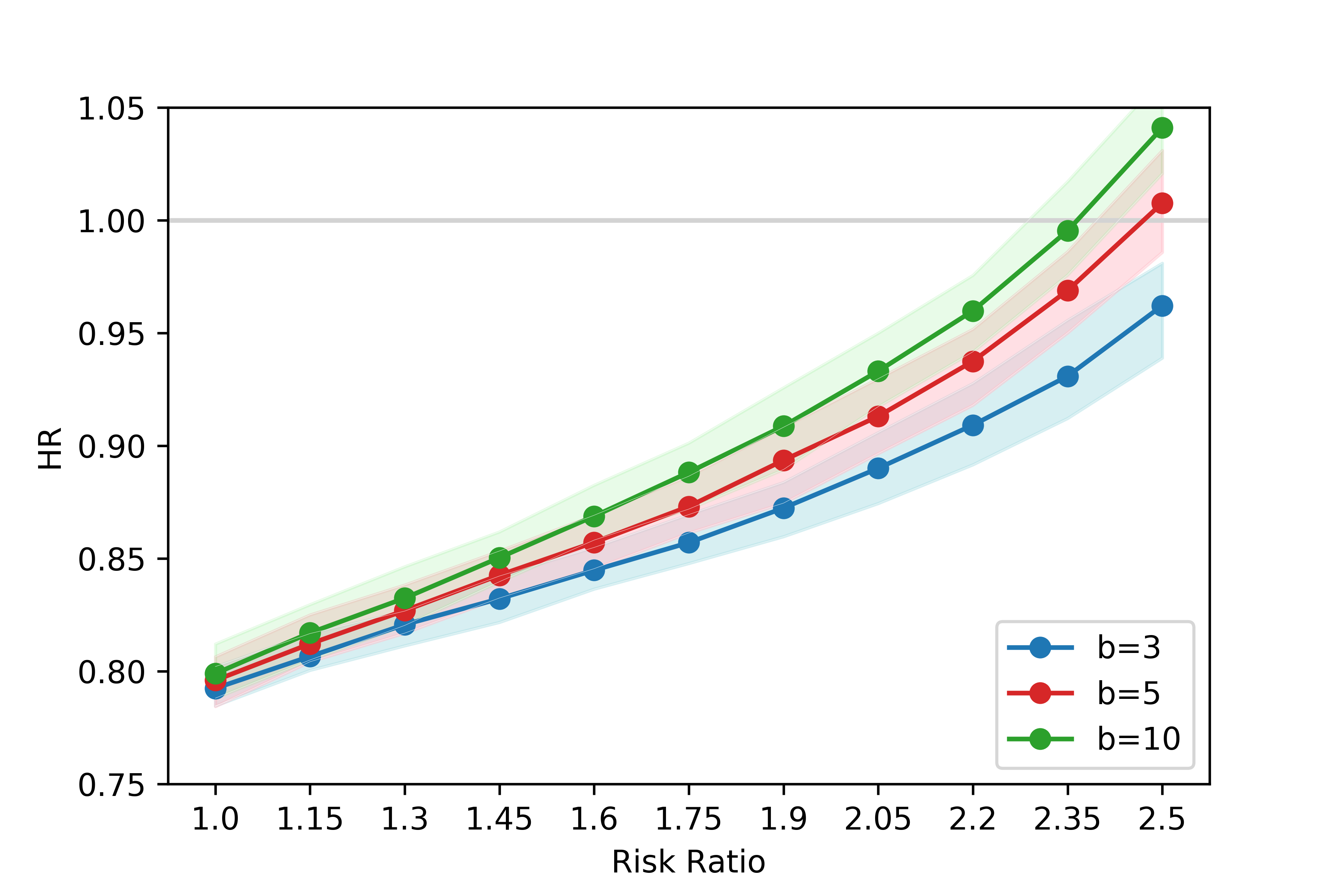}}
\caption{Sensitivity of the association between survival and concordance due to a potential unobserved confounder}
\label{fig:sensitivity}
\end{figure}

Figure \ref{fig:sensitivity} presents the results of the sensitivity analysis for four scenarios: base prevalence values of 0.1, 0.2, 0.3, and 0.4. We observe that for any base prevalence and $b$, HR increases towards 1 as $RR$ increases. Overall, only for high values of all key parameters -- $b$, $RR$, base prevalance -- does the significant association between concordance and survival vanish. For example, assuming a base prevalence of 10\%, we require a 5-10 times higher prevalence in the survived group over the died group and a 10 times higher prevalence in adjacent concordance terciles for the association to become insignificant. Given that the distribution must be very highly skewed towards the high concordance-low mortality risk individuals for our previous results to be affected, the sensitivity analysis suggests that our association is not highly sensitive to unobserved confounders.


\section{Colon Cancer Pathway Maps}
\label{app:pathway_maps}
The diagnosis and treatment pathway maps for colon cancer is shown below. Note that colon and rectum cancers (colorectal) have the same diagnosis pathway map, but the treatment pathway maps are different.



\section*{Supplementary material}

# Colorectal Cancer Diagnosis Pathway Map

Version 2020.01

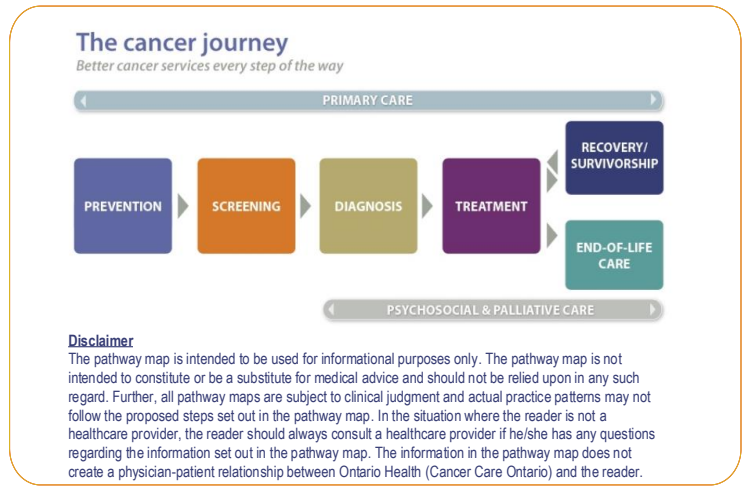

The cancer journey
*Better cancer services every step of the way*

PRIMARY CARE

PREVENTION → SCREENING → DIAGNOSIS → TREATMENT → RECOVERY/SURVIVORSHIP
END-OF-LIFE CARE

PSYCHOSOCIAL & PALLIATIVE CARE

**Disclaimer**
The pathway map is intended to be used for informational purposes only. The pathway map is not intended to constitute or be a substitute for medical advice and should not be relied upon in any such regard. Further, all pathway maps are subject to clinical judgment and actual practice patterns may not follow the proposed steps set out in the pathway map. In the situation where the reader is not a healthcare provider, the reader should always consult a healthcare provider if he/she has any questions regarding the information set out in the pathway map. The information in the pathway map does not create a physician-patient relationship between Ontario Health (Cancer Care Ontario) and the reader.

## Pathway Map Considerations

- For more information on Organized Diagnostic Assessment refer to the **Organizational Standards**
- Primary care providers play an important role in the cancer journey and should be informed of relevant tests and consultations. Ongoing care with a primary care provider is assumed to be part of the pathway map. For patients who do not have a primary care provider, **Health Care Connect**, is a government resource that helps patients find a doctor or nurse practitioner.
- Throughout the pathway map, a shared decision-making model should be implemented to enable and encourage patients to play an active role in the management of their care. For more information see **Person-Centered Care Guideline.**
- Hyperlinks are used throughout the pathway map to provide information about relevant Ontario Health (Cancer Care Ontario) tools, resources and guidance documents.
- The term 'health care provider', used throughout the pathway map, includes primary care providers and specialists, nurse practitioners, and emergency physicians.
- The pathway map is only intended for primary adenocarcinoma. Familial cancers (Lynch/non-Lynch) and cancers in the settings of inflammatory bowel disease are handled differently.

## Pathway Map Legend

### Colour Guide

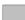 Primary Care
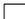 Endoscopy
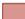 Palliative Care
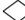 Pathology
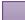 Organized Diagnostic Assessment
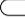 Surgery
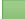 Radiation Oncology
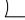 Medical Oncology
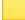 Radiology
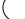 Multidisciplinary Cancer Conference (MCC)
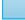 Psychosocial Oncology (PSO)

### Shape Guide

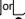 Intervention
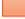 Decision or assessment point
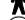 Patient (disease) characteristics
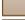 Consultation with specialist
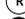 Exit pathway
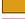 Off-page reference
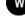 Patient/ Provider interaction
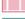 Referral
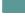 Wait time indicator time point

### Line Guide

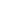 Required
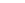 Possible

## Pathway Map Disclaimer

This pathway map is a resource that provides an overview of the treatment that an individual in the Ontario cancer system may receive.

The pathway map is intended to be used for informational purposes only. The pathway map is not intended to constitute or be a substitute for medical advice and should not be relied upon in any such regard. Further, all pathway maps are subject to clinical judgment and actual practice patterns may not follow the proposed steps set out in the pathway map. In the situation where the reader is not a healthcare provider, the reader should always consult a healthcare provider if he/she has any questions regarding the information set out in the pathway map. The information in the pathway map does not create a physician-patient relationship between Ontario Health (Cancer Care Ontario) and the reader.

While care has been taken in the preparation of the information contained in the pathway map, such information is provided on an "as-is" basis, without any representation, warranty, or condition, whether express or implied, statutory or otherwise, as to the information's quality, accuracy, currency, completeness, or reliability.

Ontario Health (Cancer Care Ontario) and the pathway map's content providers (including the physicians who contributed to the information in the pathway map) shall have no liability, whether direct, indirect, consequential, contingent, special, or incidental, related to or arising from the information in the pathway map or its use thereof, whether based on breach of contract or tort (including negligence), and even if advised of the possibility thereof. Anyone using the information in the pathway map does so at his or her own risk, and by using such information, agrees to indemnify Ontario Health (Cancer Care Ontario) and its content providers from any and all liability, loss, damages, costs and expenses (including legal fees and expenses) arising from such person's use of the information in the pathway map.

This pathway map may not reflect all the available scientific research and is not intended as an exhaustive resource. Ontario Health (Cancer Care Ontario) and its content providers assume no responsibility for omissions or incomplete information in this pathway map. It is possible that other relevant scientific findings may have been reported since completion of this pathway map. This pathway map may be superseded by an updated pathway map on the same topic.

The pathway map is intended to be used for informational purposes only. The pathway map is not intended to constitute or be a substitute for medical advice and should not be relied upon in any such regard. Further, all pathway maps are subject to clinical judgment and actual practice patterns may not follow the proposed steps set out in the pathway map. In the situation where the reader is not a healthcare provider, the reader should always consult a healthcare provider if he/she has any questions regarding the information set out in the pathway map. The information in the pathway map does not create a physician-patient relationship between Ontario Health (Cancer Care Ontario) and the reader.

**Screen for psychosocial needs, and assessment and management of symptoms. Click here for more information about symptom assessment and management tools**

**Consider the introduction of palliative care, early and across the cancer journey Click here for more information about palliative care**

**Symptomatic Patients**

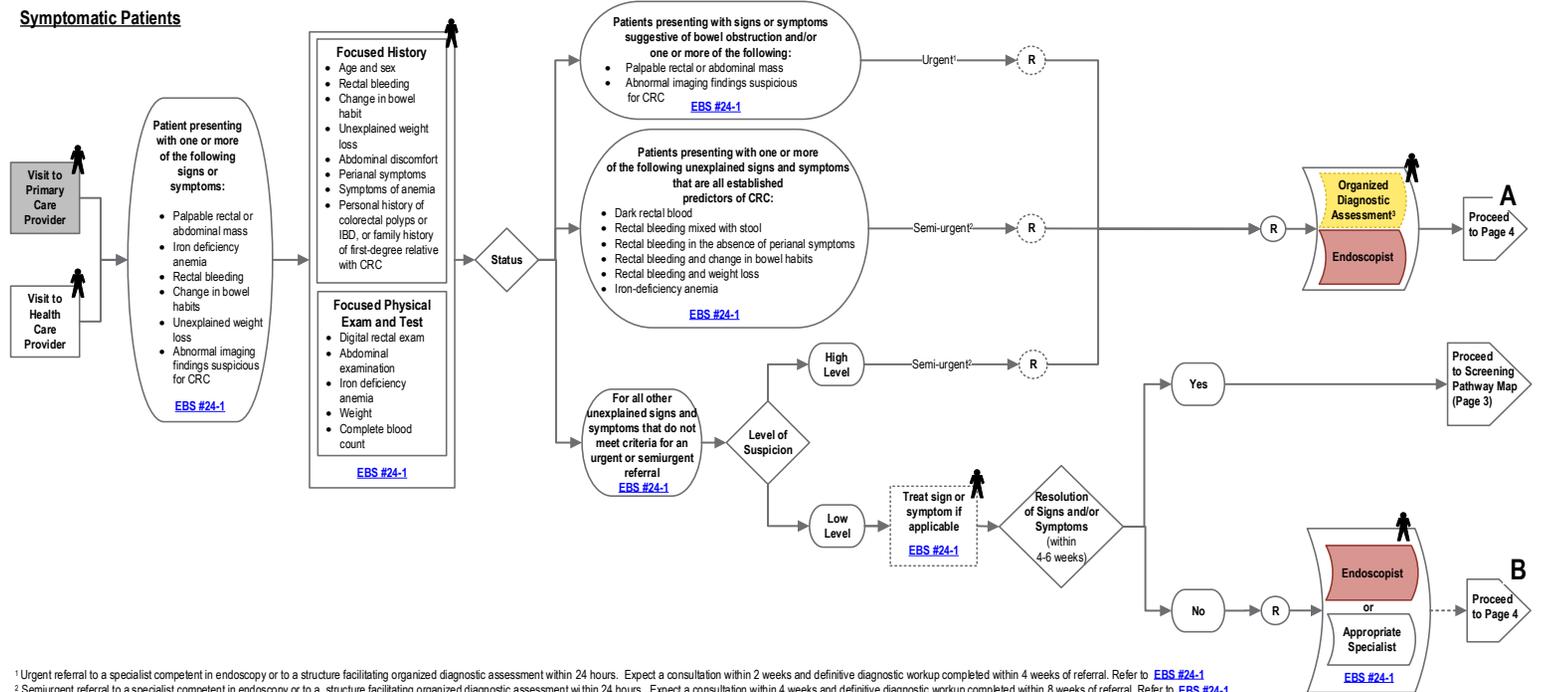

[1] Urgent referral to a specialist competent in endoscopy or to a structure facilitating organized diagnostic assessment within 24 hours. Expect a consultation within 2 weeks and definitive diagnostic workup completed within 4 weeks of referral. Refer to EBS #24-1
[2] Semiurgent referral to a specialist competent in endoscopy or to a structure facilitating organized diagnostic assessment within 24 hours. Expect a consultation within 4 weeks and definitive diagnostic workup completed within 8 weeks of referral. Refer to EBS #24-1
[3] Evaluation of patients with a high suspicion of colorectal cancer may be performed within structures facilitating organized diagnostic assessment.

Screen for psychosocial needs, and assessment and management of symptoms. Click here for more information about symptom assessment and management tools

Consider the introduction of palliative care, early and across the cancer journey Click here for more information about palliative care

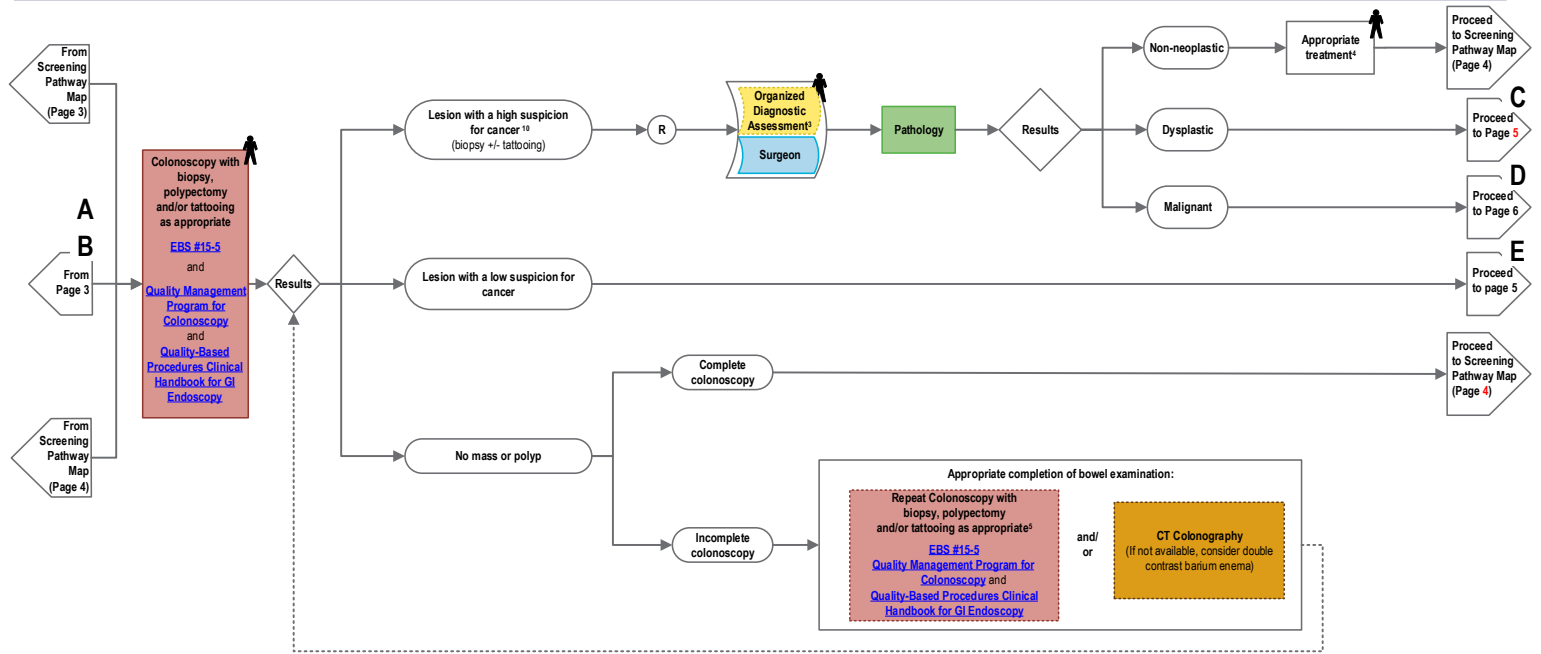

[1] Evaluation of patients with a high suspicion of colorectal cancer may be performed within structures facilitating organized diagnostic assessment.
[4] The appropriate treatment of a non-neoplastic lesion is at the discretion of the treating physician.
[5] Consider referral to a specialist endoscopist. Refer to EBS # 24-1
[10] Referral should be made for any lesions with a high suspicion for cancer regardless of inconclusive or negative biopsy result for cancer

The pathway map is intended to be used for informational purposes only. The pathway map is not intended to constitute or be a substitute for medical advice and should not be relied upon in any such regard. Further, all pathway maps are subject to clinical judgment and actual practice patterns may not follow the proposed steps set out in the pathway map. In the situation where the reader is not a healthcare provider, the reader should always consult a healthcare provider if he/she has any questions regarding the information set out in the pathway map. The information in the pathway map does not create a physician-patient relationship between Ontario Health (Cancer Care Ontario) and the reader.

Screen for psychosocial needs, and assessment and management of symptoms. Click here for more information about symptom assessment and management tools

Consider the introduction of palliative care, early and across the cancer journey Click here for more information about palliative care

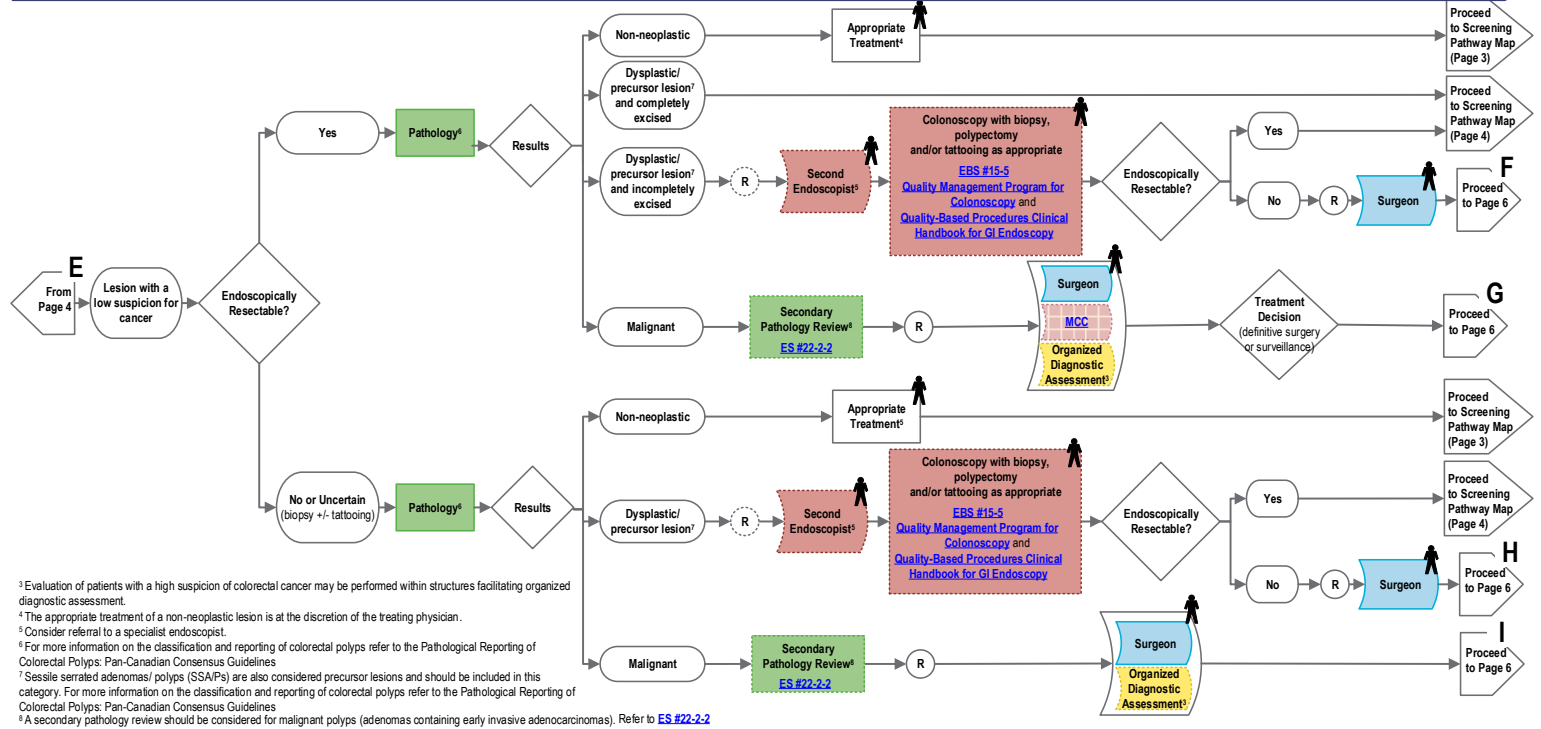

3 Evaluation of patients with a high suspicion of colorectal cancer may be performed within structures facilitating organized diagnostic assessment.

4 The appropriate treatment of a non-neoplastic lesion is at the discretion of the treating physician.

5 Consider referral to a specialist endoscopist.

6 For more information on the classification and reporting of colorectal polyps refer to the Pathological Reporting of Colorectal Polyps: Pan-Canadian Consensus Guidelines

7 Sessile serrated adenomas/ polyps (SSA/Ps) are also considered precursor lesions and should be included in this category. For more information on the classification and reporting of colorectal polyps refer to the Pathological Reporting of Colorectal Polyps: Pan-Canadian Consensus Guidelines

8 A secondary pathology review should be considered for malignant polyps (adenomas containing early invasive adenocarcinomas). Refer to ES #22-2-2

The pathway map is intended to be used for informational purposes only. The pathway map is not intended to constitute or be a substitute for medical advice and should not be relied upon in any such regard. Further, all pathway maps are subject to clinical judgment and actual practice patterns may not follow the proposed steps set out in the pathway map. In the situation where the reader is not a healthcare provider, the reader should always consult a healthcare provider if he/she has any questions regarding the information set out in the pathway map. The information in the pathway map does not create a physician-patient relationship between Ontario Health (Cancer Care Ontario) and the reader.

Screen for psychosocial needs, and assessment and management of symptoms. Click here for more information about symptom assessment and management tools

Consider the introduction of palliative care, early and across the cancer journey Click here for more information about palliative care

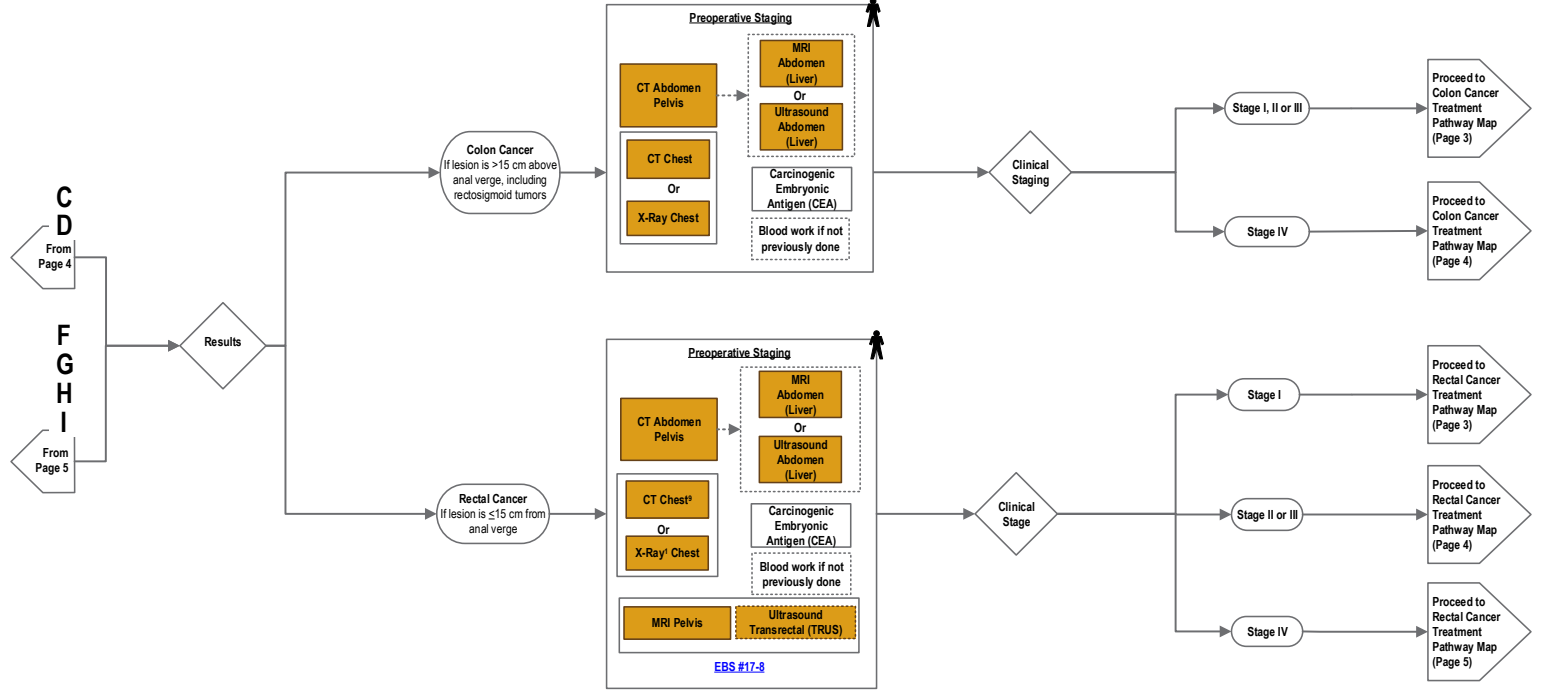

§ The choice of CT chest or chest X-ray should be consistent with the modality used for postoperative surveillance. For more information, refer to EBS #17-8

# Colon Cancer Treatment Pathway Map

Version 2020.01

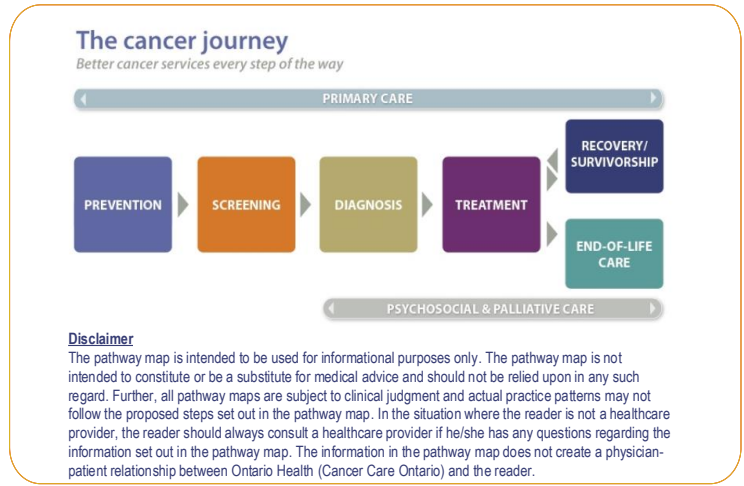

### The cancer journey
*Better cancer services every step of the way*

PRIMARY CARE

PREVENTION → SCREENING → DIAGNOSIS → TREATMENT

RECOVERY/ SURVIVORSHIP

END-OF-LIFE CARE

PSYCHOSOCIAL & PALLIATIVE CARE

**Disclaimer**
The pathway map is intended to be used for informational purposes only. The pathway map is not intended to constitute or be a substitute for medical advice and should not be relied upon in any such regard. Further, all pathway maps are subject to clinical judgment and actual practice patterns may not follow the proposed steps set out in the pathway map. In the situation where the reader is not a healthcare provider, the reader should always consult a healthcare provider if he/she has any questions regarding the information set out in the pathway map. The information in the pathway map does not create a physician-patient relationship between Ontario Health (Cancer Care Ontario) and the reader.

**Ontario Health**
Cancer Care Ontario

## Target Population

Patients with a confirmed colon cancer diagnosis who have undergone the recommended diagnostic and staging procedures as outlined in the **Colorectal Cancer Diagnosis Pathway Map.**

## Pathway Map Considerations

- All patients under consideration for an ostomy should be referred to an Enterostomal Therapy Nurse preoperatively. Patients should have access to an Enterostomal Therapy Nurse before and after ostomy surgery. Refer to:
  **Ostomy Care and Management, Clinical Best Practice Guideline, Registered Nurses Association of Ontario.**
- Primary care providers play an important role in the cancer journey and should be informed of relevant tests and consultations. Ongoing care with a primary care provider is assumed to be part of the pathway map. For patients who do not have a primary care provider, **Health Care Connect,** is a government resource that helps patients find a doctor or nurse practitioner.
- Throughout the pathway map, a shared decision-making model should be implemented to enable and encourage patients to play an active role in the management of their care. For more information see **Person-Centered Care Guideline**
- Hyperlinks are used throughout the pathway map to provide information about relevant Ontario Health (Cancer Care Ontario) tools, resources and guidance documents.
- The term 'health care provider', used throughout the pathway map, includes primary care providers and specialists, nurse practitioners, and emergency physicians.
- For more information on Multidisciplinary Cancer Conferences visit **MCC Tools**
- For more information on wait time prioritization, visit: **Surgery**
- Clinical trials should be considered for all phases of the pathway map.
- Psychosocial care should be considered an integral and standardized part of cancer care for patients and their families at all stages of the illness trajectory. For more information visit **EBS #19-3***
- The pathway map is only intended for primary adenocarcinoma. Familial cancers (Lynch/non-Lynch) and cancers in the settings of inflammatory bowel disease are handled differently.
- The following should be considered when weighing the treatment options described in this pathway map for patients with potentially life-limiting illness:
  - Palliative care may be of benefit at any stage of the cancer journey, and may enhance other types of care – including restorative or rehabilitative care – or may become the total focus of care
  - Ongoing discussions regarding goals of care is central to palliative care, and is an important part of the decision-making process. Goals of care discussions include the type, extent and goal of a treatment or care plan, where care will be provided, which health care providers will provide the care, and the patient's overall approach to care
- For more information on the systemic treatment QBP please refer to the:
  **Quality-Based Procedures Clinical Handbook for Systemic Treatment**

  *** Note.** EBS #19-3 is older than 3 years and is currently listed as 'For Education and Information Purposes'. This means that the recommendations will no longer be maintained but may still be useful for academic or other information purposes.

## Pathway Map Legend

### Colour Guide
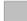 Primary Care
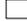 Endoscopy
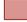 Palliative Care
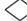 Pathology
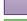 Surgery
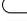 Radiation Oncology
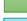 Medical Oncology
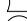 Radiology
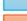 Multidisciplinary Cancer Conference (MCC)
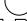 Psychosocial Oncology (PSO)

### Shape Guide
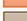 Intervention
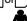 Decision or assessment point
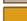 Patient (disease) characteristics
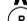 Consultation with specialist
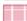 Exit pathway
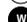 Off-page reference
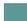 Patient/Provider interaction
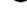 Referral
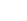 Wait time indicator time point

### Line Guide
Required
-------- Possible

## Pathway Map Disclaimer

This pathway map is a resource that provides an overview of the treatment that an individual in the Ontario cancer system may receive.

The pathway map is intended to be used for informational purposes only. The pathway map is not intended to constitute or be a substitute for medical advice and should not be relied upon in any such regard. Further, all pathway maps are subject to clinical judgment and actual practice patterns may not follow the proposed steps set out in the pathway map. In the situation where the reader is not a healthcare provider, the reader should always consult a healthcare provider if he/she has any questions regarding the information set out in the pathway map. The information in the pathway map does not create a physician-patient relationship between Ontario Health (Cancer Care Ontario) and the reader.

While care has been taken in the preparation of the information contained in the pathway map, such information is provided on an "as-is" basis, without any representation, warranty, or condition, whether express, or implied, statutory or otherwise, as to the information's quality, accuracy, currency, completeness, or reliability.

Ontario Health (Cancer Care Ontario) and the pathway map's content providers (including the physicians who contributed to the information in the pathway map) shall have no liability, whether direct, indirect, consequential, contingent, special, or incidental, related to or arising from the information in the pathway map or its use thereof, whether based on breach of contract or tort (including negligence), and even if advised of the possibility thereof. Anyone using the information in the pathway map does so at his or her own risk, and by using such information, agrees to indemnify Ontario Health (Cancer Care Ontario) and its content providers from any and all liability, loss, damages, costs and expenses (including legal fees and expenses) arising from such person's use of the information in the pathway map.

This pathway map may not reflect all the available scientific research and is not intended as an exhaustive resource. Ontario Health (Cancer Care Ontario) and its content providers assume no responsibility for omissions or incomplete information in this pathway map. It is possible that other relevant scientific findings may have been reported since completion of this pathway map. This pathway map may be superseded by an updated pathway map on the same topic.

The pathway map is intended to be used for informational purposes only. The pathway map is not intended to constitute or be a substitute for medical advice and should not be relied upon in any such regard. Further, all pathway maps are subject to clinical judgment and actual practice patterns may not follow the proposed steps set out in the pathway map. In the situation where the reader is not a healthcare provider, the reader should always consult a healthcare provider if he/she has any questions regarding the information set out in the pathway map. The information in the pathway map does not create a physician-patient relationship between Ontario Health (Cancer Care Ontario) and the reader.

**Screen for psychosocial needs, and assessment and management of symptoms.** Click here for more information about symptom assessment and management tools

**Consider the introduction of palliative care, early and across the cancer journey** Click here for more information about palliative care

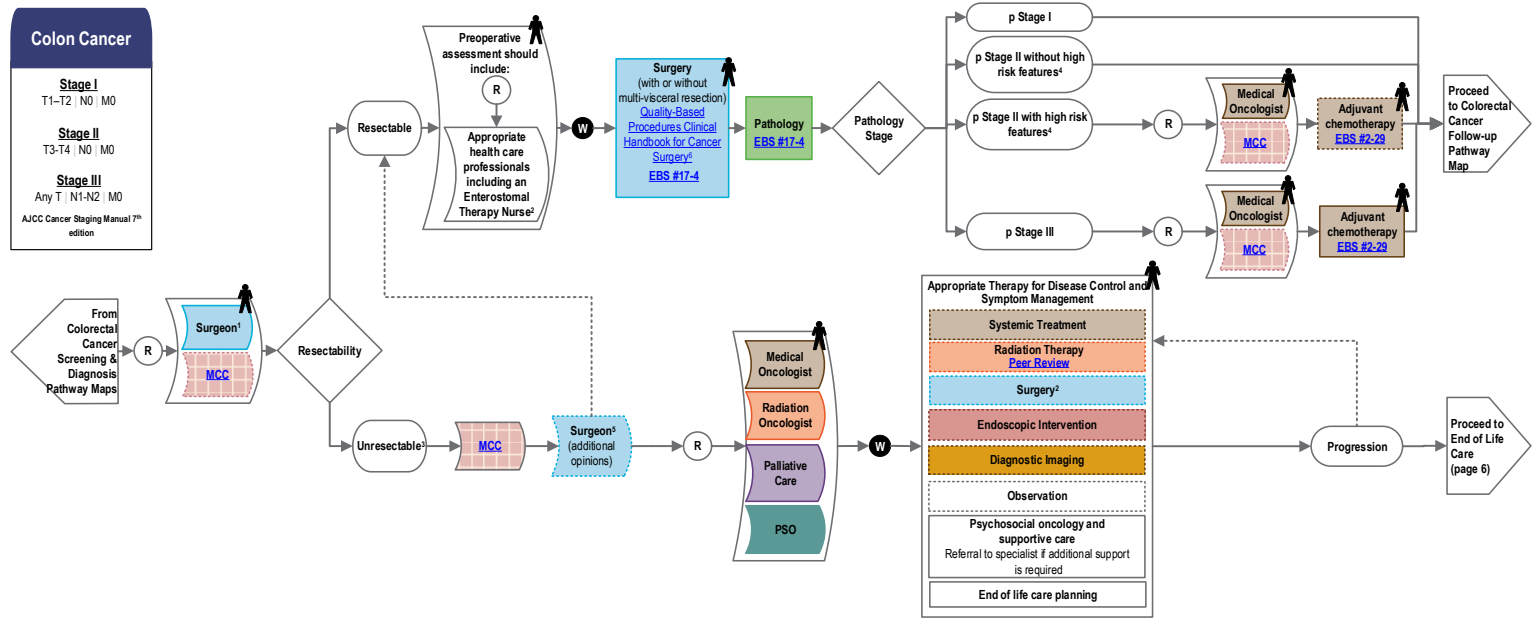

**Colon Cancer**

**Stage I**
T1–T2 : N0 : M0

**Stage II**
T3–T4 : N0 : M0

**Stage III**
Any T : N1–N2 : M0

AJCC Cancer Staging Manual 7th edition

[1] For T4 lesions, patients may also be referred to urology, plastic surgery, vascular surgery and/or hepatobiliary surgery.
[2] All patients under consideration for an ostomy should be referred to an Enterostomal Therapy Nurse preoperatively. Patients should have access to an Enterostomal Therapy Nurse before and after ostomy surgery.
[3] Unresectable refers to a tumour that cannot be completely removed even with a multivisceral resection (i.e., pelvic sidewall invasion) and/or patient is unfit for major surgery. Goals of care should be discussed. Treatment plans should be based upon MCC recommendations.
[4] High-risk features include but are not limited to: inadequate samples of nodes, T4 lesions, perforation at the site of the tumour, or poorly differentiated histology in the absence of microsatellite instability.
[5] An additional opinion from a second surgical oncologist or colorectal surgeon to reassess resectability should be considered
[6] For more information on Colorectal Cancer Surgery refer to pages 22-29 of the Quality-Based Procedures Clinical Handbook for Cancer Surgery

The pathway map is intended to be used for informational purposes only. The pathway map is not intended to constitute or be a substitute for medical advice and should not be relied upon in any such regard. Further, all pathway maps are subject to clinical judgment and actual practice patterns may not follow the proposed steps set out in the pathway map. In the situation where the reader is not a healthcare provider, the reader should always consult a healthcare provider if he/she has any questions regarding the information set out in the pathway map. The information in the pathway map does not create a physician-patient relationship between Ontario Health (Cancer Care Ontario) and the reader.

Screen for psychosocial needs, and assessment and management of symptoms. **Click here for more information about symptom assessment and management tools**

Consider the introduction of palliative care, early and across the cancer journey **Click here for more information about palliative care**

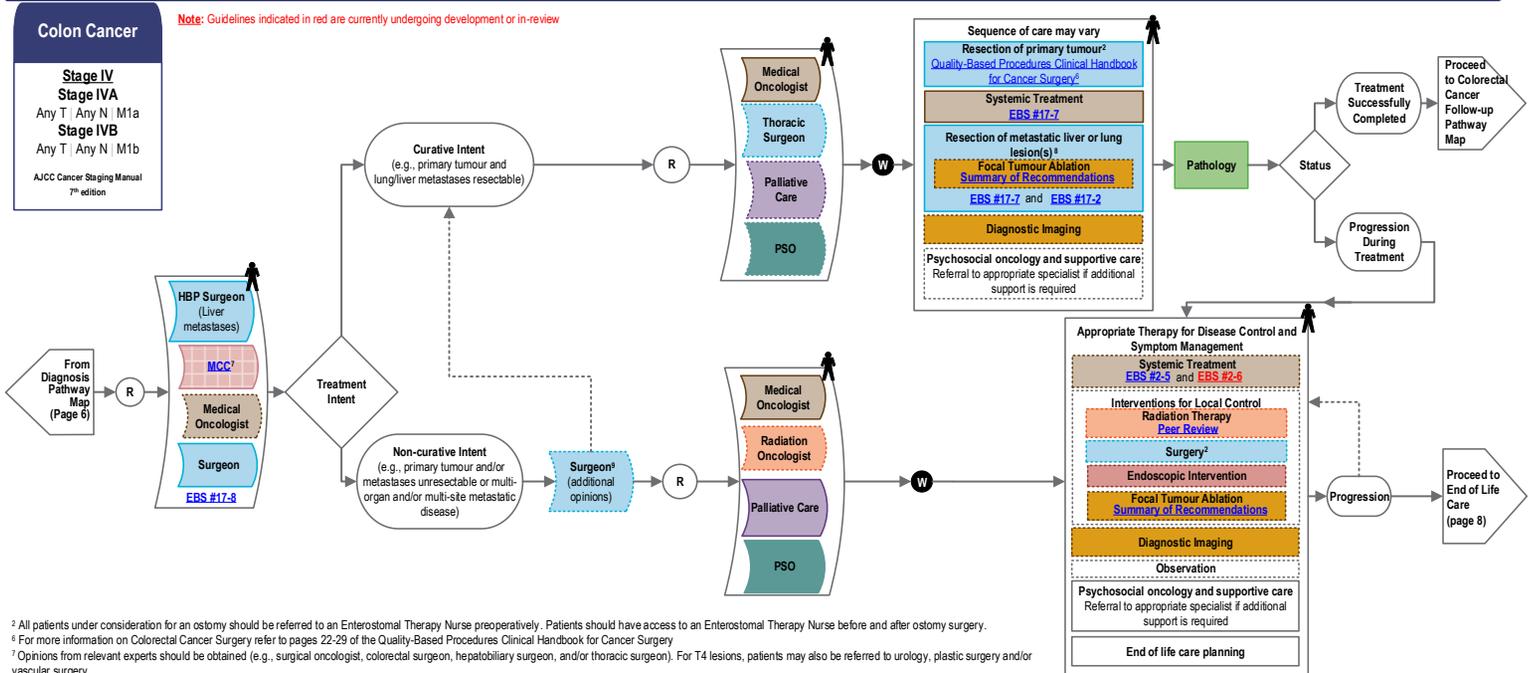

**Note:** Guidelines indicated in red are currently undergoing development or in-review

[2] All patients under consideration for an ostomy should be referred to an Enterostomal Therapy Nurse preoperatively. Patients should have access to an Enterostomal Therapy Nurse before and after ostomy surgery.

[5] For more information on Colorectal Cancer Surgery refer to pages 22-29 of the Quality-Based Procedures Clinical Handbook for Cancer Surgery

[7] Opinions from relevant experts should be obtained (e.g. surgical oncologist, colorectal surgeon, hepatobiliary surgeon, and/or thoracic surgeon). For T4 lesions, patients may also be referred to urology, plastic surgery and/or vascular surgery.

[8] Patients should be treated at a designated HPB Centre that has appropriate physical resources, staffing and a high volume of HPB surgeries. For more information on the optimum organization for the delivery of cancer-related hepatic, pancreatic, and biliary tract surgery refer to **EBS #17-2:** Hepatic, Pancreatic, and Biliary Tract Surgical Oncology Standards

[9] An additional opinion from a second surgical oncologist, colorectal surgeon, hepatobiliary surgeon or thoracic surgeon to reassess treatment intent should be considered

The pathway map is intended to be used for informational purposes only. The pathway map is not intended to constitute or be a substitute for medical advice and should not be relied upon in any such regard. Further, all pathway maps are subject to clinical judgment and actual practice patterns may not follow the proposed steps set out in the pathway map. In the situation where the reader is not a healthcare provider, the reader should always consult a healthcare provider if he/she has any questions regarding the information set out in the pathway map. The information in the pathway map does not create a physician-patient relationship between Ontario Health (Cancer Care Ontario) and the reader.

Screen for psychosocial needs, and assessment and management of symptoms. Click here for more information about symptom assessment and management tools

Consider the introduction of palliative care, early and across the cancer journey Click here for more information about palliative care

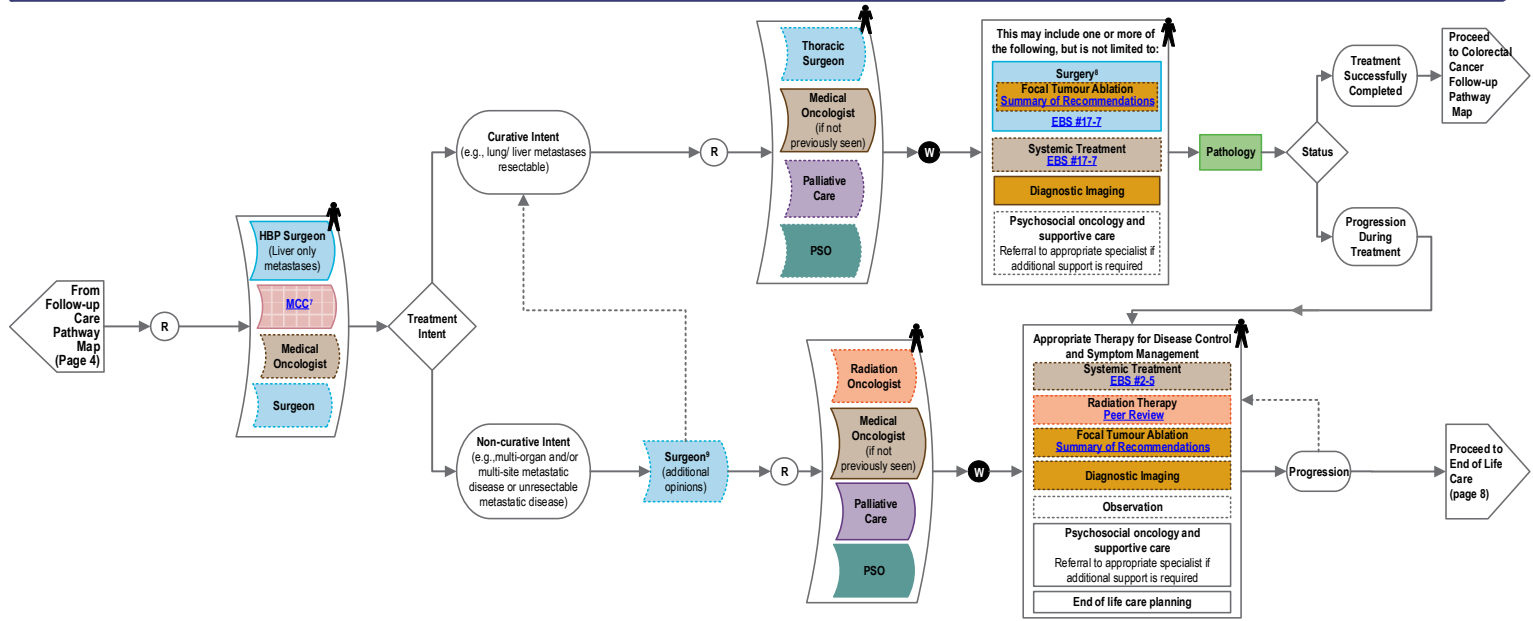

³ Opinions from relevant experts should be obtained (e.g., surgical oncologist, colorectal surgeon, hepatobiliary surgeon, and/or thoracic surgeon)
⁸ Patients should be treated at a designated HPB Centre that has appropriate physical resources, staffing  and a high volume of HPB surgeries.  For more information on the optimum organization for the delivery of cancer-related hepatic, pancreatic, and biliary tract surgery refer to EBS #17-2:  Hepatic, Pancreatic, and Biliary Tract Surgical Oncology Standards
ⁿ An additional opinion from a second surgical oncologist, colorectal surgeon, hepatobiliary surgeon or thoracic surgeon to reassess treatment intent should be considered

**Pathway Map Target Population:**

Individuals with cancer approaching the last 3 months of life and their families.

While this section of the pathway is focused on the care delivered at the end of life, palliative care should be initiated much earlier in the illness trajectory. In particular, providers can introduce a palliative approach to care as early as the time of diagnosis.

**Triggers that suggest patients are nearing the last few months and weeks of life**

- ECOG/Patient-ECOG/PRFS = 4 OR
- PPS ≤ 50
- Declining performance status/ functional ability

**Screen, Assess, Plan, Manage and Follow-Up**

➕

**End of Life Care planning and implementation** Collaboration and consultation between specialist-level care teams and primary care teams

➕

Conversations to determine where care should be provided, and who will be responsible for providing the care

## End of Life Care

☐ **Key conversations to revisit Goals of Care and to discuss and document key treatment decisions**
- Assess and address patient and family's information needs and understanding of the disease, address gaps between reality and expectation, foster realistic hope and provide opportunity to explore prognosis and life expectancy, and preparedness for death
- Explore the patient's views on medications, tests, resuscitation, intensive care and preferred location of death
- If a patient makes any treatment decisions relevant to their current condition (i.e., provides consent), these decisions can be incorporated into their Plan of Treatment
- Review Goals of Care, and patient preferences regularly, particularly when there is a change in clinical status

☐ **Screen for specific end of life psychosocial issues**
- Assess and address patient and families' loss, grief and bereavement needs including anticipatory grief, past trauma or losses, preparing children (young children, adolescents, young adults), guardianship of children, death anxiety
- Provide appropriate guidance, support and information to families, caregivers, and others, based on awareness of culture and needs, and make referrals to available resources and/or specialized services to address identified needs as required
- Identify family members at risk for abnormal/complicated grieving and connect them proactively with bereavement resources

☐ **Identify patients who could benefit from specialized palliative care services (consultation or transfer)**
- As patient and caregivers needs increase and/or change over time consult with palliative care specialists and/or other providers with additional expertise, as required. Transfer care only if/when needs become more extensive or complex than the current team can handle
- Discuss referral with the patient and their family/caregiver

☐ **Proactively develop and implement a plan for expected death**
- Explore place-of-death preferences and the resources required (e.g., home, hospice, palliative care unit, long term care or nursing home) to assess whether this is realistic
- Prepare and support the family to understand what to expect, and plan for when a loved one is actively dying, including understanding probable symptoms, as well as the processes with death certification and how to engage funeral services
- Discuss emergency plans with patient and family (including who to contact, and when to use or avoid Emergency Medical Services)

☐ **Home care planning (if this is where care will be delivered)**
- Contact the patient's primary care and home and community care providers and relevant specialist physicians to ensure an effective transfer of information related to their care. If the patient is transitioning from the hospital, this should include collaborating to develop a transition plan
- Introduce patient and family to resources in community (e.g., just in the last 2-4 weeks)
- Connect with home and community care services early (not just in the last 2-4 weeks)
- Ensure resources and services are in place to support the patient and their family/caregiver, and address identified needs
- Anticipate/plan for pain and symptom management, including consideration for a Symptom Response Kit to facilitate access to pain, dyspnea, and delirium medication for emergency purposes
- If the patient consents to withholding cardiopulmonary resuscitation, A 'Do Not Resuscitate' order must be documented in their medical record, and a Do Not Resuscitate Confirmation (DNR-C) Form should be completed. This form should be readily accessible in the home, to ensure that the patient's wishes for a natural death are respected by Emergency Medical Services

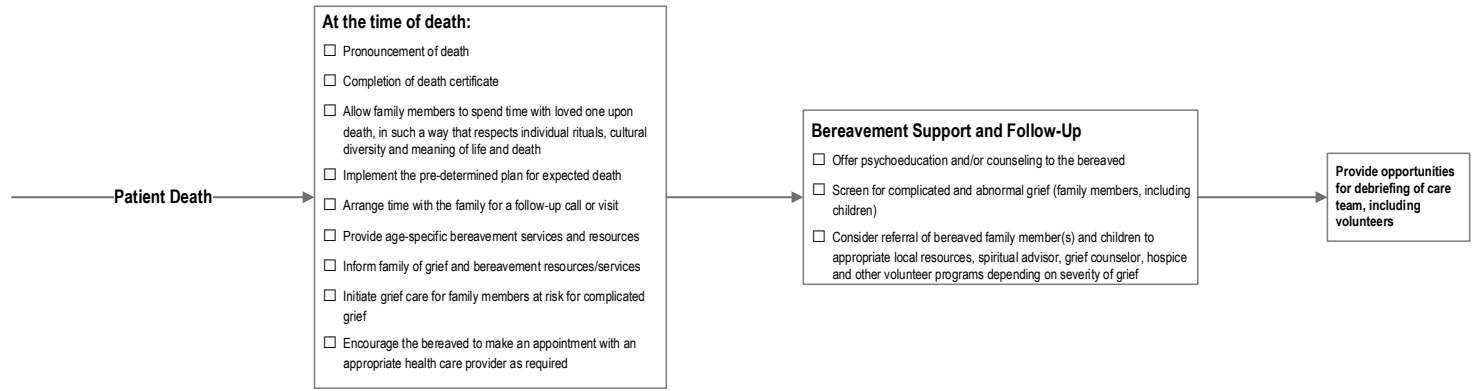

**Patient Death** →

**At the time of death:**

☐ Pronouncement of death

☐ Completion of death certificate

☐ Allow family members to spend time with loved one upon death, in such a way that respects individual rituals, cultural diversity and meaning of life and death

☐ Implement the pre-determined plan for expected death

☐ Arrange time with the family for a follow-up call or visit

☐ Provide age-specific bereavement services and resources

☐ Inform family of grief and bereavement resources/services

☐ Initiate grief care for family members at risk for complicated grief

☐ Encourage the bereaved to make an appointment with an appropriate health care provider as required

**Bereavement Support and Follow-Up**

☐ Offer psychoeducation and/or counseling to the bereaved

☐ Screen for complicated and abnormal grief (family members, including children)

☐ Consider referral of bereaved family member(s) and children to appropriate local resources, spiritual advisor, grief counselor, hospice and other volunteer programs depending on severity of grief

**Provide opportunities for debriefing of care team, including volunteers**



\bibliographystyle{informs2014}
\bibliography{references}

\end{document}